\documentclass[aps,pra,twocolumn,showpacs,showkeys,superscriptaddress,nobalancelastpage,nofootinbib,floatfix,longbibliography]{revtex4-2}
\usepackage{graphics}
\usepackage{graphicx}
\usepackage{dcolumn}
\usepackage{bm}
\usepackage{comment} 
\usepackage{xcolor}
\usepackage{bookmark}
\usepackage{tensor}
\usepackage{amsmath}
\usepackage{amssymb}
\usepackage{dsfont}
\usepackage{float}
\usepackage[utf8]{inputenc}
\usepackage{romannum}
\usepackage{moreverb}
\usepackage{subfigure}
\usepackage{hyperref}
\usepackage[capitalize]{cleveref}

\begin{document}
\pagenumbering{arabic}
\newcommand{\ket}[1]{\left| #1 \right\rangle}
\newcommand{\bra}[1]{\left\langle #1 \right|}
\newcommand{\braket}[1]{\left\langle #1 \right\rangle}
\newcommand{\expect}[1]{\langle #1 \rangle}
\newcommand{\inner}[2]{\langle{#1}|{#2}\rangle}
\newcommand{\ketbra}[2]{\mathinner{|{#1}\rangle\langle{#2}|}} 
\newcommand{\braopket}[3]{\langle{#1}|{#2}|{#3}\rangle}
\newcommand{\beq}{\begin{equation}}
\newcommand{\eeq}{\end{equation}}
\newcommand{\bea}{\begin{align}}
\newcommand{\eea}{\end{align}}
\newcommand{\bq}{\begin{quote}}
\newcommand{\eq}{\end{quote}}
\newcommand{\oem}{\color{blue}}
\newcommand{\blk}{\color{black}}
\newcommand{\Tr}{\operatorname{Tr}}
\newcommand{\erfi}{\operatorname{erfi}}
\newcommand{\dd}{\mathrm{d}}

 \newtheorem{thm}{Theorem}
 \newtheorem{cor}[thm]{Corollary}
 \newtheorem{lem}[thm]{Lemma}
 \newtheorem{prop}[thm]{Proposition}
 \newtheorem{defn}[thm]{Definition}
 \newtheorem{rem}[thm]{Remark}

\pagestyle{plain}
\title{\bf Finite-Size Effects in Quantum Metrology at Strong Coupling: Microscopic vs Phenomenological Approaches}
\author{ Ali Pedram}
\email{apedram19@ku.edu.tr}
\affiliation{Department of Physics, Koç University, Istanbul, Sarıyer 34450, Türkiye}
\affiliation{Korea Research Institute of Standards and Science, Daejeon 34113, Republic of Korea}
\affiliation{Institute for Quantum Science and Technology, University of Calgary, Alberta T3A 0E1, Canada}
\author{Özgür E. Müstecaplıoğlu}
\email{omustecap@ku.edu.tr}
\affiliation{Department of Physics, Koç University, Istanbul, Sarıyer 34450, Türkiye}
\affiliation{TÜBİTAK Research Institute for Fundamental Sciences (TBAE), 41470 Gebze, Türkiye}
\pagebreak

\begin{abstract}
Equilibrium quantum metrology is often formulated under assumptions of weak system-bath coupling and thermodynamic limit scaling, which may not be valid for realistic finite-size (FS) quantum devices. Here, we establish how strong coupling (SC) and FS effects jointly modify achievable precision bounds for equilibrium quantum probes. Considering a transverse-field anisotropic XY spin chain coupled to a thermal environment, first utilize we derive a full polaron transform to derive the effective Hamiltonian at equilibrium. Then, using the bare Hamiltonian and the obtained effective Hamiltonian and taking into account the role of FS effects in the calculation of partition function, we derive an analytic expressions for the quantum Fisher information (QFI) for a general parameter for the equilibrium state in weak coupling (WC) and SC regimes, respectively. In contrast to this microscopic approach, we also utilize the Hill's phenomenological nanothermodynamics to calculate an effective QFI expression at SC. Our results, presented for specific cases of magnetometry and thermometry, reveal that considering FS and SC effects lead to substantial and systematic deviations in QFI, even for moderate system sizes, and that commonly used thermodynamic-limit approximations can significantly misestimate precision. This work also highlights the inadequacy of phenomenological approaches in describing the metrological capability and thermodynamic behavior of systems at SC. Our work establishes a consistent framework for evaluating metrological performance of realistic, finite quantum devices operating beyond the weak-coupling paradigm.
\end{abstract}
\maketitle


\section{Introduction}
Quantum metrology, a cornerstone of quantum information science, leverages quantum resources like entanglement and squeezing to surpass the precision limits of classical measurement schemes~\cite{Liu_2020,Tóth_2014}. This promise has driven extensive theoretical and experimental efforts across diverse platforms, including atomic, optical, superconducting, and solid-state systems, consistently demonstrating the superior measurement capabilities of quantum probes~\cite{RevModPhys.89.035002,10.1116/5.0007577,PRXQuantum.3.010202,RevModPhys.90.035005,MONTENEGRO20251,PhysRevResearch.4.033060,PhysRevX.10.031003, Pedram_2024,https://doi.org/10.1002/qute.202400059,PhysRevLett.132.250204}. Metrology protocols generally fall into two categories: dynamical, where the initial state, interaction time, and measurement operators are optimized; and steady-state, where the probe equilibrates before measurement. In practical scenarios, probes inevitably interact with their environment. When the probe's relaxation timescale is shorter than other relevant timescales, such as interaction or measurement, equilibrium metrology becomes not just relevant, but essential for accurately quantifying achievable precision.

For small systems in equilibrium, the system's interface plays a non-negligible role compared to its bulk in its thermodynamic behavior~\cite{10.1116/5.0073853,RevModPhys.92.041002}. For such systems peculiar properties such as non-additivity of thermodynamic potentials, temperature dependence of energy levels and negative heat capacity can be observed~\cite{RevModPhys.92.041002,Carignano_2010,Kolar2019,CAMPISI2010187,E_W_Elcock_1957,C5CP02332G,Tang2023,PhysRevB.111.024108}. ``Smallness'' of a system can manifest in two ways: either the interaction energy between the system and its environment becomes comparable to the system's bare energy~\cite{RevModPhys.92.041002}, or the characteristic interaction lengths within the system itself become comparable to its overall size~\cite{10.1093/acprof:oso/9780199581931.001.0001}. In the former perspective for the small systems, the equilibrium state can no longer be described by a canonical Gibbs state and a microscopic perspective is used to obtain the Hamiltonian of mean force (HMF) or an effective Hamiltonian for the system at strong coupling~\cite{10.1116/5.0073853,RevModPhys.92.041002,PhysRevLett.127.250601, PhysRevB.108.115437} to describe its equilibrium properties. In the latter perspective, research is mainly focused on using a phenomenological approach, e.g. Hill's nanothermodynamics~\cite{hill1994thermodynamics,10.1063/1.1732447,Hill2001,e17010052,deMiguel2016, nano10122471, doi:10.1021/acs.jpcb.3c01525}, Tsallis' non-extensive thermodynamics~\cite{Tsallis1988,tsallis2009introduction, GARCIAMORALES200582,MANIOTIS2025116285} or Kac prescription~\cite{10.1063/1.1703946,RevModPhys.95.035002} to capture non-additive nature of nanosystems. Additionally for systems with few particles, since thermodynamic fluctuations scale as $1/\sqrt{N}$ and surface to volume ratio scales as $1/N$, it is important to take finite-size (FS) effects into account~\cite{Guisbiers01012019}.

Several approaches have been developed to describe the dynamics and steady-states of the quantum systems beyond the weak-coupling regime. For equilibrium properties, the Hamiltonian of mean force provides the formally exact reduced thermal state at arbitrary coupling~\cite{RevModPhys.92.041002,PhysRevLett.102.210401}, while polaron and variational-polaron transformations absorb part of the system-bath interaction into renormalized system parameters~\cite{10.1063/1.447055,Xu2016}. Reaction-coordinate mappings incorporate a collective environmental mode into an enlarged system, transferring strong coupling and bath memory to this super-system; the mapping is exact, although standard RC master-equation treatments subsequently treat the residual bath perturbatively~\cite{PhysRevA.90.032114,Nazir2018}. Similarly, star-to-chain mappings exactly transform bosonic reservoirs into one-dimensional chains suitable for tensor-network methods~\cite{PhysRevLett.105.050404,10.1063/1.3490188}. Finite-temperature extensions include thermofield chain mapping~\cite{PhysRevA.92.052116} and thermalized TEDOPA~\cite{PhysRevLett.123.090402}, with applications to strongly coupled non-equilibrium transport~\cite{PhysRevE.102.012155}. Another important class comprises methods based on the Feynman-Vernon influence functional, including the hierarchical equations of motion (HEOM)~\cite{doi:10.1143/JPSJ.58.101,10.1063/5.0011599} and path-integral schemes such as the quasi-adiabatic propagator path integral (QUAPI) and its tensor-network extension TEMPO~\cite{MAKAROV1994482,Strathearn2018}. Other non-perturbative approaches include stochastic formulations~\cite{PhysRevLett.88.170407} and pseudomode constructions~\cite{PhysRevA.55.2290,PhysRevLett.120.030402}. More recently, the minimal-dissipation principle has been used to formulate nonequilibrium thermodynamics at arbitrary coupling from the exact time-local reduced generator~\cite{PhysRevA.105.052216}. In~\cite{PhysRevB.108.115437, PhysRevLett.132.266701} by sequential application of reaction coordinate mapping and polaron transformation on the reaction coordinate, an effective Hamiltonian is obtained. 

SC between a quantum system and its environment is attainable in a variety of physical platforms, including quantum dots~\cite{PhysRevLett.83.4152,PhysRevLett.104.017402, PhysRevB.85.115309,Khosla2018,PhysRevLett.89.186802,PhysRevB.78.153301, Xu2016,Vora2015,PhysRevLett.129.227702}, trapped ions~\cite{Colla2025}, superconducting qubits~\cite{Jones2013,3v35-y762}, and NV centers. Although NV centers generally operate in the WC regime when diamond lattice vibrations are regarded as the thermal bath, local vibrational modes have been shown to contribute strongly to their environmental spectrum~\cite{Huxter2013}. Moreover, phonons are not the only environment relevant to NV-center relaxation. The spin bath arising from naturally occurring \(^{13}\mathrm{C}\) nuclei can couple strongly to NV centers and, in particular to P1 centers, significantly modifying their relaxation dynamics~\cite{PhysRevB.90.075201,PhysRevB.110.L220302}. More generally, when a quantum system is strongly coupled to an auxiliary degree of freedom which itself interacts weakly with a thermal reservoir, neither its reduced dynamics nor its equilibrium state are in general described by a conventional GKLS master equation or by the Gibbs state of the bare system Hamiltonian. This is closely related to the reaction-coordinate mapping commonly used to treat strong system-environment coupling~\cite{Nazir2018}. Engineered environments are widely used to investigate open-system dynamics of circuit-QED systems beyond the WC approximation~\cite{PhysRevA.84.043832,Pilar2020thermodynamicsof,https://doi.org/10.1002/andp.202200502,Tuorila2017}. Finally, while the assumption of an effectively infinite heat bath is accurate for many open quantum systems, it can break down for mesoscopic environments~\cite{PRXQuantum.2.010340}. In such cases, generalized Hamiltonian of mean force approaches have been shown to provide an accurate description of equilibrium states in the presence of finite baths~\cite{PhysRevE.111.024139,10.1063/5.0254029}. 

A considerable body of literature exists on the metrological capability of equilibrium states of quantum systems at weak coupling (WC)~\cite{PhysRevA.82.011611,PhysRevLett.114.220405,Paris_2016,Hauke2016,PhysRevA.94.042121,PhysRevLett.120.260503,Carollo_2019,PhysRevLett.133.040802, Abiuso_2024,PhysRevA.109.L050601,PhysRevLett.134.010801}. However, a comprehensive understanding of metrological precision in the strong coupling (SC) regime~\cite{PhysRevA.96.062103,Potts2019fundamentallimits,PhysRevLett.128.040502,PhysRevA.107.042614, PhysRevA.108.032220,Ravell_Rodriguez_2024,PhysRevA.111.052623,PhysRevResearch.7.023235,PRXQuantum.6.020320}, particularly when accounting for finite-size effects and non-classical equilibrium states, remains a critical outstanding problem. This work addresses this fundamental gap. By employing a full polaron transform to rigorously derive the effective Hamiltonian for a spin chain system strongly coupled to a heat bath, we move beyond perturbative methods. We then derive analytical expressions for the quantum Fisher information (QFI) for equilibrium states in both WC and SC regimes, explicitly incorporating FS effects. Furthermore, to investigate the feasibility of utilizing phenomenological methods in SC thermodynamics as suggested in~\cite{nano10122471,doi:10.1021/acs.jpcb.3c01525}, we utilize Landsberg's framework for temperature-dependent energy levels (TDEL)~\cite{E_W_Elcock_1957} and Hill's nanothermodynamics~\cite{hill1994thermodynamics,10.1063/1.1732447,Hill2001}, to obtain an effective expression for QFI. Our findings demonstrate that by controlling anisotropy parameter one can enhance magnetometric precision for the spin chain. Moreover, we show that SC reduces thermometry precision for higher temperatures while it can improve it for lower temperature values. Our work demonstrates that it is crucial to take into account FS effects to accurately quantify QFI.
\section{The Model}
\label{sec:model}
\begin{figure}[htbp!]
  \centering
  \includegraphics[width=0.65\linewidth]{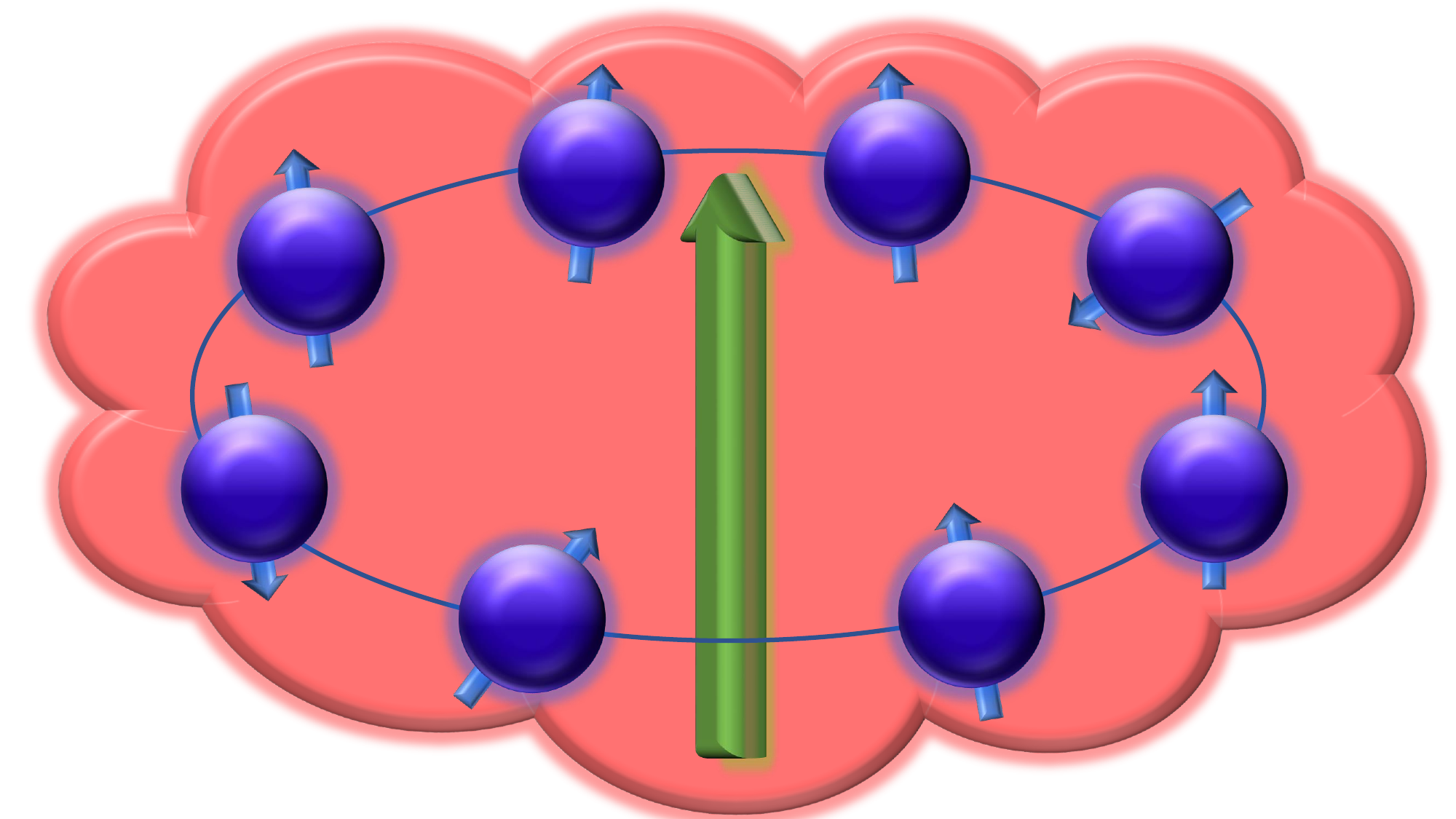}
  \caption{Schematic sketch of a spin system in equilibrium with a heat bath. The spin system under consideration is an anisotropic XY model with a transverse magnetic field.}\label{fig:schematic_model}
\end{figure}

We take the one-dimensional spin-$1/2$ anisotropic XY model in a transverse field with periodic boundary conditions.
\begin{equation}\label{xy_hamiltonian} 
\begin{split}
    \hat{H}_S=&-\frac{J}{2}\sum_{n=1}^{N}\left[(1+\gamma)\hat{\sigma}_n^{x}\hat{\sigma}_{n+1}^{x}+(1-\gamma)\hat{\sigma}_n^{y}\hat{\sigma}_{n+1}^{y}\right]\\
    &-h\sum_{n=1}^{N}\hat{\sigma}_n^{z}.
\end{split}
\end{equation}
in which $N$ is the number of sites (spins), $J$ is the exchange interaction for the neighbouring spins and $\gamma$ is the anisotropy parameter. The spin chain is taken to be in equilibrium with a heat bath at inverse temperature $\beta$. A schematic representation of the system is given in~\cref{fig:schematic_model}.
For a small system, as explained in~\cref{sec:small_thermo}, the equilibrium state will be given by mean force Gibbs state (MFG). Here we aim to find an approximation to the HMF by incorporating the effect of the bath and system-bath interaction in the system Hamiltonian. The system nodes can couple to a single heat bath collectively (global coupling) or each system can couple to a different independent heat baths of the same temperature (local coupling). However, if the spacing between nodes of the chain is larger than the spatial correlation length of the environment, we can effectively assume that the system couples locally to the bath~\cite{PhysRevLett.101.175701,PhysRevB.80.024302,PhysRevA.87.052138}. Under this local coupling assumption, the total Hamiltonian for the system and its environment is given by $\hat{H}_{\text{tot}} = \hat{H}_S+\hat{H}_B+\hat{H}_I$, where the bath and system-bath interaction Hamiltonians are
\begin{eqnarray}\label{loc_hamiltonian} 
	\hat{H}_{\text{B}} &=& \sum_{n=1}^N\sum_k \nu_{n,k} \hat{b}_{n,k}^{\dag}\hat{b}_{n,k},\\
	\hat{H}_{\text{I}} &=& \sum_{n=1}^N\sum_k
	t_{n,k}\hat{\sigma}_n^{x}(\hat{b}_{n,k}^{\dag}+\hat{b}_{n,k}).
\end{eqnarray}
in which $\hat{b}_{n,k}$ and $\nu_{n,k}$ are the annihilation operators and frequencies of the $k$-th bath mode for site $n$. $t_{n,k}$ are the coupling system-bath coupling constants with the coupling operator being $\hat{\sigma}_n^{x}$ for each site. The effective Hamiltonian for the system after full polaron transformation via the unitary operators $\hat{W}_n=\text{exp}(-i\sigma_n^x\hat{B}_n/2)$ on each site with $\hat{B}_n=2i\sum_{k}t_{n,k}/\nu_k(\hat{b}_k^{\dag}-\hat{b}_k)$~\cite{PhysRevB.108.115437,PhysRevLett.132.266701}. Note that the polaron transformation is independent of the parameters of the system and temperature. Therefore, the QFI in the qubit and polaron basis for $h$ and $\beta$ will give the same results. The effective Hamiltonian becomes 
\begin{equation}\label{pol_hamiltonian}
\begin{split}
  \hat{H}_S^{\flat} = &-\frac{J}{2}\sum_{n=1}^{N}\left[(1+\gamma)\hat{\sigma}_n^{x}\hat{\sigma}_{n+1}^{x}+(1-\gamma)\expect{\hat{\mathcal{C}}}^2\hat{\sigma}_n^{y}\hat{\sigma}_{n+1}^{y}\right]\\ &- h\expect{\hat{\mathcal{C}}}\sum_{n=1}^{N}\hat{\sigma}_n^{z}.
\end{split}
\end{equation}
Our goal is to approximate the MFG as the thermal state of the effective Hamiltonian of the system after polaron transformation, $\hat{H}_S^{\flat}$. For this approximation to hold, the system-bath coupling needs to be weak in the polaron frame, and as a result the second order perturbation theory must be valid in this frame. In~\cref{sec:polaron_valid} we provide a condition for this approximation to be valid. Using full polaron transform we derive the effective Hamiltonian as shown in~\cref{pol_hamiltonian}. We derived~\cref{pol_hamiltonian} using the formalism described in~\cite{PhysRevLett.132.266701}. For simplicity we assume that the system-bath coupling is identical for all sites. The average of $\hat{\mathcal{C}}$ calculated over the bath Hamiltonian is found as
\begin{equation}\label{decay_mag}
\begin{split}
  \expect{\hat{\mathcal{C}}} = \text{exp}\left(-2\sum_{k}\frac{t_{k}^2}{\nu_k^2}\text{coth}(\frac{\beta\nu_k}{2}) \right) = \\
\text{exp}\left(-2\int_{0}^{\infty}d\omega\frac{K(\omega)}{\omega^2}\text{coth}(\frac{\beta\omega}{2}) \right)=\\
\text{exp}\left(2g-\frac{4g}{(\beta\omega_c)^2}\psi^{(1)}(\frac{1}{\beta\omega_c})\right).
\end{split}
\end{equation}
In~\cref{decay_mag}, $K(\omega) = g\omega^3/\omega_c^2\text{exp}(-\omega/\omega_c)$ is the spectral density function for the super-Ohmic bath. $\omega_c$ and $g$ are the cut-off frequency of the bath modes and the dimensionless site-bath coupling constant respectively. The microscopic system bath coupling parameters $t_{n,k}$ and the dimensionless collective coupling parameter $g$ are related through the bath mode density. The integral in~\cref{decay_mag} results in an expression involving the trigamma function $\psi^{(1)}(z)=d^2/dz^2\text{ln}\Gamma(z)$. The methodology that we have laid out so far restricts us from using any general spectral density function of the form $K_{\text{SD}}(\omega) = g\omega^s/\omega_c^{s-1}\text{exp}(-\omega/\omega_c)$ with $s\leq 2$ due to the fact that the integral in~\cref{decay_mag} diverges for such cases~\cite{10.1116/5.0073853}. One can mitigate this issue by using variational polaron transform to regularize the infrared divergence in the low-frequency bath modes by replacing $t_{n,k}$ with the variational parameters $f_{n,k}$ and finding their optimal values via minimalization of Gibbs-Bogoliubov-Feynman upper bound on the free energy~\cite{PhysRevB.108.115437,PhysRevLett.132.266701}. However, this is out of scope of this work. Moreover, it has been shown that if a vibrational motion of a system couples to a three dimensional acoustic phonon bath, the spectral density will be a super-Ohmic one with $s=3$~\cite{Ooi2022}. 

We can re-write the Hamiltonian~\cref{pol_hamiltonian} as an anisotropic XY chain, exactly in the same form as~\cref{xy_hamiltonian} by replacing $(h,\gamma,J)$ with $(h^{\flat},\gamma^{\flat},J^{\flat})$ defined below.
\begin{equation}\label{new_params}
\begin{split}
  &h^{\flat} = \expect{\hat{\mathcal{C}}}h,\quad \gamma^{\flat}=\frac{(1+\gamma)-\expect{\hat{\mathcal{C}}}^2(1-\gamma)}{(1+\gamma)+\expect{\hat{\mathcal{C}}}^2(1-\gamma)},\\
  &J^{\flat} = \frac{J}{2}\left((1+\gamma)+\expect{\hat{\mathcal{C}}}^2(1-\gamma) \right).
\end{split}
\end{equation}
This class of Hamiltonians can be diagonalized by Jordan-Wigner transformation followed by Fourier transformation and Bogoliubov-Valatin transformation~\cite{LIEB1961407,PFEUTY197079,10.21468/SciPostPhysLectNotes.82,10.21468/SciPostPhys.11.1.013}. More information on the diagonalization procedure and spectrum of these Hamiltonians is provided in~\cref{sec:hamiltonian_diag}. Both~\cref{xy_hamiltonian} and~\cref{pol_hamiltonian} have parity symmetry, therefore upon diagonalization, both of them can be written as a direct sum of their positive and negative parity sectors~\cite{10.21468/SciPostPhysLectNotes.82,10.21468/SciPostPhys.11.1.013}. The partition function $\mathcal{Z}$ for such a system is derived in~\cite{PhysRev.127.1508,Kapitonov1998,PhysRevResearch.1.033175,10.21468/SciPostPhys.11.1.013,varizi_thesis} and the result for the equilibrium state of~\cref{xy_hamiltonian} becomes
\begin{equation}\label{partition_total}
\begin{aligned}
  \mathcal{Z}=&\biggl[\prod_{k\in \mathbf{K^+}}2\text{cosh}(\frac{\beta\epsilon_k}{2}) + \prod_{k\in \mathbf{K^+}}2\text{sinh}(\frac{\beta\epsilon_k}{2})\\
 & + \prod_{k\in \mathbf{K^-}}2\text{cosh}(\frac{\beta\epsilon_k}{2}) - \prod_{k\in \mathbf{K^-}}2\text{sinh}(\frac{\beta\epsilon_k}{2})\biggr].
\end{aligned}
\end{equation}
In~\cref{partition_total}, $\epsilon_k$ is the energy of the quasiparticle at mode $k$ and $\mathbf{K^+}$ and $\mathbf{K^-}$ are the sets containing the $k$-modes in the negative and positive parity sectors. We provide an alternative derivation for the partition function in~\cref{sec:finite_partition}. Finally, we can obtain a similar expression for the partition function for the system at SC, $\mathcal{Z}^{\flat}$, by replacing $(h,\gamma,J) \rightarrow (h^{\flat},\gamma^{\flat},J^{\flat})$ to obtain the quasi-particle energy at SC $\epsilon_k^{\flat}$ and replacing in~\cref{partition_total}. 
\section{Methods}
\label{sec:methods}
The fundamental bound on the estimation precision of an unbiased estimator in the asymptotic limit is given by the QFI.
\begin{equation}\label{qfi}
  \mathcal{F}(\alpha) = \sum_{n}\frac{(\partial_{\alpha} p_n)^2}{p_n} + 2\sum_{n,m}\frac{(p_n - p_m)^2}{p_n + p_m}|\inner{m}{\dot{n}}|^2.
\end{equation} 
In~\cref{qfi}, $p_n$ are the parameter dependent eigenvalues of the probe state and $\ket{n}$ are their corresponding parameter dependent eigenvectors and the summations are over all terms with non-vanishing denominators. The first and second terms in~\cref{qfi} are the classical and quantum contributions to the QFI and are denoted by $\mathcal{F}^c$ and $\mathcal{F}^q$ respectively. Note that the Bogoliubov transformation is parameter dependent and one needs to take this into account for calculating the QFI given in~\cref{qfi}. A more detailed discussion on QFI and its relationship with systems at equilibrium is provided in~\cref{sec:intro_metrology} and~\cref{sec:fisher_thermo} respectively.

With the knowledge of the eigenvalues and eigenvectors of the bare Hamiltonian~\cref{xy_hamiltonian} and the effective Hamiltonian~\cref{pol_hamiltonian} one can use~\cref{qfi} to obtain QFI for equilibrium states at WC ($\mathcal{F}(\alpha)$) and SC ($\mathcal{F}^{\flat}(\alpha)$) respectively. We have provided a detailed calculation of the quantum and classical contributions to the QFI in~\cref{sec:qfi_calculation}. The general expression for $\mathcal{F}(\alpha)$ becomes
\begin{equation}\label{qfi_body}
\begin{aligned}
 \mathcal{F}&(\alpha)=\frac{9}{4\mathcal{Z}}\left[\sum_{k\in\mathbf{K^+}}\frac{e^{-\beta\epsilon_k}}{2}\mathds{Z}_k^+\frac{(e^{-2\beta\epsilon_k}-1)^2}{e^{-2\beta\epsilon_k}+1}(\frac{\partial\theta_k}{\partial\alpha})^2\right.\\ &\left.+\sum_{k\in\mathbf{K^-}\setminus\{0,\pi\}}\frac{e^{-\beta\epsilon_k}}{2}\mathds{Z}_k^-\frac{(e^{-2\beta\epsilon_k}-1)^2}{e^{-2\beta\epsilon_k}+1}(\frac{\partial\theta_k}{\partial\alpha})^2\right]\\
 &+\frac{\partial^2\psi}{\partial\alpha^2} + \tilde{\mathcal{F}}^c,
\end{aligned}
\end{equation}
in which $\psi=\text{ln}\mathcal{Z}$ is the free entropy and $\mathds{Z}_k^{\pm}$ is the positive/negative parity sector partition function not including the pair $(-k,k)$. The angle $\theta_k$ is called the Bogoliubov angle which is defined in~\cref{bog_angle}. The first and last two terms in~\cref{qfi_body} constitute $\mathcal{F}^q$ and $\mathcal{F}^c$ respectively. One can write a similar expression for $\mathcal{F}^{\flat}(\alpha)$ by replacing all necessary variables in~\cref{qfi_body} by their corresponding variables in SC. The last term in~\cref{qfi_body} is zero for both magnetometry and thermometry problems at WC and also for magnetometry problem at SC. For thermometric QFI at SC, an expression for $\tilde{\mathcal{F}}^c$ is derived in~\cref{sec:subdiv_fisher_extra}.

A crucial point to consider is the FS effects, which play an important role in the accurate description of the thermodynamic properties of the systems away from thermodynamic limit. The partition function $\mathcal{Z}$ given in~\cref{partition_total}, takes into account the FS effects. It is argued that, in the limit $N\rightarrow\infty$, the second and fourth term in~\cref{partition_total}, cancel and the first and the third term become equal~\cite{PhysRev.127.1508}. Hence, in this limit we can calculate the thermodynamic functions assuming the positive parity approximation (PPA) and neglecting the negative parity sector. With this assumption, the partition function becomes
\begin{equation}\label{partition_ppa}
  \mathcal{Z}_{\text{PPA}} = \prod_{k\in \mathbf{K^+}}2\text{cosh}(\frac{\beta\epsilon_k}{2}).
\end{equation}
This assumption is widely used in both thermodynamics and metrology literature~\cite{PhysRevLett.120.260503,Carollo_2019,Abiuso_2024, PhysRevLett.127.080504,PhysRevE.79.031101,PhysRevLett.109.160601} and we aim to characterize its influence on QFI. The QFI calculated with this assumption is dubbed as $\mathcal{F}_{\text{PPA}}$ ($\mathcal{F}^{\flat}_{\text{PPA}}$) for the WC (SC) case. It is important to note that, in a considerable number of works on equilibrium metrology FS effects and the term $\tilde{\mathcal{F}}^c$ are ignored and in some works on SC thermometry, $\mathcal{F}^q$ is neglected~\cite{MONTENEGRO20251,PhysRevA.82.011611,PhysRevA.94.042121,Carollo_2019, PhysRevLett.133.040802,Abiuso_2024, PhysRevLett.134.010801, PhysRevA.96.062103,Potts2019fundamentallimits,PhysRevLett.128.040502, PhysRevA.107.042614,Ravell_Rodriguez_2024,PhysRevA.111.052623}.

The methodology explained so far utilizes the microscopic details of the system, bath and their interaction to obtain the effective equilibrium state of the system, using which we can calculate the measurement sensitivity limits for different physical parameters. Utilizing the connection established between the energy level corrections in SC and Hill's nanothermodynamics in~\cite{doi:10.1021/acs.jpcb.3c01525}, we can also give a phenomenological expression for the QFI of the system using its subdivision potential and partition function. Subdivision potential is a key thermodynamic variable in Hill's framework which characterizes nanosystem's non-additivity and it vanishes in thermodynamic limit. We provided a brief description of Hill's nanothermodynamics in~\cref{sec:small_thermo}. The phenomenological QFI is given as 
\begin{equation}\label{qfic_finite_size_body}
  \mathcal{F}'(\alpha) = \frac{\partial^2\psi'}{\partial \alpha^2} + \frac{\partial^2 (\beta\mathcal{E})}{\partial \alpha^2}.
\end{equation}
In~\cref{qfic_finite_size_body} $\psi'=\text{ln}\mathcal{Z}'$ is the effective free entropy and $\mathcal{E}$ is the subdivision potential. Considering that the free energy and free entropy are proportional by a factor of $-k_bT$, we can extract $\mathcal{E}$ via a linear ansatz $F(\beta,N) = NF_b(\beta)+\mathcal{E}(\beta)$, in which $F$ and $F_b$ are the total and extensive contributions to the free energy~\cite{hill1994thermodynamics}. Definition of $\psi'$ along with the detailed derivation of~\cref{qfic_finite_size_body} is given in~\cref{sec:qfi_hill}.\\
\section{Results}
\label{sec:results}
\begin{figure}[htbp!]
  \centering
  \includegraphics[width=\linewidth]{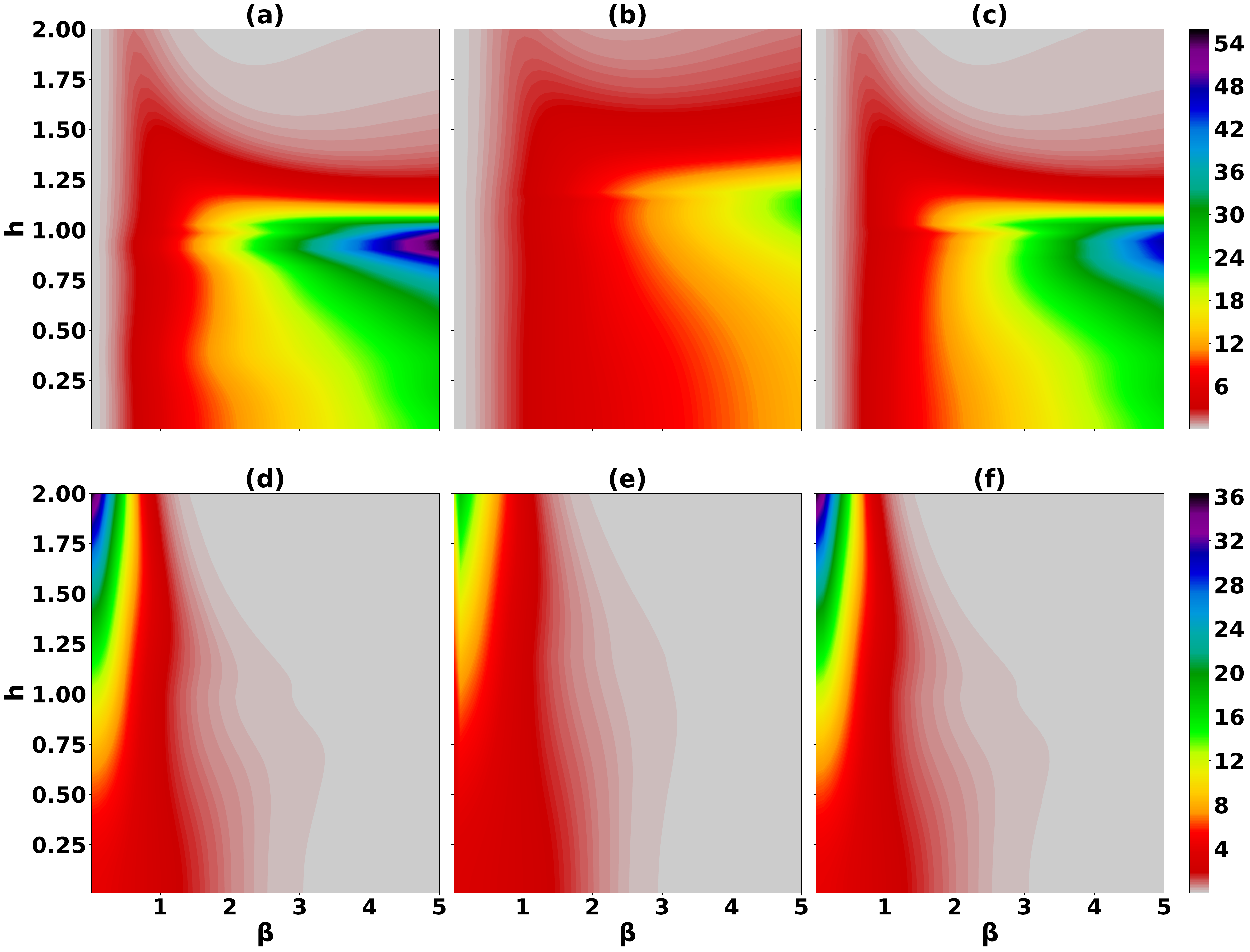}
  \caption{Comparison between QFI calculated for 3 cases. Top panel: QFI calculated for $h$ (a) at WC $\mathcal{F}(h)$, (b) at SC $\mathcal{F}^{\flat}(h)$, (c) using the phenomenological approach $\mathcal{F}'(h)$. Bottom panel: QFI calculated for $\beta$ (d) at WC $\mathcal{F}(\beta)$, (e) at SC $\mathcal{F}^{\flat}(\beta)$, (f) using the phenomenological approach $\mathcal{F}'(\beta)$. The parameters are $N=8$, $J=1$, $\gamma=0.25$. For calculating $\mathcal{F}^{\flat}(h)$ and $\mathcal{F}^{\flat}(\beta)$ we set $g=0.2$.}\label{fig:full_qfi}
\end{figure}

In this section we present the results regarding the sensitivity limits of the FS probe beyond the WC limit and compare them to their counterparts for a weakly-coupled probe in thermodynamic limit. To make sure that the thermal equilibrium state of the polaron transformed system's Hamiltonian gives an accurate approximation of the MFG (see~\cref{sec:polaron_valid}), in all our calculations in this section the cut-off frequency is set to $\omega_c=50$. Additionally, we assumed $J=1$, which sets the energy scale of the problem. In~\cref{fig:full_qfi} we present the results obtained for QFI at both WC and SC and compare the effective QFI derived in~\cref{qfic_finite_size_body} with the microscopically derived expression~\cref{qfi_body} for the case of SC.
The top row of~\cref{fig:full_qfi} suggests that for all three cases, the maximum QFI is obtained at higher values of $\beta$ (lower temperatures) and around the phase transition point $h=J=1$. Comparing~\cref{fig:full_qfi}(a) and~\cref{fig:full_qfi}(b) it is clear that in the SC regime, the QFI for $h$ is reduced substantially for a wide range of parameter range except for larger values of $h$ and $\beta$ in which we see an increase in QFI compared with the WC case. The phenomenological $\mathcal{F}'(h)$ only gives a lower QFI compared to $\mathcal{F}(h)$ only around the phase transition point and at lower temperatures. The microscopically derived QFI at SC, $\mathcal{F}^{\flat}(h)$ predicts lower values across a wide parameter range as compared with $\mathcal{F}'(h)$ and unlike the phenomenological case, it also suggests that SC shifts the maximum value for QFI to a higher value of $h$ than that the original phase transition point. We present a more detailed analysis for the effect of SC on phase transition in~\cref{sec:add_res}. The influence of system-bath coupling on the phase transition properties is also studied in~\cite{PhysRevLett.132.266701,PhysRevLett.94.047201,PhysRevB.90.224401,PhysRevB.104.L060410,PhysRevA.105.012431} and references therein.

Comparing~\cref{fig:full_qfi}(d) and~\cref{fig:full_qfi}(f) we see that for QFI calculated for $\beta$, the phenomenological approach yields almost identical result compared with the WC case. However~\cref{fig:full_qfi}(e) shows that although the thermometric precision at SC is reduced for lower $\beta$ at SC compared with WC, the maximum of $\mathcal{F}^{\flat}(\beta)$ shifts to a higher value of $\beta$ and we observe an increase in thermometric precision at lower temperatures due to SC.
\begin{figure}[htbp!]
  \centering
  \includegraphics[width=\linewidth]{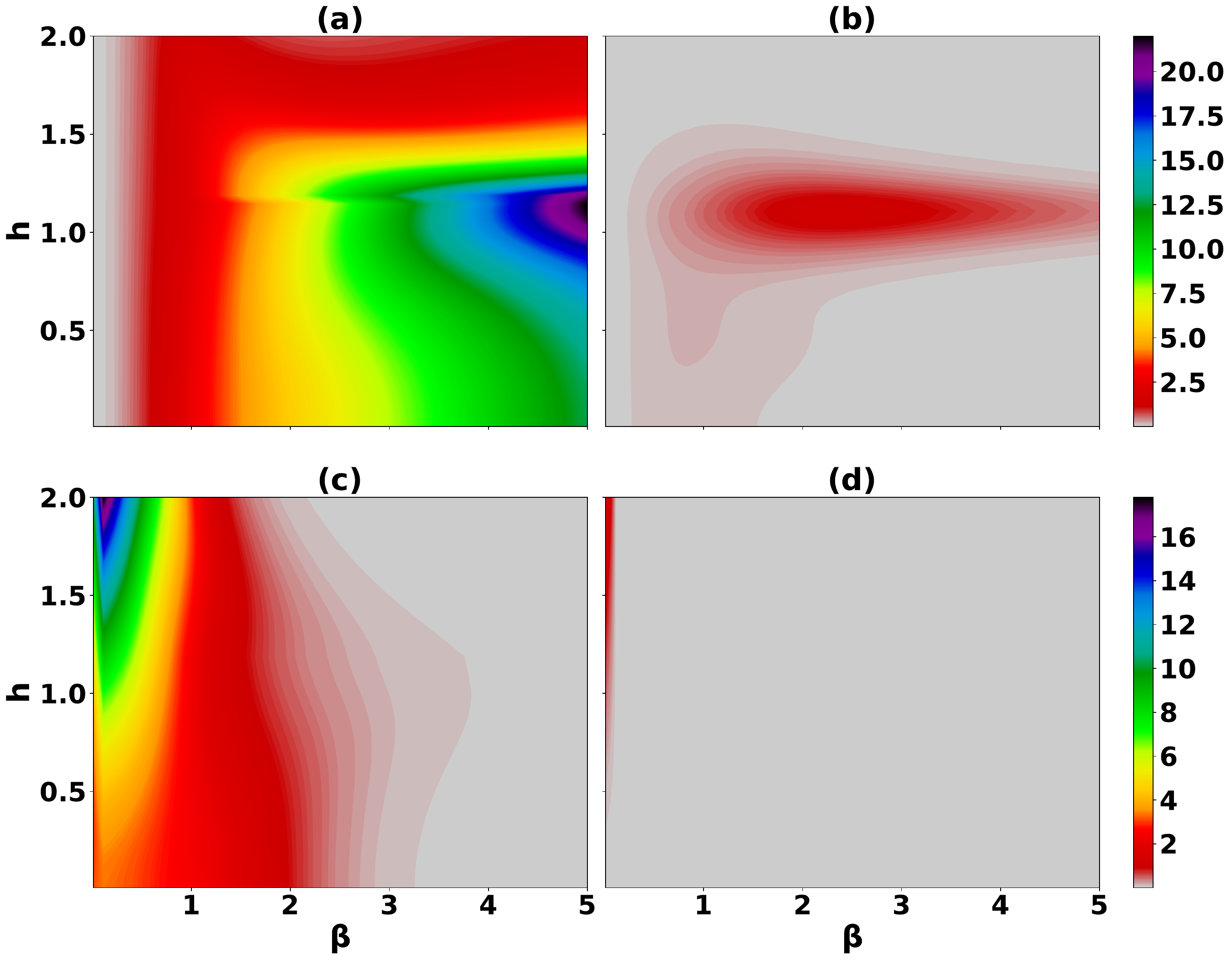}
  \caption{Classical and quantum contributions to the QFI at SC for $\gamma=0.25$, $N=8$, $J=1$. For calculating $\mathcal{F}^{\flat}(h)$ and $\mathcal{F}^{\flat}(\beta)$ we set $g=0.2$. In (a) and (b) $\mathcal{F}^{\flat c}(h)$ and $\mathcal{F}^{\flat q}(h)$ are shown respectively. In (c) and (d) $\mathcal{F}^{\flat c}(\beta)$ and $\mathcal{F}^{\flat q}(\beta)$ are shown respectively.}\label{fig:qfi_cq1}
\end{figure}
Our results demonstrate that the phenomenological approach is incapable of quantifying the precision limits accurately. This is understandable since for the results presented in this section, both the effective free entropy and subdivision potential in~\cref{qfic_finite_size_body} are independent of both bath's and interaction Hamiltonian's parameters. Our calculations reveal that even if we utilize the temperature-dependent effective Hamiltonian~\cref{pol_hamiltonian} to calculate the effective free entropy and subdivision potential, the QFI obtained using the phenomenological approach would largely agree with the microscopic approach (see~\cref{sec:add_res}) but will still yield different results for a certain parameter domain. Additionally, we demonstrate in~\cref{sec:subdiv_fisher_extra} that the subdivision potential calculated using the effective Hamiltonian is drastically different than the one calculated phenomenologically by imposing a linear ansatz on thermodynamic potentials. This can be understood from the fact that in SC, due to extreme nonlinearity of the free energy in some parameter range, using a linear ansatz to extract the subdivision potential (refer to~\cref{sec:small_thermo}) is doomed to fail and that in this regime the corrections to the energy levels induced by the bath become non-negligible. Moreover, if we have access to the microscopic description of the system at SC, (HMF or effective Hamiltonian) there is no need to resort to the phenomenological approach. Therefore we can justifiably ignore the phenomenological approach in the remainder of this section and focus on further analyzing the results for microscopic calculations at WC and SC. 

In~\cref{fig:qfi_cq1} we give the quantum and classical contributions to QFI at SC. We see that overall for both thermometry and magnetometry cases, the quantum contribution $\mathcal{F}^{\flat q}(h)$ ($\mathcal{F}^{\flat q}(\beta)$) is substantially smaller than its classical counterpart $\mathcal{F}^{\flat c}(h)$ ($\mathcal{F}^{\flat c}(\beta)$). $\mathcal{F}^{\flat q}(\beta)$ is only non-negligible for small values of $\beta$ and large values of $h$, but for $\mathcal{F}^{\flat q}(h)$ this range is much larger and its values peak around the shifted phase transition point at SC.
\begin{figure}[htbp!]
  \centering
  \includegraphics[width=\linewidth]{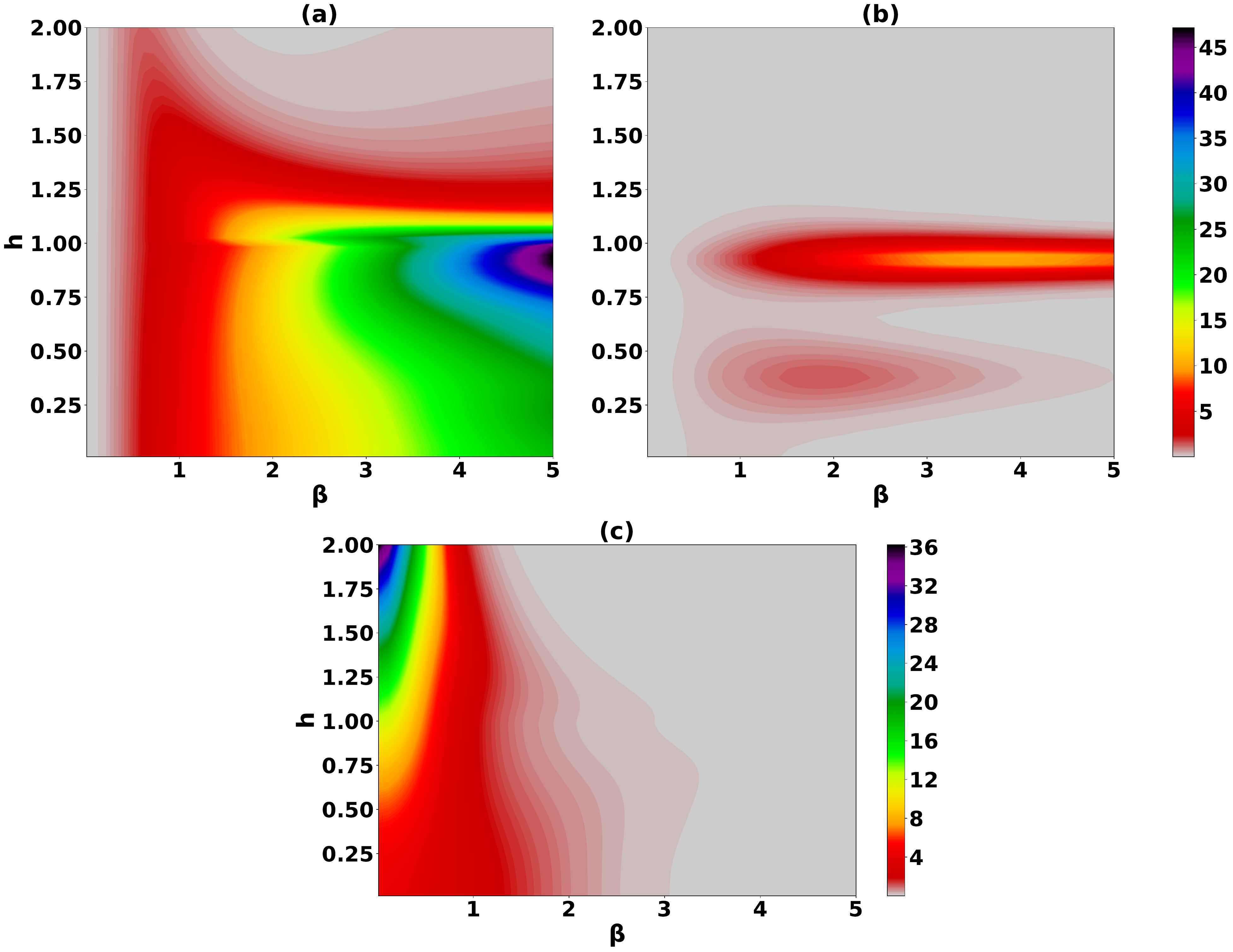}
  \caption{Classical and quantum contributions to the QFI at WC for $\gamma=1$, $N=8$, $J=1$. In (a) and (b) $\mathcal{F}^{\flat c}(h)$ and $\mathcal{F}^{\flat q}(h)$ are shown respectively. In (c) and (d) $\mathcal{F}^{\flat c}(\beta)$ and $\mathcal{F}^{\flat q}(\beta)$ are shown respectively.}\label{fig:qfi_cq2}
\end{figure}
In~\cref{fig:qfi_cq2}, we present the classical and quantum contributions to the QFI in the WC limit. Comparing~\cref{fig:qfi_cq2} and \cref{fig:qfi_cq1} we see that for the parameter regime under investigation, the quantum contribution to the QFI with respect to the magnetic field at WC ($\mathcal{F}^{q}(h)$), attains a larger maximum value and is non-negligible for a larger range of $h$ and $\beta$. In the WC limit, the system is in the thermal equilibrium state of its bare Hamiltonian~\cref{xy_hamiltonian}, which is independent of the temperature. Therefore in this regime the quantum contribution to the $\mathcal{F}^{q}(\beta)$ is identically zero. Moreover, our calculations show that as we decrease $\gamma$ (keeping all other parameters fixed), the share of the quantum contribution to the QFI to the total QFI increases. These and additional results, including the influence of dependence of the phase transition point on the system-bath coupling strength and temperature, are presented in~\cref{sec:add_res}. 

Now we turn our attention to the contribution of FS effects in QFI. We define $\mathcal{R}(\alpha) = \mathcal{F}_{\text{PPA}}(\alpha)/\mathcal{F}(\alpha)$ as the ratio of the QFI assuming PPA to the total QFI. $\mathcal{R}(\alpha)\neq 1$ would suggest a discrepancy between the FS and PPA calculations for QFI.  
\begin{figure}[htbp!]
  \centering
  \includegraphics[width=\linewidth]{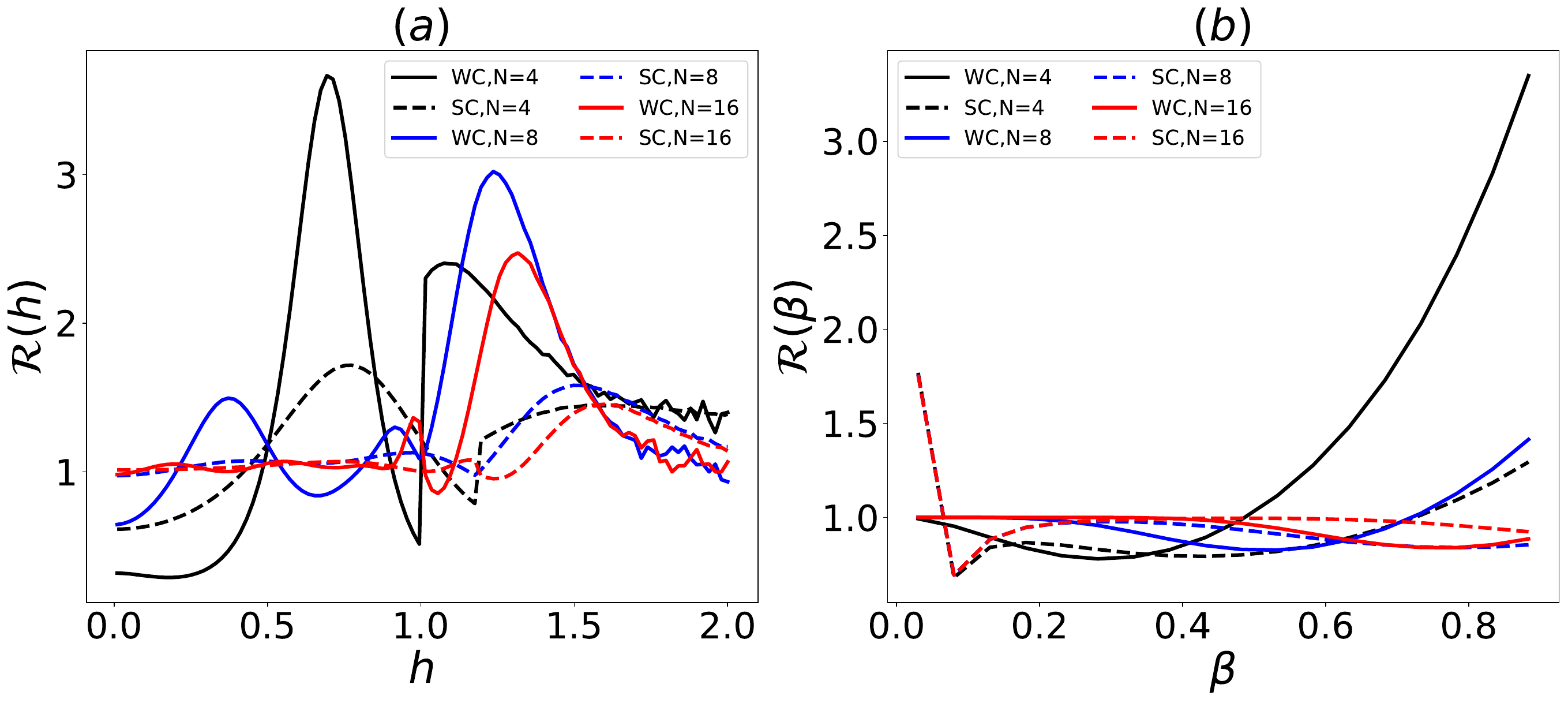}
  \caption{The ratios of the QFI calculated assuming PPA to the full QFI for different $N$ at WC and SC. The parameters are $J=1$, $\gamma=0.25$. For QFI calculations at SC we take $g=0.2$. (a) $\mathcal{R}(h)$ at WC and SC for $\beta = 5$. (b) $\mathcal{R}(\beta)$ at WC and SC for $h = 2$.}\label{fig:ppa_full}
\end{figure}
It is clear from~\cref{fig:ppa_full} that for both thermometry and magnetometry problems, PPA assumptions causes a drastic deviation (either as an overestimation or underestimation) from the QFI obtained considering FS effects. However, from both~\cref{fig:ppa_full}(a) and~\cref{fig:ppa_full}(b) we can see that this deviation tends to get smaller with larger values of $N$. This is in line with the fact that at the thermodynamic limit PPA assumption is valied and it can be used to obtain QFI, which makes the calculations more tractable numerically.
\begin{figure}[htbp!]
  \centering
  \includegraphics[width=\linewidth]{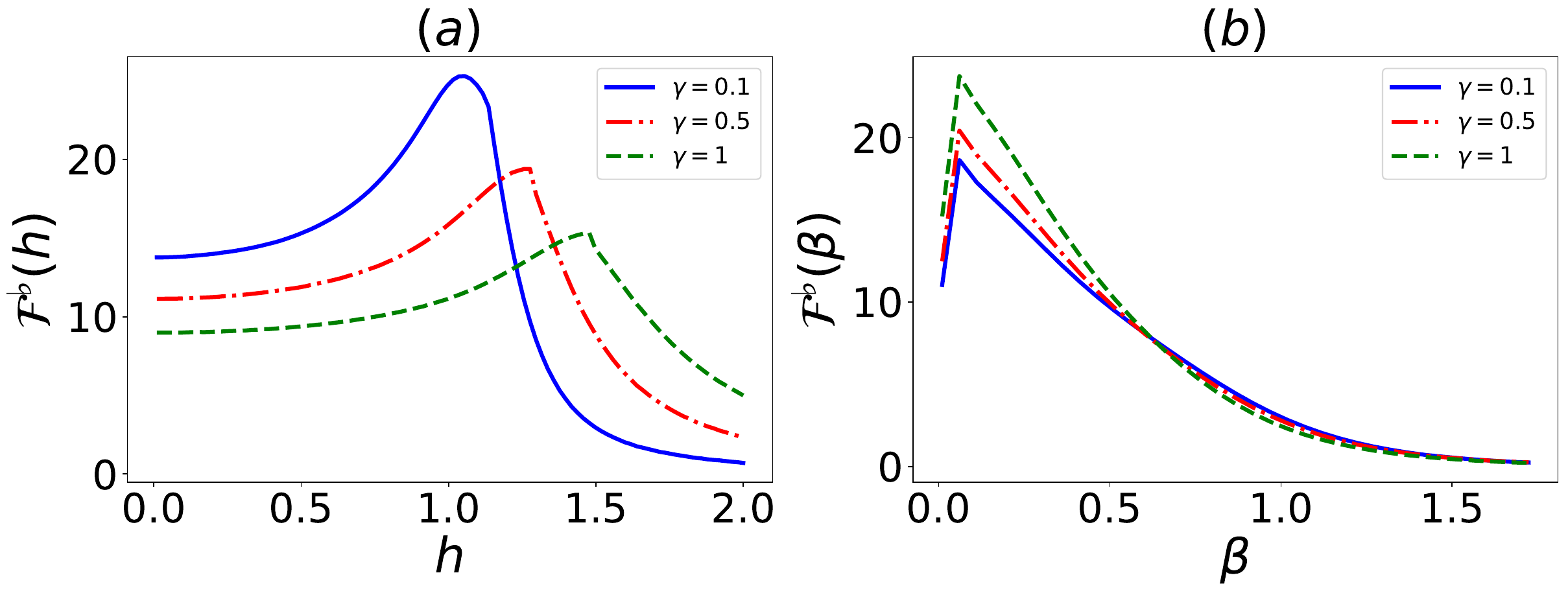}
  \caption{QFI at SC calculated for different values of anisotropy parameter. The parameters are $N=8$, $J=1$ and $g=0.2$. (a) $\mathcal{F}^{\flat}(h)$ for $\beta = 5$. (b) $\mathcal{F}^{\flat}(\beta)$ for $h = 2$.}\label{fig:qfi_gamma}
\end{figure}

Now we aim to examine the behavior of QFI at SC in more detail around the parameter range where it is maximal. From~\cref{fig:qfi_gamma}(a) we can see that for large values of $\beta$, increasing $\gamma$ while keeping the other parameters fixed, generally decreases/increases $\mathcal{F}^{\flat}(h)$ for lower/higher values of $h$ but this change is not monotonic for all $h$, as there is a range of $h$ in which an intermediate value for $\gamma$ yields a higher $\mathcal{F}^{\flat}(h)$ a larger value of $\gamma$. However, upon examining~\cref{fig:qfi_gamma}(b) it is evident that for large values of $h$, increasing $\gamma$ increases/decreases $\mathcal{F}^{\flat}(\beta)$ for lower/higher values of $\beta$ but after a certain point in $\beta$, $\mathcal{F}^{\flat}(\beta)$ calculated for any $\gamma$ value tends to zero.
\begin{figure}[htbp!]
  \centering
  \includegraphics[width=\linewidth]{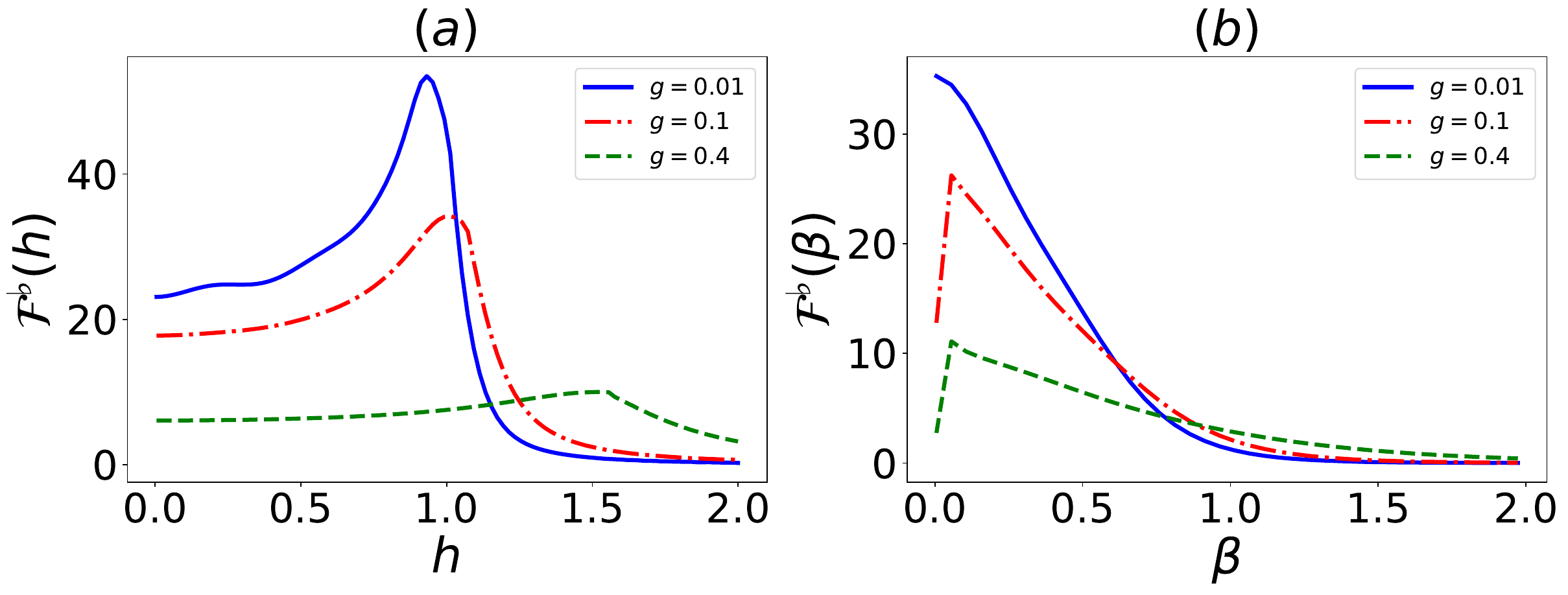}
  \caption{QFI at SC calculated for different values of system-bath coupling strength. The parameters are $N=8$, $J=1$ and $\gamma=0.25$. (a) $\mathcal{F}^{\flat}(h)$ for $\beta = 5$. (b) $\mathcal{F}^{\flat}(\beta)$ for $h = 2$.}\label{fig:qfi_g}
\end{figure}

In~\cref{fig:qfi_g} we study the effect of system-bath coupling $g$ around the parameter regime where QFI attains its maximal value. \cref{fig:qfi_g}(a) demonstrates that for large values of $\beta$, increasing $g$ while keeping the other parameters fixed, generally decreases/increases $\mathcal{F}^{\flat}(h)$ for lower/higher values of $h$ but this change is not monotonic as argued earlier. Our calculations for $\mathcal{F}^{\flat}(h)$ suggests a pronounced peak near the critical point $h^{\flat} = J^{\flat}$, consistent with the divergence of fidelity susceptibility and enhanced state distinguishability at criticality. From both~\cref{fig:qfi_g}(a) and~\cref{fig:qfi_g}(b) we can see that SC with bath reduces the QFI peak height but broadens it. In the paramagnetic phase (after the critical field), where the system is gapped and polarized along the transverse field direction, the QFI is typically suppressed. However, increasing the system-bath coupling $g$ leads to a partial recovery of field sensitivity. This enhancement arises from environment-induced population mixing across energy eigenstates, which effectively broadens the spectral response and restores sensitivity to $h$ even in the absence of criticality. By contrast, in the ferromagnetic phase (before the critical field), where the ground state manifold exhibits near-degeneracy and long-range quantum correlations, the QFI decreases monotonically with $g$. Here, the high sensitivity originates from intrinsic quantum coherence and correlations present in the nearly degenerate ground-state manifold. The environment acts destructively in this regime by introducing decoherence and suppressing long-range order, leading to a monotonic degradation of sensitivity. This asymmetry between the two phases highlights a complementary role of environmental interactions: they can either activate or degrade metrological performance, depending on the structure of the underlying quantum phase. 

\cref{fig:qfi_g}(b) demonstrates that increasing $g$ boosts $\mathcal{F}^{\flat}(\beta)$ for large values of $\beta$ (small temperature values $T$). The location of the largest peak for $\mathcal{F}^{\flat}(\beta)$ shifts toward higher $\beta$ as we increase $g$, accompanied by a broadening of the low-temperature tail. The enhanced sensitivity at higher $\beta$ and $g$ is attributed to environment-assisted transitions that repopulate thermally active states. As a result, the QFI grows in the low temperature regime with increasing $g$, even as the system moves away from criticality. These features suggest a coupling-tunable thermometric protocol in which the peak sensitivity can be shifted to target lower temperatures, offering operational flexibility in regimes where quantum thermometry is typically limited by vanishing thermal fluctuations.

\begin{figure}[htbp!]
  \centering
  \includegraphics[width=\linewidth]{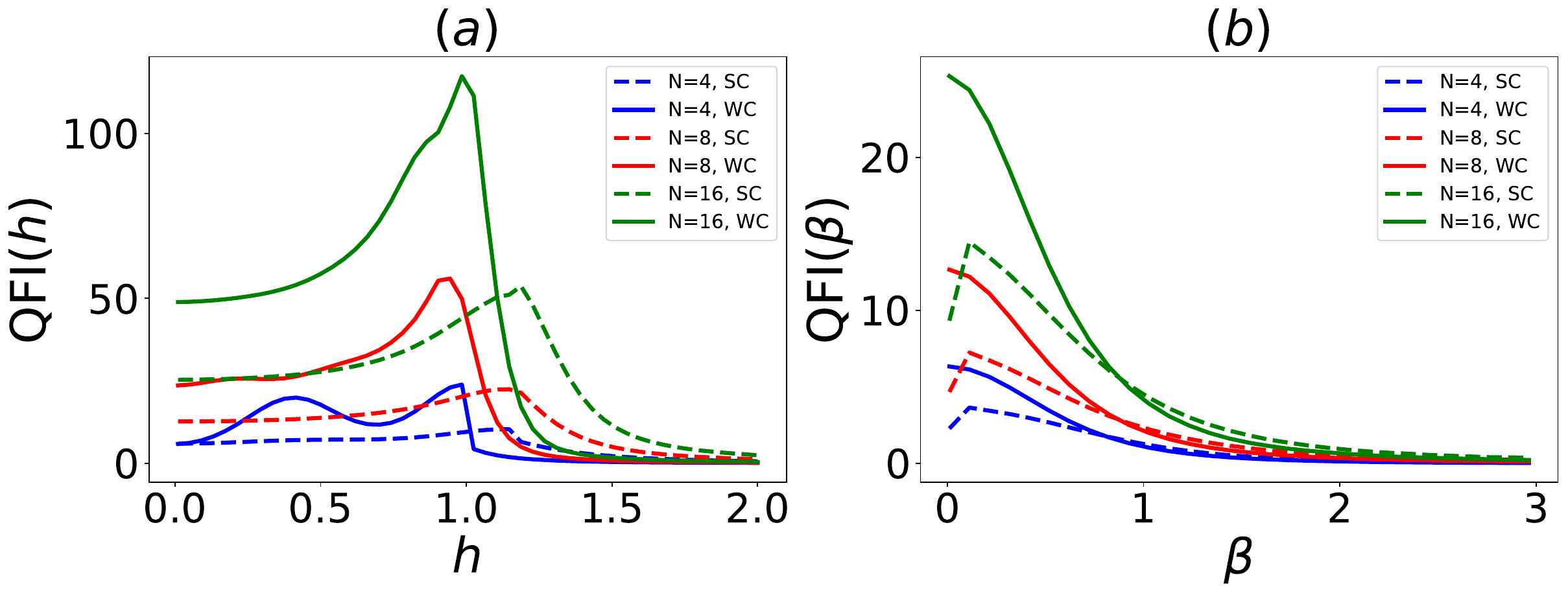}
  \caption{(a) $\mathcal{F}(h)$ and $\mathcal{F}^{\flat}(h)$ vs $h$ at $\beta=5$ and (b) $\mathcal{\beta}(h)$ and $\mathcal{F}^{\flat}(\beta)$ vs $\beta$ at $h=1$ for various values of $N$. The remaining parameters are $J=1$, $\gamma=0.25$ and $g=0.2$.}\label{fig:compare_n}
\end{figure}

In~\cref{fig:compare_n} we investigate the behavior of QFI near the critical point for various values of $N$. From both~\cref{fig:compare_n}(a) and~\cref{fig:compare_n}(b) it is clear that the maximum value for QFI shrinks considerably in the SC regime compared to the WC case. 
Additionally, both figures clearly demonstrate that due to the SC effects, the value of $h$/$\beta$ for which $\mathcal{F}^{\flat}(h)$/$\mathcal{F}^{\flat}(\beta)$ attains its maximum increases compared to $\mathcal{F}(h)$/$\mathcal{F}(\beta)$ in the WC case. \cref{fig:compare_n}(a) shows that for smaller values of $N$, there are two local minima for $\mathcal{F}(h)$ within the parameter regime of investigation. As we increase $N$, we are only left with one local minimum in the critical point. However, regardless of the value of $N$,in the SC regime we only see a single local minimum. In contrast, from~\cref{fig:compare_n}(b) we see that $\mathcal{F}(\beta)$ shows a monotonic decrease with $\beta$ but $\mathcal{F}^{\flat}(\beta)$ displays two local minima.\\

Finally, we investigate scaling behavior of the QFI, both at WC and SC. To do this, for each $N$ we calculate the QFI for a range of values of $h$ and $\beta$ and find the maximum value of QFI ($\mathcal{F}_{\text{max}}$) at that value of $N$. 
\begin{figure}[htbp!]
  \centering
  \includegraphics[width=\linewidth]{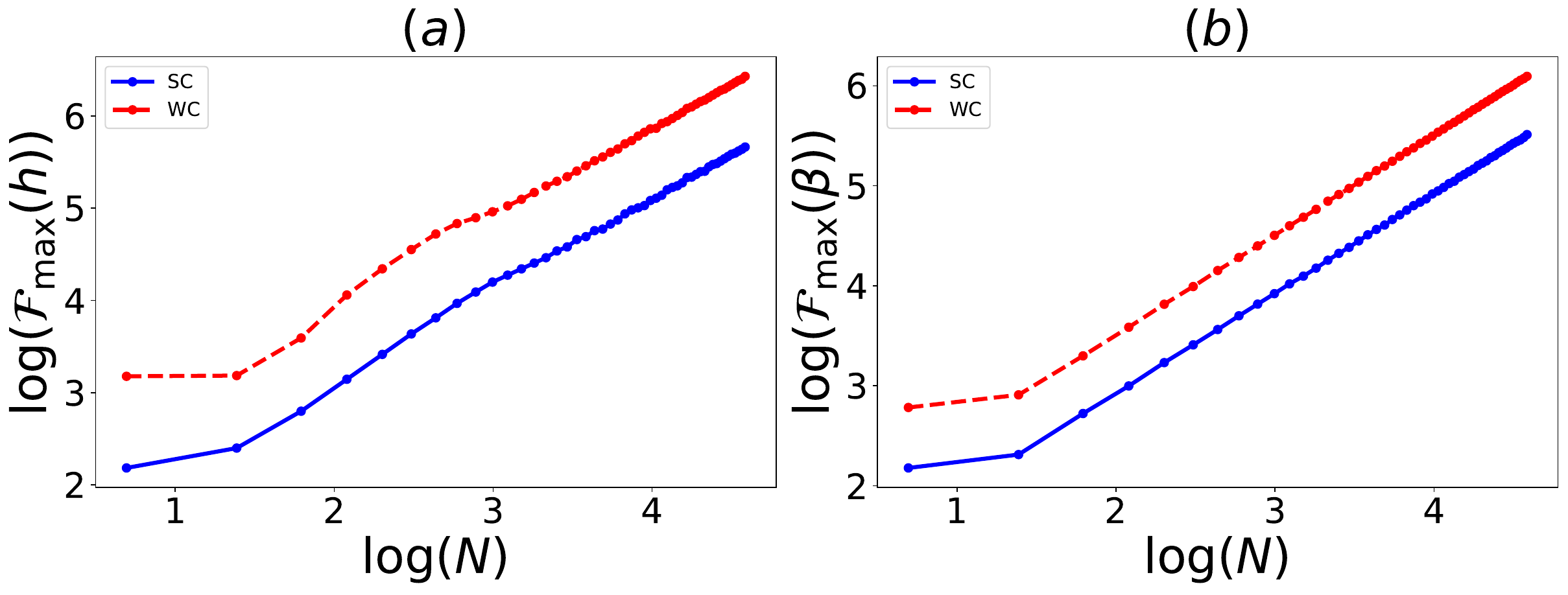}
  \caption{(a) Log-log plot for the maximum of QFI for each value of $N$, with respect to $h$ at WC and SC. (b) Log-log plot for the maximum of QFI for each value of $N$, with respect to $\beta$ at WC and SC. $N$ takes even values from 2 to 102.}\label{fig:scaling_n}
\end{figure}
After repeating this procedure for a range of values of $N$, for the case of WC, we numerically fit the values for $\mathcal{F}_{\text{max}}$ to the ansatz of form
\begin{equation}\label{scaling_ansatz}
  \mathcal{F}_{\text{max}}(\alpha) = aN^{\xi}+\frac{b}{N}+c.
\end{equation}
The derivation of the ansatz~\cref{scaling_ansatz} is given in~\cref{sec:ansatz} using the FS scaling theory~\cite{PhysRev.185.832, PhysRevLett.28.1516,PhysRevB.32.1720,PhysRevB.30.322,PhysRevB.89.094516,ARDOUREL202399,PhysRevB.81.064418, 10.1093/oxfordhb/9780195392043.013.0005, goldenfeld2019lectures, 10.1093/acprof:oso/9780199577224.001.0001}. Using an identical approach we can also extract the parameters of the scaling function at SC regime using $\mathcal{F}^{\flat}_{\text{max}}(\alpha)$. From both figures in~\cref{fig:scaling_n} it is clear that for FS, although we get a power-law behavior in both WC and SC regimes, the log-log curve doesn't result in a single line with a fixed slope throughout the range of considered values of $N$ but for larger values of $N$ a clear linear relationship emerges. We have performed our calculations for even values of $N$ ranging from 2 to 100. The remaining parameters are identical to the ones given in~\cref{fig:full_qfi}. We see that the values of the scaling exponents are very close to the theoretical value of 1 found in~\cref{sec:ansatz}, which indicates that in both cases the probe operates in the standard quantum limit. The parameters which encode the FS contributions from the scaling perspective are the $b$ and $c$ values. Our results show that the FS effects play a non-negligible role in the scaling function away from the thermodynamic limit.
\begin{table}[htbp!]
\begin{center}
\begin{tabular*}{0.8\linewidth}{@{\extracolsep{\fill}}|c|c|c|c|c|}
  \hline
  $ $ & $\xi$ & $a$ & $b$ & $c$ \\ \hline
  $\mathcal{F}_{\text{max}}(h)$ & 0.999 &  6.093 & -27.930 & 18.290 \\ \hline
  $\mathcal{F}_{\text{max}}(\beta)$ & 0.989 & 4.794 & 18.671 & -3.514 \\ \hline
  $\mathcal{F}^{\flat}_{\text{max}}(h)$ & 0.962 & 3.480 & -1.188 & 0.527 \\ \hline
  $\mathcal{F}^{\flat}_{\text{max}}(\beta)$ & 0.984 & 2.728 & 10.668 & -2.292 \\
  \hline
\end{tabular*}
\end{center}
      \caption{The fitting parameters for the ansatz given in~\cref{scaling_ansatz}.}\label{tab1}
\end{table}

\section{Conclusion}
\label{sec:conclusion}
We have investigated the ultimate precision limits of equilibrium quantum probes beyond the standard assumptions of WC and thermodynamic-limit scaling. By combining a microscopic SC treatment based on full polaron transform (to approximate the Hamiltonian of mean force) with an exact FS formulation of the partition function, we derived analytic expressions for the quantum Fisher information (QFI) of a spin-chain probe and analyzed their behavior in both WC and SC regimes.\\

Our results establish three main conclusions. First, FS introduce significant and systematic corrections to QFI that persist well beyond very small system sizes. Approximations based on thermodynamic-limit arguments, such as PPA, can therefore lead to substantial misestimation of achievable precision. Second, strong system–bath coupling qualitatively modifies metrological performance. While it generally suppresses sensitivity, it can enhance thermometric precision in low-temperature regimes and shift optimal operating points. Utility of SC in low-temperature thermometry is also shown in~\cite{PhysRevA.96.062103,Potts2019fundamentallimits,PhysRevLett.128.040502,PhysRevA.108.032220, Ravell_Rodriguez_2024,PhysRevA.111.052623, PhysRevResearch.7.023235}. Third, phenomenological thermodynamic approaches, including those based on subdivision potentials and temperature-dependent energy levels, fail to reliably capture these effects, highlighting the necessity of a fully microscopic description.\\

Taken together, our findings show that realistic quantum sensors, being finite and often operating outside the WC regime, cannot be accurately characterized within conventional frameworks. Instead, precision limits must be evaluated using approaches that explicitly incorporate both FS and SC physics. This work provides such a framework and clarifies the regimes in which environmental interactions can be either detrimental or beneficial for quantum metrology. 

The predictions presented here could be tested in programmable quantum simulators, with semiconductor quantum dot arrays and trapped-ion platforms providing a particularly promising route. Exchange-coupled quantum-dot spin chains have already been demonstrated experimentally~\cite{Kandel2021,PhysRevLett.126.017701, 10.1063/5.0055908}. The 3D acoustic-phonon bath in these systems intrinsically yields a super-Ohmic spectral density with $s=3$, matching the form of the bath spectral density function assumed here~\cite{PhysRevB.65.195313,PhysRevLett.104.017402,PhysRevLett.105.177402, PhysRevB.87.195306,STEFANATOS2026170628}. Furthermore, the experiments on trapped-ions have demonstrated independently controllable XX and YY spin couplings, enabling the realization of anisotropic XY-type Hamiltonians~\cite{PhysRevResearch.6.033038, PhysRevA.111.042607}. Recent experiments on these platforms have implemented spin-boson models with programmable bath spectral densities and temperatures~\cite{Sun2025} and finite-temperature bosonic reservoirs with independently tunable temperature and dissipation rates~\cite{mf8d-91ws}. These capabilities provide the ingredients needed to approximate the super-Ohmic spectral density considered here and to explore the dependence on system-bath coupling, temperature, and cutoff frequency. Together, these platforms span the parameter regimes relevant to the present theory and provide complementary tests of its predictions.\\

Additional work needs to be done to quantify the precision limits of the spin chain for the case of global coupling to a bath. It is shown that in this case the bath will induce long-range interactions in the chain~\cite{PhysRevLett.132.266701}, which rules out the possibility of analytical diagonalization of the effective Hamiltonian. Moreover, utilizing variational polaron transform to study the QFI in strong coupling with an Ohmic and sub-Ohmic bath would bring additional insight in the role of bath's characteristics in system's precision limits at SC. Furthermore, it would be interesting to study the effect of SC using the methodology of polaron transformation beyond including higher order perturbation terms (beyond ultra-weak coupling regime) in the polaron frame. Finally, studying multiparameter estimation for an equilibrium probe in the SC regime and characterizing the incompatibility of the parameters is a natural extension of the present study. 

\acknowledgments
This work is funded by the Scientific and Technological Research Council of Türkiye (TÜBİTAK), grant No. 123F150. A.P. gratefully acknowledges financial support from the Korea Research Institute of Standards and Science and the National Research Council of Canada.

%

\clearpage
\onecolumngrid
\appendix
\section{Error Analysis of the Second-order Perturbation Theory in the Polaron Frame}
\label{sec:polaron_valid}
For the polaron transformation approach to be useful in obtaining the equilibrium state of the system beyond weak system-bath coupling limit, one needs to be able to 
\begin{equation}\label{polaron_u}
  \tilde{\rho}_{\text{eq}} = \frac{\text{Tr}_B[e^{-\beta \tilde{H}_{\text{tot}}}]}{\text{Tr}_{S+B}[e^{-\beta \tilde{H}_{\text{tot}}}]},
\end{equation}
in which $\tilde{H}_{\text{tot}}$ is the polaron transformed total Hamiltonian. The bath Hamiltonian $\hat{H}_B$ doesn't change under polaron transformation. $\hat{U}_{\text{pol}}$ can be written as~\cite{PhysRevLett.132.266701} 
\begin{equation}\label{u_pol}
  \hat{U}_{\text{pol}} = \prod_{n}\hat{W}_n,\qquad \hat{W}_n=\text{exp}(-i\sigma_n^x\hat{B_n}/2),\quad \hat{B}_n=2i\sum_{k}t_{k}/\nu_k(\hat{b}_{n,k}^{\dag}-\hat{b}_{n,k}).
\end{equation}
We define
\begin{equation}\label{def_supp}
  \hat{H}_S^{\flat} \equiv \langle \tilde{H}_S\rangle_B,\quad \hat{H}_I^{\flat} \equiv \hat{V} \equiv \tilde{H}_S - \hat{H}_S^{\flat}.
\end{equation}
With these definitions we get 
\begin{equation}\label{def_supp_cons}
  \tilde{H}_{\text{tot}} = \hat{H}_0 + \hat{V},\quad  \hat{H}_0 = \hat{H}_S^{\flat} + \hat{H}_B.
\end{equation}
This splitting ensures that the average of the interaction Hamiltonian in the polaron frame vanishes, $\langle\hat{V}\rangle_B=0$. We can expand $\tilde{H}_{\text{tot}}$ as,
\begin{equation}\label{expand_polaron}
  e^{-\beta (\hat{H}_0+\hat{V})}\approx e^{-\beta \hat{H}_0}\left[1-\int_{0}^{\beta} d\beta' e^{\beta' \hat{H}_0}\hat{V} e^{-\beta'\hat{H}_0} + \int_{0}^{\beta}\int_{0}^{\beta'}d\beta'd\beta''e^{\beta' \hat{H}_0}\hat{V} e^{-(\beta'-\beta'') \hat{H}_0}\hat{V} e^{-\beta''\hat{H}_0} \right],
\end{equation}
Using the results in~\cite{PhysRevLett.132.266701}, the interaction Hamiltonian in the polaron frame becomes
\begin{equation}\label{h_int_pol}
\begin{aligned}
  \hat{V} &= -h\sum_n \left[(\hat{\mathcal{C}}_n-\expect{\hat{\mathcal{C}}_n})\hat{\sigma}_n^z - \hat{\mathcal{S}}_n\hat{\sigma}_n^y \right] - \frac{J}{2}(1-\gamma) \sum_n\left[(\hat{\mathcal{C}}_n \hat{\mathcal{C}}_{n+1}-\expect{\hat{\mathcal{C}}_n}\expect{\hat{\mathcal{C}}_{n+1}})\hat{\sigma}_n^y\hat{\sigma}_{n+1}^y + \hat{\mathcal{S}}_n \hat{\mathcal{S}}_{n+1} \hat{\sigma}_n^z\hat{\sigma}_{n+1}^z \right.\\
  &\left. +\hat{\mathcal{C}}_n\hat{\mathcal{S}}_{n+1}\hat{\sigma}_n^y\hat{\sigma}_{n+1}^z +\hat{\mathcal{S}}_n\hat{\mathcal{C}}_{n+1}\hat{\sigma}_n^z\hat{\sigma}_{n+1}^y \right].
\end{aligned}
\end{equation}
In~\cref{h_int_pol}, $\hat{\mathcal{C}}_n = \text{cos}(\hat{B}_n)$, $\hat{\mathcal{S}}_n = \text{sin}(\hat{B}_n)$, and $\expect{\hat{\mathcal{C}}_n}$ is the thermal average of $\hat{\mathcal{C}}$ over the bath degrees of freedom, given in~\cref{decay_mag}. 
We next quantify the accuracy of retaining only the zeroth-order contribution in the expansion of the reduced equilibrium state.  The appropriate comparison is between the zeroth and second-order reduced states. We introduce the imaginary-time interaction-picture operator
\begin{equation}
    \hat V(\tau)=e^{\tau \hat H_0}\hat V e^{-\tau \hat H_0},
    \qquad 0\leq \tau\leq\beta,
    \label{eq:Vtau_cont}
\end{equation}
and the zeroth-order reduced equilibrium state
\begin{equation}
    \hat\rho_0=\frac{e^{-\beta \hat H_S^{\flat}}}{Z_S^{\flat}},
    \qquad
    Z_S^{\flat}=\Tr_S\!\left[e^{-\beta \hat H_S^{\flat}}\right].
    \label{eq:rho0_cont}
\end{equation}
Because the splitting in the preceding equations was chosen such that \(\expect{\hat V}_B=0\), stationarity of the thermal bath implies
\begin{equation}
    \expect{\hat V(\tau)}_B
    =\frac{1}{Z_B}\Tr_B\!\left(e^{-\beta \hat H_B}e^{\tau\hat H_B}\hat V e^{-\tau\hat H_B}\right)=0.
    \label{eq:first_order_zero_cont}
\end{equation}
Consequently, the first-order contribution vanishes identically after tracing over the bath.  Defining
\begin{equation}
    \hat{\mathcal K}_2
    =\int_0^\beta \dd\tau\int_0^\tau \dd\tau'\,
    \expect{\hat V(\tau)\hat V(\tau')}_B,
    \label{eq:K2_cont}
\end{equation}
the unnormalized reduced Gibbs operator takes the form
\begin{equation}
    \Tr_B\!\left[e^{-\beta(\hat H_0+\hat V)}\right]
    =Z_B e^{-\beta\hat H_S^{\flat}}
    \left[\hat I+\hat{\mathcal K}_2+O(\hat V^3)\right].
    \label{eq:unnormalized_second_cont}
\end{equation}
The same second-order contribution appears in the total partition function.  Taking the trace of \cref{eq:unnormalized_second_cont} over the system gives
\begin{equation}
\begin{aligned}
    Z_{S+B}
    &\equiv \Tr_{S+B}\!\left[e^{-\beta(\hat H_0+\hat V)}\right]\\
    &=Z_B\Tr_S\!\left\{e^{-\beta\hat H_S^{\flat}}
    \left[\hat I+\hat{\mathcal K}_2+O(\hat V^3)\right]\right\}\\
    &=Z_B Z_S^{\flat}\left[1+\expect{\hat{\mathcal K}_2}_0+O(\hat V^3)\right],
\end{aligned}
\label{eq:partition_second_cont}
\end{equation}
where \(\expect{\hat O}_0\equiv\Tr_S(\hat\rho_0\hat O)\). Multiplying~\cref{eq:unnormalized_second_cont} by the reciprocal of~\cref{eq:partition_second_cont} gives
\begin{equation}
    \tilde\rho_{\mathrm{eq}}
    =\hat\rho_0
    +\hat\rho_0\left[
      \hat{\mathcal K}_2-
      \expect{\hat{\mathcal K}_2}_0
    \right]
    +O(\hat V^3),
    \label{eq:normalized_second_cont}
\end{equation}
where \(\expect{\hat O}_0\equiv\Tr_S(\hat\rho_0\hat O)\).  In particular, for any system observable \(\hat O\),
\begin{equation}
    \delta\expect{\hat O}^{(2)}
    =\expect{\hat O\hat{\mathcal K}_2}_0
    -\expect{\hat O}_0\expect{\hat{\mathcal K}_2}_0.
    \label{eq:connected_correction_cont}
\end{equation}
\Cref{eq:connected_correction_cont} shows explicitly that only the connected part of the second-order correction changes normalized equilibrium expectation values.  Terms in \(\hat{\mathcal K}_2\) that are disconnected from the observable cancel against the partition-function correction.\\

A spectrum-independent sufficient bound follows by writing the residual interaction as \(\hat V=\sum_\mu a_\mu\hat A_\mu\otimes\hat B_\mu\), where the system operators \(\hat A_\mu\) are Pauli strings and therefore satisfy \(\lVert\hat A_\mu\rVert=1\).  Generalized H\"older inequalities for the Kubo--Martin--Schwinger ordered thermal correlators imply, for \(0\leq\tau'\leq\tau\leq\beta\),
\begin{equation}
    \left|\expect{\hat O\hat A_\mu(\tau)\hat A_\nu(\tau')}_0\right|
    \leq \lVert\hat O\rVert\,\lVert\hat A_\mu\rVert\,\lVert\hat A_\nu\rVert
    =\lVert\hat O\rVert.
    \label{eq:holder_bound_cont}
\end{equation}
The same bound applies to the disconnected product in \cref{eq:connected_correction_cont}. Using triangle inequality we can see that the connected system factor is bounded by \(2\lVert\hat O\rVert\).  Consequently, the integrated bath correlation functions derived below provide a sufficient perturbative control parameter without requiring any information about the excitation spectrum of \(\hat H_S^{\flat}\).\\

We now derive the bath correlation functions for a Gaussian local reservoir.  For a single bosonic mode of frequency \(\nu\), the thermal characteristic function of the displacement operator is
\begin{equation}
    \expect{e^{\alpha \hat b^\dag-\alpha^*\hat b}}_{\beta,\nu}
    =\exp\!\left[-\frac{|\alpha|^2}{2}\coth\!\left(\frac{\beta\nu}{2}\right)\right].
    \label{eq:thermal_displacement_characteristic_cont}
\end{equation}
The operator \(\hat B_n\) is a sum of independent single-mode displacement generators.  Since operators belonging to different bath modes commute, its exponential factorizes as
\begin{equation}
    e^{\pm i\hat B_n}
    =\prod_k\exp\!\left[\mp\frac{2t_k}{\nu_k}
    \left(\hat b_{n,k}^{\dag}-\hat b_{n,k}\right)\right],
    \label{eq:displacement_product_cont}
\end{equation}
in which we have assumed that the interaction parameters for all sites are equal, $t_{n,k}=t_k$. Thus each factor in \cref{eq:displacement_product_cont} is of the form appearing in \cref{eq:thermal_displacement_characteristic_cont}, with \(\alpha_k=\mp2t_k/\nu_k\).  Because the free thermal bath state factorizes over its normal modes, the corresponding multimode thermal average is
\begin{equation}
    \expect{e^{i\hat B_n}}_B
    =\expect{e^{-i\hat B_n}}_B
    =\exp\!\left[-2\sum_k\frac{t_k^2}{\nu_k^2}
    \coth\!\left(\frac{\beta\nu_k}{2}\right)\right].
    \label{eq:single_displacement_average_cont}
\end{equation}
Consequently, using \(\hat{\mathcal C}_n=(e^{i\hat B_n}+e^{-i\hat B_n})/2\) and \(\hat{\mathcal S_n}=(e^{i\hat B_n}-e^{-i\hat B_n})/(2i)\) we get

\begin{equation}
    c=\expect{\hat{\mathcal C}_n}_B
    =\exp\!\left[-2\sum_k\frac{t_k^2}{\nu_k^2}
    \coth\!\left(\frac{\beta\nu_k}{2}\right)\right],
    \qquad
    \expect{\hat{\mathcal S}_n}_B=0.
    \label{eq:c_discrete_cont}
\end{equation}
Before calculating the two-time bath correlations that enter \cref{eq:K2_cont}, it is useful to make the system--bath structure of \cref{h_int_pol} explicit.  The Pauli operators act only on the spin-chain Hilbert space, whereas \(\hat{\mathcal C}_n\) and \(\hat{\mathcal S}_n\) are functions only of the bosonic operators of bath \(n\).  With $J_y=J(1-\gamma)/2$ and $c = \expect{\hat{\mathcal C}_n} = \expect{\hat{\mathcal C}_{n+1}}$, the residual interaction can therefore be written as
\begin{equation}
    \hat V
    =h\sum_n\sum_{r=1}^{2}\hat A^{(h)}_{n,r}\otimes\hat R^{(h)}_{n,r}
    +J_y\sum_n\sum_{r=1}^{4}\hat A^{(J)}_{n,r}\otimes\hat R^{(J)}_{n,r},
    \label{eq:V_system_bath_split_cont}
\end{equation}
where the system operators are
\begin{subequations}\label{eq:system_operators_cont}
\begin{align}
    \hat A^{(h)}_{n,1}&=-\hat\sigma_n^z,
    &\hat A^{(h)}_{n,2}&=\hat\sigma_n^y,
    \label{eq:system_field_ops_cont}\\
    \hat A^{(J)}_{n,1}&=-\hat\sigma_n^y\hat\sigma_{n+1}^y,
    &\hat A^{(J)}_{n,2}&=-\hat\sigma_n^z\hat\sigma_{n+1}^z,
    \label{eq:system_bond_ops1_cont}\\
    \hat A^{(J)}_{n,3}&=-\hat\sigma_n^y\hat\sigma_{n+1}^z,
    &\hat A^{(J)}_{n,4}&=-\hat\sigma_n^z\hat\sigma_{n+1}^y,
    \label{eq:system_bond_ops2_cont}
\end{align}
\end{subequations}
and the corresponding bath operators are
\begin{subequations}\label{eq:bath_operators_cont}
\begin{align}
    \hat R^{(h)}_{n,1}&=\hat{\mathcal C}_n-c,
    &\hat R^{(h)}_{n,2}&=\hat{\mathcal S}_n,
    \label{eq:bath_field_ops_cont}\\
    \hat R^{(J)}_{n,1}&=\hat{\mathcal C}_n\hat{\mathcal C}_{n+1}-c^2,
    &\hat R^{(J)}_{n,2}&=\hat{\mathcal S}_n\hat{\mathcal S}_{n+1},
    \label{eq:bath_bond_ops1_cont}\\
    \hat R^{(J)}_{n,3}&=\hat{\mathcal C}_n\hat{\mathcal S}_{n+1},
    &\hat R^{(J)}_{n,4}&=\hat{\mathcal S}_n\hat{\mathcal C}_{n+1}.
    \label{eq:bath_bond_ops2_cont}
\end{align}
\end{subequations}
All operators \(\hat A^{(h,J)}\) are Pauli strings and act only on the system, while all operators \(\hat R^{(h,J)}\) act only on one or two local baths.  Equation~\cref{eq:c_discrete_cont}, together with the statistical independence of different baths, gives \(\expect{\hat R^{(h)}_{n,r}}_B=\expect{\hat R^{(J)}_{n,r}}_B=0\), as required by the definition of \(\hat V\).  Since \(\hat H_0=\hat H_S^{\flat}+\hat H_B\), the imaginary-time evolution of each tensor-product term factorizes,
\begin{equation}
    [\hat A\otimes\hat R](\tau)
    =\hat A(\tau)\otimes\hat R(\tau),
    \quad
    \hat A(\tau)=e^{\tau\hat H_S^{\flat}}\hat A e^{-\tau\hat H_S^{\flat}},
    \quad
    \hat R(\tau)=e^{\tau\hat H_B}\hat R e^{-\tau\hat H_B}.
    \label{eq:imaginary_time_factorization_cont}
\end{equation}
Accordingly, the bath trace in \cref{eq:K2_cont} acts only on products of the \(\hat R\)'s; the corresponding Pauli strings remain system operators.

We now introduce the bath kernel that controls all of these two-time correlations.  For one local reservoir,
\begin{equation}
    \hat B_n(\tau)
    =e^{\tau\hat H_B}\hat B_n e^{-\tau\hat H_B}
    =2i\sum_k\frac{t_k}{\nu_k}
    \left(e^{\nu_k\tau}\hat b_{n,k}^{\dag}-e^{-\nu_k\tau}\hat b_{n,k}\right),
    \label{eq:B_tau_cont}
\end{equation}
where \(e^{\tau\hat H_B}\hat b_{n,k}e^{-\tau\hat H_B}=e^{-\nu_k\tau}\hat b_{n,k}\).  With \(\bar n_k=(e^{\beta\nu_k}-1)^{-1}\), the only nonzero same-mode thermal contractions are \(\expect{\hat b_{n,k}^{\dag}\hat b_{n,k}}_B=\bar n_k\) and \(\expect{\hat b_{n,k}\hat b_{n,k}^{\dag}}_B=\bar n_k+1\).  Substitution into \(\expect{\hat B_n(\tau)\hat B_n(0)}_B\) gives
\begin{equation}
\begin{aligned}
    \frac{1}{4}\expect{\hat B_n(\tau)\hat B_n(0)}_B
    &=\sum_k\frac{t_k^2}{\nu_k^2}
    \left[\bar n_k e^{\nu_k\tau}+(\bar n_k+1)e^{-\nu_k\tau}\right]\\
    &=\sum_k\frac{t_k^2}{\nu_k^2}
    \frac{\cosh[(\beta/2-\tau)\nu_k]}{\sinh(\beta\nu_k/2)}.
\end{aligned}
    \label{eq:B_correlation_cont}
\end{equation}
We therefore define the discrete-mode displacement kernel as
\begin{equation}
    \mathcal A_\beta^{\mathrm{disc}}(\tau)
    \equiv\frac{1}{4}\expect{\hat B_n(\tau)\hat B_n(0)}_B
    =\sum_k\frac{t_k^2}{\nu_k^2}
    \frac{\cosh[(\beta/2-\tau)\nu_k]}{\sinh(\beta\nu_k/2)},
    \qquad 0\leq\tau\leq\beta.
    \label{eq:A_beta_discrete_cont}
\end{equation}

At this point we pass directly to the continuum representation used in the remainder of the calculation.  The spectral density is defined by
\begin{equation}
    K(\omega)=\sum_k t_k^2\delta(\omega-\nu_k),
    \qquad
    \sum_k t_k^2F(\nu_k)=\int_0^\infty\dd\omega\,K(\omega)F(\omega).
    \label{eq:spectral_replacement_cont}
\end{equation}
Applying \cref{eq:spectral_replacement_cont} to \cref{eq:A_beta_discrete_cont} gives
\begin{equation}
    \mathcal A_\beta(\tau)
    \equiv\frac{1}{4}\expect{\hat B_n(\tau)\hat B_n(0)}_B
    =\int_0^\infty \dd\omega\,
    \frac{K(\omega)}{\omega^2}
    \frac{\cosh[(\beta/2-\tau)\omega]}{\sinh(\beta\omega/2)},
    \qquad
    \mathcal A_0\equiv\mathcal A_\beta(0).
    \label{eq:A_beta_cont}
\end{equation}
For the super-Ohmic reservoir considered here,
\begin{equation}
    K(\omega)=g\frac{\omega^3}{\omega_c^2}e^{-\omega/\omega_c}.
    \label{eq:superohmic_K_cont}
\end{equation}
Substituting \cref{eq:superohmic_K_cont} into \cref{eq:A_beta_cont} immediately yields
\begin{equation}
    \mathcal A_\beta(\tau)
    =\frac{g}{\omega_c^2}\int_0^\infty\dd\omega\,
    \omega e^{-\omega/\omega_c}
    \frac{\cosh[(\beta/2-\tau)\omega]}{\sinh(\beta\omega/2)}.
    \label{eq:A_superohmic_integral_cont}
\end{equation}
The thermal factor can be expanded as
\begin{equation}
    \frac{\cosh[(\beta/2-\tau)\omega]}{\sinh(\beta\omega/2)}
    =\frac{e^{-\tau\omega}+e^{-(\beta-\tau)\omega}}{1-e^{-\beta\omega}}
    =\sum_{m=0}^\infty\left[
      e^{-(m\beta+\tau)\omega}
      +e^{-((m+1)\beta-\tau)\omega}
    \right].
    \label{eq:thermal_kernel_series_cont}
\end{equation}
Using \(\int_0^\infty\omega e^{-a\omega}\dd\omega=a^{-2}\), \cref{eq:A_superohmic_integral_cont} becomes
\begin{equation}
    \mathcal A_\beta(\tau)
    =\frac{g}{\omega_c^2}
    \sum_{m=0}^\infty
    \left[
      \frac{1}{(\omega_c^{-1}+m\beta+\tau)^2}
      +\frac{1}{(\omega_c^{-1}+(m+1)\beta-\tau)^2}
    \right].
    \label{eq:A_series_cont}
\end{equation}
Introducing $x\equiv\beta\omega_c$ and $u\equiv\frac{\tau}{\beta}$ and using $\psi_1(z)=\sum_{m=0}^\infty(m+z)^{-2}$, the sum can be written exactly as
\begin{equation}
    \mathcal A_x(u)
    =\frac{g}{x^2}
    \left[
      \psi_1\!\left(u+\frac{1}{x}\right)
      +\psi_1\!\left(1-u+\frac{1}{x}\right)
    \right].
    \label{eq:A_trigamma_cont}
\end{equation}
Setting \(u=0\) and using the recurrence relation \(\psi_1(z+1)=\psi_1(z)-z^{-2}\) gives
\begin{equation}
    \mathcal A_0
    =g\left[
      1+\frac{2}{x^2}\psi_1\!\left(1+\frac{1}{x}\right)
    \right].
    \label{eq:A0_exact_cont}
\end{equation}
Since \(\mathcal A_0\) is the continuum representation of the mode sum in \cref{eq:c_discrete_cont}, the renormalization factor is
\begin{equation}
    c=\expect{\hat{\mathcal C}_n}_B = \expect{\hat{\mathcal{C}}} = e^{-2\mathcal A_0}.
    \label{eq:c_A0_cont}
\end{equation}
Combining \cref{eq:A0_exact_cont,eq:c_A0_cont} and applying the same trigamma recurrence relation gives
\begin{equation}
    c
    =\exp\!\left[
      2g-\frac{4g}{x^2}\psi_1\!\left(\frac{1}{x}\right)
    \right],
    \qquad x=\beta\omega_c,
    \label{eq:c_decay_equiv_cont}
\end{equation}
which is precisely the expression given in~\cref{decay_mag}.  We have therefore determined both the full unequal-imaginary-time kernel \(\mathcal A_\beta(\tau)\) and its equal-separation value \(\mathcal A_0\) before evaluating any of the correlation functions entering the second-order expansion.

We can now calculate the required two-time bath correlations.  Because \(\hat B_n(\tau)\) is linear in the bosonic creation and annihilation operators, all of its thermal moments are Gaussian.  For zero-mean linear bosonic operators \(\hat X\) and \(\hat Y\) whose commutator is a c-number, Gaussian averaging together with the Baker-Campbell-Hausdorff identity gives
\begin{equation}
    \expect{e^{\hat X}e^{\hat Y}}_B
    =\exp\!\left[
      \frac{1}{2}\expect{\hat X^2}_B
      +\frac{1}{2}\expect{\hat Y^2}_B
      +\expect{\hat X\hat Y}_B
    \right].
    \label{eq:gaussian_two_exp_cont}
\end{equation}
For \(\hat X=i\hat B_n(\tau)\) and \(\hat Y=\mp i\hat B_n(0)\), stationarity gives \(\expect{\hat B_n(\tau)^2}_B=\expect{\hat B_n^2}_B=4\mathcal A_0\), while \cref{eq:A_beta_cont} gives \(\expect{\hat B_n(\tau)\hat B_n(0)}_B=4\mathcal A_\beta(\tau)\).  Hence
\begin{subequations}\label{eq:disp_corr_cont}
\begin{align}
    \expect{e^{i\hat B_n(\tau)}e^{-i\hat B_n(0)}}_B
    &=e^{-4[\mathcal A_0-\mathcal A_\beta(\tau)]},
    \label{eq:disp_minus_cont}\\
    \expect{e^{i\hat B_n(\tau)}e^{i\hat B_n(0)}}_B
    &=e^{-4[\mathcal A_0+\mathcal A_\beta(\tau)]}.
    \label{eq:disp_plus_cont}
\end{align}
\end{subequations}
The averages obtained by changing both displacement signs are identical.  Expanding \(\hat{\mathcal C}=(e^{i\hat B}+e^{-i\hat B})/2\) and \(\hat{\mathcal S}=(e^{i\hat B}-e^{-i\hat B})/(2i)\) therefore yields
\begin{subequations}\label{eq:CS_corr_cont}
\begin{align}
    \expect{\hat{\mathcal C}(\tau)\hat{\mathcal C}(0)}_B
    &=c^2\cosh[4\mathcal A_\beta(\tau)],
    \label{eq:CC_corr_cont}\\
    \expect{\hat{\mathcal S}(\tau)\hat{\mathcal S}(0)}_B
    &=c^2\sinh[4\mathcal A_\beta(\tau)],
    \label{eq:SS_corr_cont}\\
    \expect{\hat{\mathcal C}(\tau)\hat{\mathcal S}(0)}_B
    &=\expect{\hat{\mathcal S}(\tau)\hat{\mathcal C}(0)}_B=0.
    \label{eq:CS_zero_cont}
\end{align}
\end{subequations}
These are single-bath correlation functions.  We now use them to evaluate separately the field sector, the bond sector, and the mixed field--bond sector of \cref{eq:V_system_bath_split_cont}.

For the field-dependent part \(\hat H_h^{\flat}\), the two bath factors are \(\hat R^{(h)}_{n,1}=\hat{\mathcal C}_n-c\) and \(\hat R^{(h)}_{n,2}=\hat{\mathcal S}_n\).  Using \cref{eq:CS_corr_cont},
\begin{equation}
\begin{aligned}
    \expect{[\hat{\mathcal C}(\tau)-c][\hat{\mathcal C}(0)-c]}_B
    &=c^2\left\{\cosh[4\mathcal A_\beta(\tau)]-1\right\},\\
    \expect{\hat{\mathcal S}(\tau)\hat{\mathcal S}(0)}_B
    &=c^2\sinh[4\mathcal A_\beta(\tau)],
\end{aligned}
    \label{eq:field_channel_components_cont}
\end{equation}
while the cross correlation between these two bath factors vanishes.  Their summed autocorrelation scale is therefore
\begin{equation}
    G_h(\tau)
    \equiv\sum_{r=1}^{2}
    \expect{\hat R^{(h)}_{n,r}(\tau)\hat R^{(h)}_{n,r}(0)}_B
    =c^2\left[e^{4\mathcal A_\beta(\tau)}-1\right].
    \label{eq:Gh_cont}
\end{equation}
The factor \(h^2\) multiplying this bath function in the second-order expansion comes from the two system-side coefficients in \cref{eq:V_system_bath_split_cont}.

For the nearest-neighbour part \(\hat H_J^{\flat}\), each interaction vertex contains one of the four two-bath operators in \cref{eq:bath_bond_ops1_cont,eq:bath_bond_ops2_cont}.  Since reservoirs \(n\) and \(n+1\) are independent, a two-bath correlation factorizes into a product of the corresponding single-bath correlations.  For example,
\begin{align}
    &\expect{
      [\hat{\mathcal C}_n(\tau)\hat{\mathcal C}_{n+1}(\tau)-c^2]
      [\hat{\mathcal C}_n(0)\hat{\mathcal C}_{n+1}(0)-c^2]
    }_B\nonumber\\
    &\qquad=\expect{\hat{\mathcal C}(\tau)\hat{\mathcal C}(0)}_B^2-c^4
    =c^4\sinh^2[4\mathcal A_\beta(\tau)],
    \label{eq:CCCC_corr_cont}
\end{align}
and
\begin{equation}
    \expect{
      \hat{\mathcal S}_n(\tau)\hat{\mathcal S}_{n+1}(\tau)
      \hat{\mathcal S}_n(0)\hat{\mathcal S}_{n+1}(0)
    }_B
    =c^4\sinh^2[4\mathcal A_\beta(\tau)].
    \label{eq:SSSS_corr_cont}
\end{equation}
The two remaining channels are
\begin{equation}
\begin{aligned}
    &\expect{
      \hat{\mathcal C}_n(\tau)\hat{\mathcal S}_{n+1}(\tau)
      \hat{\mathcal C}_n(0)\hat{\mathcal S}_{n+1}(0)
    }_B\\
    &\quad=\expect{
      \hat{\mathcal S}_n(\tau)\hat{\mathcal C}_{n+1}(\tau)
      \hat{\mathcal S}_n(0)\hat{\mathcal C}_{n+1}(0)
    }_B\\
    &\quad=c^4\cosh[4\mathcal A_\beta(\tau)]\sinh[4\mathcal A_\beta(\tau)].
\end{aligned}
    \label{eq:CSCS_corr_cont}
\end{equation}
Cross correlations between distinct bond channels vanish because they contain at least one single-bath \(\mathcal C\)-\(\mathcal S\) correlator from \cref{eq:CS_zero_cont}.  Summing the four nonzero autocorrelation channels gives
\begin{equation}
\begin{aligned}
    G_J(\tau)
    &\equiv\sum_{r=1}^{4}
    \expect{\hat R^{(J)}_{n,r}(\tau)\hat R^{(J)}_{n,r}(0)}_B\\
    &=c^4\left[e^{8\mathcal A_\beta(\tau)}-1\right].
\end{aligned}
    \label{eq:GJ_cont}
\end{equation}
The corresponding second-order coefficient is \(J_y^2\).

Mixed field--bond contractions occur only when the field vertex and the bond vertex share a bath.  For a field vertex on site \(n\) and the bond \((n,n+1)\), the two nonzero contractions are
\begin{equation}
\begin{aligned}
    &\expect{[\hat{\mathcal C}_n(\tau)-c]
      [\hat{\mathcal C}_n(0)\hat{\mathcal C}_{n+1}(0)-c^2]}_B
      =c^3\left\{\cosh[4\mathcal A_\beta(\tau)]-1\right\},\\
    &\expect{\hat{\mathcal S}_n(\tau)
      \hat{\mathcal S}_n(0)\hat{\mathcal C}_{n+1}(0)}_B
      =c^3\sinh[4\mathcal A_\beta(\tau)].
\end{aligned}
    \label{eq:mixed_channel_components_cont}
\end{equation}
Their sum defines
\begin{equation}
    G_{hJ}(\tau)
    =c^3\left[e^{4\mathcal A_\beta(\tau)}-1\right].
    \label{eq:GhJ_cont}
\end{equation}
Since \(\mathcal A_\beta(\tau)\geq0\) for \(0\leq\tau\leq\beta\),
\begin{equation}
    G_{hJ}(\tau)^2\leq G_h(\tau)G_J(\tau).
    \label{eq:mixed_bound_cont}
\end{equation}
Thus the mixed term, proportional to \(hJ_y\), is controlled by the geometric mean of the field and bond fluctuation scales and does not introduce a third independent small parameter.

We now integrate the bath correlation functions over imaginary-time as it appears in~\cref{eq:K2_cont}.  It is convenient to define
\begin{equation}
    \mathcal G_p(g,x)
    \equiv
    \int_0^1\dd u\,(1-u)
    e^{-p\mathcal A_0}
    \left[e^{p\mathcal A_x(u)}-1\right],
    \qquad p>0.
    \label{eq:Fp_cont}
\end{equation}
Stationarity implies that a bath correlator depends only on the separation \(s=\tau-\tau'\), and therefore
\begin{equation}
    \int_0^\beta\dd\tau\int_0^\tau\dd\tau'\,G(\tau-\tau')
    =\int_0^\beta\dd s\,(\beta-s)G(s).
    \label{eq:triangle_to_separation_cont}
\end{equation}
Using \cref{eq:Gh_cont,eq:GJ_cont} together with \(s=\beta u\) gives
\begin{subequations}\label{eq:integrated_channels_cont}
\begin{align}
    \int_0^\beta \dd s\,(\beta-s)G_h(s)
    &=\beta^2\mathcal G_4(g,\beta\omega_c),
    \label{eq:int_Gh_cont}\\
    \int_0^\beta \dd s\,(\beta-s)G_J(s)
    &=\beta^2\mathcal G_8(g,\beta\omega_c).
    \label{eq:int_GJ_cont}
\end{align}
\end{subequations}
Restoring the system-side coefficients \(h\) and \(J_y\) from \cref{eq:V_system_bath_split_cont}, the two independent dimensionless second-order bath-controlled parameters are
\begin{equation}
    \epsilon_h
    \equiv \beta^2h^2\mathcal F_4(g,\beta\omega_c),
    \qquad
    \epsilon_J
    \equiv \beta^2J_y^2\mathcal F_8(g,\beta\omega_c).
    \label{eq:epsilon_channels_cont}
\end{equation}
The mixed contribution is bounded by a quantity proportional to \(\sqrt{\epsilon_h\epsilon_J}\) as a consequence of \cref{eq:mixed_bound_cont}. A sufficient perturbative condition can be given as
\begin{equation}
    \epsilon_h\ll1,
    \qquad
    \epsilon_J\ll1.
    \label{eq:general_criterion_cont}
\end{equation}
Equivalently,
\begin{equation}
    \max\!\left\{
      \beta^2h^2\mathcal G_4(g,\beta\omega_c),
      \beta^2J_y^2\mathcal G_8(g,\beta\omega_c)
    \right\}\ll1.
    \label{eq:general_criterion_max_cont}
\end{equation}
We finally examine the limiting forms of the general criterion.  For \(x=\beta\omega_c\gg1\),
\begin{equation}
    \mathcal A_0=g+O(x^{-2}),
    \qquad
    c=e^{-2g}\left[1+O(x^{-2})\right].
    \label{eq:lowT_A0_cont}
\end{equation}
The integral in \cref{eq:Fp_cont} is dominated by the boundary layer \(u=O(x^{-1})\), for which
\begin{equation}
    \mathcal A_x(u)
    =\frac{g}{(1+xu)^2}+O(x^{-2}).
    \label{eq:boundary_layer_cont}
\end{equation}
Setting \(y=xu\) then gives
\begin{equation}
    \mathcal G_p(g,x)
    =\frac{f(pg)}{x}+O(x^{-2}),
    \label{eq:Fp_lowT_cont}
\end{equation}
where
\begin{equation}
    f(a)
    \equiv
    e^{-a}\int_0^\infty\dd y\,
    \left[e^{a/(1+y)^2}-1\right].
    \label{eq:f_integral_cont}
\end{equation}
With \(z=(1+y)^{-1}\) followed by an integration by parts,
\begin{equation}
    f(a)=e^{-a}-1
    +e^{-a}\sqrt{\pi a}\,\erfi(\sqrt a).
    \label{eq:f_closed_cont}
\end{equation}
Thus, for \(\beta\omega_c\gg1\),
\begin{equation}
    \frac{\beta h^2}{\omega_c}f(4g)\ll1,
    \qquad
    \frac{\beta J_y^2}{\omega_c}f(8g)\ll1.
    \label{eq:criterion_lowT_cont}
\end{equation}
At fixed temperature and coupling strength, increasing \(\omega_c\) therefore suppresses the integrated residual interaction as \(1/\omega_c\).

The weak- and strong-coupling limits of \cref{eq:f_closed_cont} are
\begin{equation}
    f(a)=a+O(a^2),\qquad a\ll1,
    \label{eq:f_weak_cont}
\end{equation}
and
\begin{equation}
    f(a)=\frac{1}{2a}+O(a^{-2}),\qquad a\gg1.
    \label{eq:f_strong_cont}
\end{equation}
Consequently, for \(g\ll1\),
\begin{equation}
    \epsilon_h\simeq4g\frac{\beta h^2}{\omega_c},
    \qquad
    \epsilon_J\simeq8g\frac{\beta J_y^2}{\omega_c},
    \label{eq:weak_g_eps_cont}
\end{equation}
whereas for \(g\gg1\),
\begin{equation}
    \epsilon_h\simeq\frac{\beta h^2}{8g\omega_c},
    \qquad
    \epsilon_J\simeq\frac{\beta J_y^2}{16g\omega_c}.
    \label{eq:strong_g_eps_cont}
\end{equation}
The latter behavior reflects the increasingly rapid decay of the residual polaron-frame fluctuations at strong displacement.

In the opposite limit \(x=\beta\omega_c\ll1\), expansion of \cref{eq:A_trigamma_cont} gives
\begin{equation}
    \mathcal A_x(u)
    =\frac{2g}{x}
    +gx\left(\frac{1}{3}-2u+2u^2\right)
    +O(x^3),
    \label{eq:A_highT_cont}
\end{equation}
and therefore
\begin{equation}
    \mathcal A_0
    =\frac{2g}{x}+\frac{gx}{3}+O(x^3).
    \label{eq:A0_highT_cont}
\end{equation}
For fixed nonzero \(g\), \(e^{-p\mathcal A_0}\) is exponentially small as \(x\to0\), while
\begin{equation}
    e^{-p[\mathcal A_0-\mathcal A_x(u)]}
    =e^{-2pgx\,u(1-u)+O(x^3)}.
    \label{eq:highT_exp_cont}
\end{equation}
Substitution into \cref{eq:Fp_cont} yields
\begin{equation}
    \mathcal G_p(g,x)
    =\frac{1}{2}-\frac{pg}{6}x+O(x^2).
    \label{eq:Fp_highT_cont}
\end{equation}
The leading high-temperature condition is therefore
\begin{equation}
    \frac{\beta^2h^2}{2}\ll1,
    \qquad
    \frac{\beta^2J_y^2}{2}\ll1,
    \label{eq:criterion_highT_cont}
\end{equation}
with the first cutoff-dependent corrections of relative order \(g\beta\omega_c\).  The absence of \(\omega_c\) from the leading term follows from the fact that, when \(\beta\omega_c\ll1\), the complete imaginary-time interval is shorter than the bath correlation time \(\omega_c^{-1}\).

\cref{eq:general_criterion_cont} provide the finite-temperature validity criterion without requiring a high- or low-temperature approximation.  Numerically, \(\mathcal G_4\) and \(\mathcal G_8\) are one-dimensional integrals and can be evaluated directly for the chosen \(g\), \(\beta\), and \(\omega_c\).  

\section{Structure of the Hamiltonian in its Diagonal Basis}
\label{sec:hamiltonian_diag}
In what follows, we will discuss the details regarding the eigenvalues and eigenvectors of $\hat{H}_S$ and related topics. All of the contents that will follow until the end of this section is also applicable to $\hat{H}_S^{\flat}$ upon substituting the parameters $(h,\gamma,J) \rightarrow (h^{\flat},\gamma^{\flat},J^{\flat})$. Jordan-Wigner transformation maps Pauli operators to fermionic operators as
\begin{equation}\label{jw_ops}
\begin{aligned}
&\hat{\sigma}_n^x = (\hat{c}_n^{\dag}+\hat{c}_n)\prod_{m<n}(\hat{I}_m-2\hat{c}_m^{\dag}\hat{c}_m),\\
&\hat{\sigma}_n^y = -i(\hat{c}_n^{\dag}-\hat{c}_n)\prod_{m<n}(\hat{I}_m-2\hat{c}_m^{\dag}\hat{c}_m),\\
&\hat{\sigma}_n^z = \hat{I}_n - 2\hat{c}_n^{\dag}\hat{c}_n,
\end{aligned}  
\end{equation}
in which $\hat{I}_n$ is the identity operator acting on site $n$ and $\hat{c}_n$ is the fermionic annihilation operator at site $n$. Fourier transformation maps the fermionic operators from site basis to Fourier basis as
\begin{equation}\label{fourier_ops}
\hat{c}_{n}=\frac{e^{-i\pi/4}}{\sqrt{N}}\sum_{k}\hat{c}_{k}e^{ink},\quad \hat{c}_{n}^{\dagger}=\frac{e^{i\pi/4}}{\sqrt{N}}\sum_{k}\hat{c}_{k}^{\dagger}e^{-ink}.
\end{equation}
in which $\hat{c}_{k}$ is the fermionic annihilation operator for mode $k$. We name the quasi-particles in this basis as $c$-fermions. Finally, Bogoliubov-Valatin transformation maps the fermionic operators from Fourier basis to Bogoliubov basis, in which the Hamiltonian becomes diagonal. This transformation is given as 
\begin{equation}\label{bogoliubov_ops}
  \hat{\eta}_k = \text{cos}\left(\frac{\theta_k}{2}\right)\hat{c}_k - i\text{sin}\left(\frac{\theta_k}{2}\right)\hat{c}_{-k}^{\dag},\quad \hat{\eta}_{-k}^{\dag} = \text{cos}\left(\frac{\theta_k}{2}\right)\hat{c}_{-k}^{\dag} - i\text{sin}\left(\frac{\theta_k}{2}\right)\hat{c}_{k},
\end{equation}
in which $\hat{\eta}_{k}$ is the fermionic annihilation operator for mode $k$ for a quasi-particle in Bogoliubov basis, which we call $\eta$-fermion. The Bogoliubov angle $\theta_k$ is defined as
\begin{equation}\label{bog_angle}
  \theta_k = \text{arctan}\left(\frac{J\gamma\text{sin}(k)}{h-J\text{cos}(k)}\right),
\end{equation}
Note that for the case of SC, the Bogoliubov angle $\theta_k^{\flat}$ depends on $\beta$ through temperature-dependence of $h^{\flat}$ and $J^{\flat}$. Assuming a periodic boundary condition and even $N$ along with the fact that parity is conserved for the Hamiltonian~\cref{xy_hamiltonian}, we can diagonalize it by consecutively applying the transformations~\cref{jw_ops},~\cref{fourier_ops} and~\cref{bogoliubov_ops}. The diagonalized Hamiltonian can be written as a direct sum of two Hamiltonians, one in positive and the other in negative parity sector~\cite{LIEB1961407,PFEUTY197079,10.21468/SciPostPhysLectNotes.82, 10.21468/SciPostPhys.11.1.013}.
\begin{equation}\label{diag_hamiltonian}
\begin{aligned}
  &\hat{\mathcal{H}}^+ = \sum_{k\in \mathbf{K^+}} \epsilon_k (\hat{\eta}_k^{\dag}\hat{\eta}_k - \frac{1}{2}) -Nh\hat{\Pi}^+,\\
  &\hat{\mathcal{H}}^- = \sum_{k\in \mathbf{K^-}\setminus \{0,\pi\}} \epsilon_k (\hat{\eta}_k^{\dag}\hat{\eta}_k - \frac{1}{2})+\epsilon_0(\hat{\eta}_0^{\dag}\hat{\eta}_0 - \frac{1}{2})+\epsilon_{\pi}(\hat{\eta}_{\pi}^{\dag}\hat{\eta}_{\pi} - \frac{1}{2}) -Nh\hat{\Pi}^-
\end{aligned}
\end{equation}
The parity operator $\hat{\Pi}$ and the projection operators on the positive and negative parity sectors are defined as
\begin{equation}\label{parity_operator}
  \hat{\Pi} = \prod_{i=1}^{N}\hat{\sigma}_i^z,\quad \hat{\Pi}^{\pm} = \frac{1}{2}(1\pm \hat{\Pi}).
\end{equation}
In~\cref{diag_hamiltonian}, $\hat{\eta}_k$ is the annihilation operator for the fermionic Bogoliubov quasiparticle in mode $k$ and its energy is
\begin{equation}\label{epsilon_k}
  \epsilon_k = 2\sqrt{(h-J\text{cos}k)^2+J^2\gamma^2\text{sin}^2k}.
\end{equation}
The modes $k=0$ and $k=\pi$ are not paired with any mode and their energies are $\epsilon_0 = 2(h-J)$ and $\epsilon_{\pi} = 2(h+J)$, respectively. The energy eigenvalues of the Hamiltonian can be written as a sum of $\epsilon_k$ values for different quasiparticle modes taking into account the parity 
symmetry of the Hamiltonian. The reader can refer to~\cref{sec:finite_partition} for more details. Defining the sets $\mathbf{k^+}$ and 
$\mathbf{k^-}$ as
\begin{equation}\label{small_k_modes}
\begin{aligned}
  &\mathbf{k^+} = \left\{\frac{(2\ell + 1)\pi}{N}\middle| \ell = 0,1,...,\frac{N}{2}-1 \right\},\\
  &\mathbf{k^-} = \left\{\frac{2\ell\pi}{N}\middle| \ell = 1,...,\frac{N}{2}-1 \right\},
\end{aligned}
\end{equation}
we can express the set of momentum modes corresponding to the positive and negative parity sectors $\mathbf{K^+}$ and $\mathbf{K^-}$ for even $N$ as
\begin{equation}\label{big_k_modes}
  \mathbf{K^+} = \mathbf{k^+}\cup \{-\mathbf{k^+}\},\quad \mathbf{K^-} = \mathbf{k^-}\cup \{-\mathbf{k^-}\}\cup\{0,\pi \}.
\end{equation}
the ground state can be written as a product state in the basis of $c$-fermions as
\begin{equation}\label{bcs_ground}
\begin{aligned}
  &\ket{0^+}_{\eta}=\prod_{k\in \mathbf{k^+}}\left(\text{cos}(\frac{\theta_k}{2})+i\text{sin}(\frac{\theta_k}{2}) \hat{c}_k^{\dag}\hat{c}_{-k}^{\dag}\right)\ket{0_{k}0_{-k}}_c,\\
  &\ket{0^-}_{\eta}=\hat{c}_0^{\dag}\prod_{k\in \mathbf{k^-}}\left(\text{cos}(\frac{\theta_k}{2})+i\text{sin}(\frac{\theta_k}{2}) \hat{c}_k^{\dag}\hat{c}_{-k}^{\dag}\right)\ket{0_{k}0_{-k}}_c.
\end{aligned}
\end{equation}
For the sake of convenience, for $k\in \mathbf{k^{\pm}}$ we define $(\text{cos}(\theta_k/2)+i\text{sin}(\theta_k/2) \hat{c}_k^{\dag}\hat{c}_{-k}^{\dag})\ket{0_{k}0_{-k}}_c = \ket{0_k^{\pm}}_{\eta}$. 
Using this we can write the ground states of each parity sector as product states $\ket{0^+}_{\eta} = \bigotimes_{k\in \mathbf{k^+}}\ket{0_k^+}_{\eta}$ and $\ket{0^-}_{\eta} = \hat{c}_0^{\dag}\bigotimes_{k\in \mathbf{k^+}}\ket{0_k^-}_{\eta}$. 
The Eigenvectors of the Hamiltonian in each parity sector can be obtained by acting on the ground state of that parity sector with the creation operators of its corresponding modes. Note that for $\mathcal{H}_S^{\flat}$, we define $\mathcal{H}^{\flat}$, $\epsilon_k^{\flat}$, $\mathcal{Z}^{\flat}$ and $\theta_k^{\flat}$ similar to the definitions provided in this section simply by replacing the parameters of the Hamiltonian as mentioned earlier.
\section{Derivation of the Finite-Size Partition Function}
\label{sec:finite_partition}
The equilibrium state can be written in its diagonal basis as
\begin{equation}\label{eq_state}
  \hat{\rho}_{\text{eq}}=\frac{\hat{\rho}}{\mathcal{Z}} = \frac{\text{exp}(-\beta\hat{\mathcal{H}})}{\mathcal{Z}},
\end{equation}
in which $\mathcal{Z}=\text{Tr}[\text{exp}(-\beta\hat{\mathcal{H}})]$ is the partition function that we intend to calculate. The diagonalized Hamiltonian is given in~\cref{diag_hamiltonian}. Using this we can write
\begin{equation}\label{eq_sectors}
  \hat{\rho} = \text{exp}\left[-\beta(\hat{\Pi}^+\hat{\mathcal{H}}^+\hat{\Pi}^+ + \hat{\Pi}^-\hat{\mathcal{H}}^-\hat{\Pi}^-)\right] = \hat{\rho}^+ \oplus \hat{\rho}^-,
\end{equation}
in which the equilibrium state is written as a direct sum of contributions from positive and negative parity sectors. With these considerations, we can write the partition function as
\begin{equation}\label{partition_sectors}
  \text{Tr}(\hat{\rho}) = \text{Tr}(\hat{\rho}^+) + \text{Tr}(\hat{\rho}^-) \Rightarrow \mathcal{Z} = \mathcal{Z}^+ + \mathcal{Z}^-.
\end{equation}
This suggests that to calculate the total partition function, we need to find the partition functions for each parity sector. We can write the equilibrium state in terms of the eigenstates of the diagonalized Hamiltonian.
\begin{equation}\label{eq_diag}
  \hat{\rho}_{\text{eq}} = \sum_{n} \left(\frac{e^{-\beta E_n^+}}{\mathcal{Z}}\ketbra{E_n^+}{E_n^+} + \frac{e^{-\beta E_n^-}}{\mathcal{Z}}\ketbra{E_n^-}{E_n^-} \right).
\end{equation}
Here, $\ket{E_n^{\pm}}$ are the eigenstates of $\hat{\mathcal{H}}^{\pm}$ with eigenenergies $E_n^{\pm}$ and the probabilities in each parity sector are given by $p_n^{\pm}=\text{exp}(-\beta E_n^{\pm})/\mathcal{Z}$. The eigenenergies can be written as the summation of the ground state energy and energetic contributions from adding fermionic quasiparticles. For now, we neglect the constant terms $-Nh\hat{\Pi}^+$ and $-Nh\hat{\Pi}^-$ in~\cref{diag_hamiltonian}. Let's first focus on the positive parity sector. Upon inspection of~\cref{diag_hamiltonian} it is evident that the ground state energy of $\hat{\mathcal{H}}^+$ is given by 
\begin{equation}\label{gs_energy_p}
  E_g^+ = -\frac{1}{2}\sum_{k\in \mathbf{K^+}}\epsilon_k.
\end{equation}
Due to parity conservation, the full spectrum of $\hat{\mathcal{H}}^+$ can be obtained by adding an even number of fermionic quasiparticles to the ground state.
\begin{equation}\label{energy_p}
  E_k^+ = E_g^+ + \sum_{k\in \mathcal{P}_{e}(\mathbf{K^+})} \epsilon_k.
\end{equation}
Here, $\mathcal{P}_{e}(\mathbf{K^+})$ is the subset of the power set of $\mathbf{K^+}$ with an even number of terms. For the negative parity sector assuming even $N$, the ground state contains the mode $k=0$ and doesn't contain the mode $k=\pi$. Upon 
inspecting~\cref{diag_hamiltonian} we can write the ground state energy as
\begin{equation}\label{gs_energy_n}
  E_g^- = -\frac{1}{2}\sum_{k\in \mathbf{K^-}}s(k)\epsilon_k
\end{equation}
in which $s(k)=0$ if $k=0$ and $s(k)=1$ for all other values of $k$. The full spectrum is found by
\begin{equation}\label{energy_n}
  E_k^- = E_g^- + \sum_{k\in \mathcal{P}_{e}(\mathbf{K^-})} \epsilon_k.
\end{equation}
The partition function for the positive parity sector can be written as
\begin{equation}\label{partition_p_def}
  \mathcal{Z}^+ = \text{exp}\left[-\beta(E_g^+ + \sum_{k\in \mathcal{P}_{e}(\mathbf{K^+})} \epsilon_k) \right].
\end{equation}
Assume that $\vec{n}^+ = (n_1,...,n_{N/2})$ is a vector in which $n_k=\{0,1\}$, $k\in \mathbf{K^+}$ and each $n_k$ shows if the mode $k$ is occupied or not. The restriction $k\in \mathcal{P}_{e}(\mathbf{K^+})$ can be re-stated in terms of the total number of quasiparticles such that $\sum_{k}n_k = M$ is an even number. Using this we write
\begin{equation}\label{partition_p_condition}
  \mathcal{Z^+} = \text{exp}(-\beta E_g^+)\text{exp}(-\beta\sum_{\text{even}M}\epsilon_k)=\mathcal{Z}_0^+\text{exp}(-\beta\sum_{\text{even}M}\epsilon_k).
\end{equation} 
In~\cref{partition_p_condition} we defined $\mathcal{Z}_0^+=\text{exp}(-\beta E_g^+)$. Using~\cref{gs_energy_p} we can write
\begin{equation}\label{z0_p}
  \mathcal{Z}_0^+ = \text{exp}(-\frac{1}{2}\beta\sum_{k\in \mathbf{K^+}}\epsilon_k) = \prod_{k\in \mathbf{K^+}}e^{-\frac{\beta\epsilon_k}{2}}
\end{equation}
 Now we turn our attention back to~\cref{partition_p_condition}. To ensure that $M$ is even we can write
\begin{equation}\label{even_m}
  \text{exp}(-\beta\sum_{\text{even}M}\epsilon_k) = \sum_{\vec{n}^+}\frac{1+(-1)^{\sum_{k\in \mathbf{K^+}}n_k}}{2}\text{exp}(-\beta\sum_{k\in \mathbf{K^+}}n_k\epsilon_k),
\end{equation}
in which $\sum_{\vec{n}^+}$ signifies sum over all configurations. We can divide the summation into two parts.
\begin{equation}\label{even_m_divide}
\text{exp}(-\beta\sum_{\text{even}M}\epsilon_k) = \frac{1}{2}\left[\sum_{\vec{n}^+}\text{exp}(-\beta\sum_{k\in \mathbf{K^+}}n_k\epsilon_k)
 + \sum_{\vec{n}^+}(-1)^{\sum_{k\in \mathbf{K^+}}n_k}\text{exp}(-\beta\sum_{k\in \mathbf{K^+}}n_k\epsilon_k) \right].
\end{equation}
Let's call the first and second terms inside the square brackets $\mathcal{Z}_1^+$ and $\mathcal{Z}_2^+$ respectively. Using properties of exponential function, 
we can factorize each term. For $\mathcal{Z}_1^+$ we get
\begin{equation}\label{z1_p}
  \mathcal{Z}_1^+ = \sum_{\vec{n}^+}\text{exp}(-\beta\sum_{k\in \mathbf{K^+}}n_k\epsilon_k)=\prod_{k\in \mathbf{K^+}}\sum_{n_k=0}^{1}\text{exp}(-\beta n_k\epsilon_k)
  = \prod_{k\in \mathbf{K^+}}(1+e^{-\beta\epsilon_k}).
\end{equation}
Similarly for $\mathcal{Z}_2^+$ we can write
\begin{equation}\label{z2_p}
  \mathcal{Z}_2^+ = \sum_{\vec{n}^+}(-1)^{\sum_{k\in \mathbf{K^+}}n_k}\text{exp}(-\beta\sum_{k\in \mathbf{K^+}}n_k\epsilon_k)=\prod_{k\in \mathbf{K^+}}\sum_{n_k=0}^{1}(-1)^{n_k}\text{exp}(-\beta n_k\epsilon_k)
  = \prod_{k\in \mathbf{K^+}}(1-e^{-\beta\epsilon_k}).
\end{equation}
Putting it all together we can write $\mathcal{Z}^+ = \mathcal{Z}_0^+[\mathcal{Z}_1^+ + \mathcal{Z}_2^+]/2$. Upon substitution we get
\begin{equation}\label{partition_p_tot}
  \mathcal{Z}^+ = \frac{1}{2}\left[\prod_{k\in \mathbf{K^+}}2\text{cosh}(\frac{\beta\epsilon_k}{2}) + \prod_{k\in \mathbf{K^+}}2\text{sinh}(\frac{\beta\epsilon_k}{2})\right].
\end{equation}
For $\mathcal{Z}^-$ all derivation steps are similar except that the ground state energy contains the factor $s(k)$ as shown in~\cref{gs_energy_n} which takes the value $-1$ only for mode $k=0$ and is equal to $1$ for all other modes. Therefore we will have
\begin{equation}\label{partition_n_tot}
  \mathcal{Z}^- = \frac{1}{2}\left[\prod_{k\in \mathbf{K^-}}2\text{cosh}(\frac{\beta\epsilon_k}{2}) + \left(e^{\frac{s(0)\beta\epsilon_0}{2}}-e^{-\frac{s(0)\beta\epsilon_0}{2}} \right) \prod_{k\in \mathbf{K^-}\setminus \{0\}}2\text{sinh}(\frac{\beta\epsilon_k}{2})\right]
  = \frac{1}{2}\left[\prod_{k\in \mathbf{K^-}}2\text{cosh}(\frac{\beta\epsilon_k}{2}) - \prod_{k\in \mathbf{K^-}}2\text{sinh}(\frac{\beta\epsilon_k}{2})\right].
\end{equation}
The total partition function is found by summing $\mathcal{Z}^+$ and $\mathcal{Z}^-$ and multiplying this summation by $\text{exp}(Nh\beta)$, which takes into account the contribution of the constant term $-Nh$. Note that for calculating thermal occupation probabilities and their derivatives, the overall factor $\text{exp}(Nh\beta)$ is irrelevant and can be neglected. The final result becomes~\cref{partition_total}. 

\section{Preliminaries in Quantum Metrology}
\label{sec:intro_metrology}
In this section we introduce some of the fundamental results for quantum single parameter estimation. Assume that through a process, the parameter $\alpha$ gets encoded on a quantum state. Quantum Cramer-Rao bound states that the ultimate precision limit for a parameterized state is given by the reciprocal of its QFI. In other words, the lower bound for the variance of any unbiased estimator $\tilde{\alpha}$ of the parameter, is given by 
the quantum Cramer-Rao bound (QCRB).
\begin{equation}\label{qcrb}
  (\Delta\tilde{\alpha})^2 \geq \frac{1}{\nu\mathcal{F}(\alpha)}.
\end{equation}
Here, $\mathcal{F}$ is the QFI and $\nu$ is the number of measurement repetitions. QFI is a generalization of the classical Fisher information (CFI) which is defined as the expectation value of the squared derivative of the logarithm of the probability distribution. The ultimate precision limit for estimating a parameter using quantum probes is dubbed the Heisenberg limit (HL), in which the estimation precision increases quadratically with the number of resources. This is in contrast with the standard quantum limit (SQL) attainable using classical probes, in which the precision scales linearly with the number of probes. Therefore, the main goal in quantum metrology is to find the optimal probe state and optimal measurement operator such that the QFI scales quadratically with the number of resources. 

For a pure state QFI is defined as,
\begin{equation}\label{pureqfi}
  \mathcal{F}(\alpha) = 4[\langle \partial_{\alpha}\psi|\partial_{\alpha}\psi \rangle - |\langle \partial_{\alpha}\psi|\psi \rangle|^2].
\end{equation}
For a mixed state, one can define QFI using the symmetric logarithmic derivative (SLD) operator as
\begin{equation}\label{mixedqf}
  \mathcal{F}(\alpha) = \text{Tr}[\hat{\rho} \hat{L}_{\alpha}^2].
\end{equation}
SLD is implicitly defined by the following equation
\begin{equation}\label{sld}
  \partial_{\alpha}\hat{\rho} = \frac{\hat{L}_{\alpha}\hat{\rho} + \hat{\rho} \hat{L}_{\alpha}}{2}.
\end{equation}
If the eigendecomposition of $\hat{\rho}$ gives $\sum_n p_n(\alpha)\ketbra{n(\alpha)}{n(\alpha)}$ and the eigenbasis set $\{\ket{n(\alpha)}\}$ is complete, we can express the SLD in this basis
\begin{equation}\label{sld_expand}
  L_{\alpha} = \sum_{(m,n)|p_m+p_n\neq 0}\frac{2|\langle m|\partial_{\alpha}\hat{\rho}|n\rangle|^2}{p_m + p_n}.
\end{equation}
Using this explicit form of SLD, we can calculate the QFI to obtain~\cref{qfi}. 

\section{Connection Between Fisher Information and Thermodynamics}
\label{sec:fisher_thermo}
For a system in thermal equilibrium, we can establish a relationship between the classical contribution to the QFI and its partition function. We can write
\begin{equation}\label{cfi_probs}
  \mathcal{F}^c(\alpha) = \sum_{n}p_n(\alpha)\left(\frac{\partial \text{ln}(p_n(\alpha))}{\partial \alpha} \right)^2 
  = \left\langle \left(\frac{\partial \text{ln}(p_n(\alpha))}{\partial \alpha} \right)^2  \right\rangle
\end{equation}
In thermal equilibrium, for canonical ensemble, the probability distribution and its parametric derivative are given by
\begin{equation}\label{thermal_dist}
p_n(\alpha) =  \frac{e^{-\beta E_n}}{\mathcal{Z}},\quad \partial_{\alpha}p_n = p_n\left[-\partial_{\alpha}(\beta E_n)-\partial_{\alpha}\text{ln}\mathcal{Z} \right] = p_n\left[-\braopket{n}{\partial_{\alpha}(\beta\hat{H})}{n}+\expect{\partial_{\alpha}(\beta \hat{H})}\right]
\end{equation}
in which $E_n$ is the energy for the microstate $n$ with probability $p_n$. Note that the derivatives in~\cref{thermal_dist} is written such that it includes the case when the parameter $\alpha=\beta$. Substituting~\cref{thermal_dist} into~\cref{cfi_probs} we get
\begin{equation}\label{thermal_fisher}
\mathcal{F}^c(\alpha) = \left[\text{Tr}\left(\hat{\rho}\left[\frac{\partial(\beta\hat{H})}{\partial \alpha}\right]^2\right)-\left(\text{Tr}\left(\hat{\rho}\left[\frac{\partial(\beta\hat{H})}{\partial \alpha}\right]\right)\right)^2 \right] = \braket{\left(\frac{\partial(\beta\hat{H})}{\partial \alpha}\right)^2}-\braket{\frac{\partial(\beta\hat{H})}{\partial \alpha}}^2
\end{equation} 
Free entropy (Massieu potential) is defined as $\psi = -\beta F = \text{ln}(\mathcal{Z})$. Taking second derivative of this equation with respect to $\alpha$
one can verify that
\begin{equation}\label{fisher_free}
\frac{\partial^2\psi}{\partial \alpha^2} = \braket{\left(\frac{\partial(\beta\hat{H})}{\partial \alpha}\right)^2}-\braket{\frac{\partial(\beta\hat{H})}{\partial \alpha}}^2 - \braket{\frac{\partial^2(\beta \hat{H})}{\partial \alpha^2}}
= \mathcal{F}^c(\alpha) - \braket{\frac{\partial^2(\beta \hat{H})}{\partial \alpha^2}}.
\end{equation} 
\section{Thermodynamics of Small Systems: Phenomenological vs Microscopic}
\label{sec:small_thermo}
Take the case of a system interacting with a thermal environment. For the super-system comprised of the system and bath, the Hamiltonian is $\hat{H}_{\text{tot}} = \hat{H}_S + \hat{H}_I + \hat{H}_B$.
For a small system (i.e. a system in which the effective interaction length is comparable with system size), we can no longer assume the equilibrium state to be in the Gibbs state of the system's bare Hamiltonian. For such a system, the interaction Hamiltonian is not negligible compared to the system Hamiltonian and we should take into account its contribution to the equilibrium state. Therefore, one should resort to SC thermodynamics not only for short-range and strong system-bath interaction potentials, but also for relatively weak long-range (compared to the typical linear dimension of the system) interactions~\cite{RevModPhys.92.041002}.
For such systems thermodynamic functions such as energy and entropy are non-extensive and non-additive. It is also worth mentioning that such non-additivity also appears for weak and short-ranged system-bath interaction but with a long-range intra-system interaction~\cite{10.1093/acprof:oso/9780199581931.001.0001}. Within the frameworks of quantum thermodynamics and stochastic thermodynamics, assuming that the interaction between the system and bath is not weak, the equilibrium state of the system is given via the mean force Gibbs state (MFG)
\begin{equation}\label{mfg}
  \hat{\rho}^{*}=\frac{e^{-\beta \hat{H}_S^*}}{\mathcal{Z}_S^*}
\end{equation}
in which $\hat{H}_S^*$ is the Hamiltonian of mean force (HMF) and $\mathcal{Z}_S^*$ is the partition function associated with the HMF. HMF is defined as
\begin{equation}\label{hmf}
  \hat{H}_S^* = -\frac{1}{\beta}\text{ln}\left(\frac{\text{Tr}_B[e^{-\beta\hat{H}_{\text{tot}}}]}{\text{Tr}_B[e^{-\beta\hat{H}_B}]} \right)
  = -\frac{1}{\beta}\text{ln}\left(\frac{\text{Tr}_B[e^{-\beta\hat{H}_{\text{tot}}}]}{\mathcal{Z}_B}\right).
\end{equation}
In~\cref{hmf}, $\text{Tr}_B[\bullet]$ stands for tracing out bath's degrees of freedom and $\mathcal{Z}_B=\text{Tr}_B[\text{exp}(-\beta \hat{H}_B)]$ is the partition function
of the bath. It is straightforward to show that in the WC limit, where contribution of the interaction Hamiltonian $\hat{H}_I$ to the equilibrium state 
becomes vanishingly small, $\hat{H}_S^*$ reduces to $\hat{H}_S$ and the equilibrium state becomes the Gibbs state. The partition function
for the HMF is defined as
\begin{equation}\label{partition_mfg}
  \mathcal{Z}_S^*=\text{Tr}_S\left[\text{exp}(-\beta \hat{H}_S^*)\right]=\frac{\mathcal{Z}_{\text{tot}}}{\mathcal{Z}_B},
\end{equation}
where $\mathcal{Z}_{\text{tot}}$ is the partition function associated with $\hat{H}_{\text{tot}}$. Using $\mathcal{Z}_S^*$ and $\hat{H}_S^*$ we can define all relevant thermodynamic
functions such as free energy, internal energy and entropy.
\begin{align}
  &F^*= -\frac{1}{\beta}\text{ln}(\mathcal{Z}_S^*),\label{mf_free_energy}\\
  &U^*= \braket{\hat{H}^*}_S + \beta \braket{\frac{\partial \hat{H}^*}{\partial \beta}}_S,\label{mf_internal_energy}\\
  &S^*= \mathfrak{S}(\hat{\rho}^*) + k_B \beta^2 \braket{\frac{\partial \hat{H}^*}{\partial \beta}}_S.\label{mf_entropy}
\end{align}
In~\cref{mf_internal_energy}, $\braket{\bullet}_S = \text{Tr}_S[\bullet \hat{\rho}^*]$ signifies the expected value of an operator with respect to the MFG and 
in~\cref{mf_entropy} $\mathfrak{S}(\hat{\rho}^*)$ is the von Neumann entropy of the MFG.\\

The SC thermodynamics via MFG presented above describes the thermodynamics of the system microscopically. In other words, thermodynamical description of the system is built from the ground up using the Hamiltonians for the system, bath and their interaction, which govern the microscopic dynamics of the constituents of the entire super-system. There are alternative ways to describe the thermodynamic behavior of small systems which take a phenomenological approach. Two famous examples are Tsallis' extended thermodynamics and Hill's nanothermodynamics~\cite{hill1994thermodynamics,Hill2001,Tsallis1988,tsallis2009introduction}. Both of these methods incorporate the non-extensive nature of thermodynamic functions for small systems but using different strategies but they are closely related~\cite{GARCIAMORALES200582,MANIOTIS2025116285}. In what follows we will give a brief description of Hill's nanothermodynamics.\\

The core idea of Hill's nanothermodynamics or thermodynamics of small systems is that a large ensemble of small systems should yield the same thermodynamic behavior as a large system. Assume that we have a large system composed of many particles (in thermodynamic limit) which is interacting with a thermal bath. In canonical ensemble one can write the Euler equation as
\begin{equation}\label{euler_canonical}
  E = TS - pV.
\end{equation}
Gibbs generalized the canonical ensemble to construct the grand canonical ensemble, in which the system and the bath are allowed to exchange particles as well. For grand canonical ensemble we have
\begin{equation}\label{euler_grand}
  E = TS - pV + \mu N,
\end{equation}
in which for simplicity we assumed that only one type of particle is being exchanged with the bath and $\mu$ is the chemical potential. Now assume that we have a nanosystem in equilibrium with a bath which does not satisfy the thermodynamic limit condition. However, if we take a large ensemble consisting of many replicas of such nanosystems that are not interacting with each other, the entire ensemble will be a large system and regular thermodynamics can be applied. We denote the thermodynamic variables of the ensemble of FS systems using the subscript ``$t$'' ($E_t$, $S_t$, ...) and for a single FS system we use regular symbols ($E$, $S$, ...). Assuming that the total number of nanosystems in the ensemble is $\mathcal{N}$, we can write the euler equation for the ensemble as
\begin{equation}\label{euler_nano_tot}
  E_t = TS_t - pV_t+ \mu N_t + \mathcal{E}\mathcal{N},
\end{equation}
in which $\mathcal{E}$ is called the subdivision potential or replica energy, which is a composite intensive property at the ensemble level and it quantifies changes in energy due to change of the number of the replicas. In short, similar to what Gibbs did to generalize from canonical to grand canonical ensemble to allow for fluctuation of number of particles, Hill's strategy is to add the pair of conjugate thermodynamic variables $(\mathcal{E},\mathcal{N})$ to allow for fluctuation of number of subsystems. We can divide~\cref{euler_nano_tot} by $\mathcal{N}$ and write for a single nanosystem
\begin{equation}\label{euler_nano}
  E = TS - pV + \mu N + \mathcal{E} = U + \mathcal{E},
\end{equation}
in which $E=E_t/\mathcal{N}$, $V=V_t/\mathcal{N}$, $N=N_t/\mathcal{N}$ and $U$ is the internal energy without considering the FS effects. The natural ensemble for such systems is called the unconstrained (also completely open or nanocanonical) ensemble for which all environmental variables are intensive $(\mu,p,T)$. Unlike for the case of macroscopic thermodynamics, thermodynamic description of small systems is ensemble dependent~\cite{hill1994thermodynamics}. The subdivision potential $\mathcal{E}$ accounts for non-additivity of the nanosystem due to it being FS and as it is clear from~\cref{euler_nano}, the average energy is not extensive in the number of particles for such a system. For a large system, subdivision potential approaches zero in the thermodynamic limit. The subdivision potential is temperature dependent and owing to this fact, there exists a close relationship between Landsberg's TDEL and Hill's nanothermodynamics~\cite{C5CP02332G,deMiguel2016,nano10122471,doi:10.1021/acs.jpcb.3c01525}. For such a system with energy levels $E_n(\beta)$ we can write 
\begin{align}
  &F= -\frac{1}{\beta}\text{ln}\mathcal{Z}=-\frac{1}{\beta}\text{ln}\left(\sum_n e^{-\beta E_n}\right),\label{tdel_free_energy}\\
  &U= -\frac{\partial}{\partial \beta}\text{ln}\mathcal{Z} = \sum_n p_n (E_n+\beta\frac{\partial E_n}{\partial \beta}),\label{tdel_internal_energy}\\
  &S= -k_B\sum_n p_n\left[\text{ln}(p_n) + \frac{\partial E_n}{\partial \beta} \right].\label{tdel_entropy}
\end{align}
Note that the definitions given in~\cref{tdel_free_energy} and~\cref{tdel_internal_energy} preserve the Legendre transformation relation between the internal energy and free energy. In~\cref{tdel_internal_energy} we can see that unlike macroscopic systems with temperature-independent energy levels, the thermodynamic internal energy $U$ calculated using the derivative of the partition function for a FS system does not give the system's energy $\expect{\hat{H}}=E$. Using~\cref{tdel_internal_energy} and~\cref{euler_nano} we can establish a relationship between Hill's nanothermodynamics and Landsberg's theory.
\begin{equation}\label{tdel_hill}
  \mathcal{E}(\beta) = -\beta\braket{\frac{\partial\hat{H}}{\partial \beta}} = -\beta \sum_n p_n \frac{\partial E_n}{\partial \beta},
\end{equation}
Note that~\cref{tdel_hill} is not a rigorous equivalence between Landsberg and Hill's theories. It is derived based on an analogy of the mathematical expressions for the internal energy, in which it was assumed that spectral perturbations are the sole source of non-additivity. That being said~\cref{tdel_hill} suggests that for a system with TDEL the subdivision potential $\mathcal{E}$, quantifies the average of contribution of each energy level's temperature dependence. 

\section{Calculating QFI Using TDEL and Nanothermodynamics}
\label{sec:qfi_hill}
For a spin chain in canonical ensemble, a natural choice for the environmental variables are $N$ and $T$. The Euler equation is~\cite{hill1994thermodynamics}
\begin{equation}\label{chain_euler}
  E = TS + \mu N + \mathcal{E} = TS + \hat{\mu} N = TS + F,
\end{equation}
in which $\hat{\mu}$ is the integral chemical potential, $F$ is the Helmholtz free energy and $\mathcal{E}$ is the subdivision potential and $U = TS + \mu N$ is the internal energy. It is evident from~\cref{chain_euler} that the free energy $F$ contains the term $\mathcal{E}$ which makes it non-additive.\\

It is asserted that Hill's theory, not only can describe non-additive thermodynamis of small systems, but it can also shed light on the SC thermodynamic behavior of such systems~\cite{nano10122471,doi:10.1021/acs.jpcb.3c01525}. Different suggestions have been made to capture the non-additive nature of small systems using different definitions for the subdivision potential. The first approach is to define $\mathcal{E}$ using the HMF to establish a microscopic foundation for Hill's nanothermodynamics as applied to systems strongly coupled to environment~\cite{nano10122471}. The second approach is to define $\mathcal{E}$ using the first order perturbation of HMF in temperature~\cite{doi:10.1021/acs.jpcb.3c01525}. The first definition requires microscopic knowledge of the system, bath and their interaction and assumes a priori knowledge of the HMF. The second definition also requires either the full HMF or at the very least a temperature dependent effective Hamiltonian. However, if we know the HMF, there is no need to resort to Hill's nanothermodynamics to describe the thermodynamic behavior of the system. Instead, one can calculate MFG and utilize the methodology of SC thermodynamics in the first part of~\cref{sec:small_thermo} to accurately quantify the thermodynamic behavior of the small system, including its non-additivity. Assuming that we don't have the HMF or an effective temperature dependent Hamiltonian, we must utilize the phenomenological nature of Hill's theory to carry on our calculations.\\

One can connect the thermodynamic relation in~\cref{chain_euler} with statistical mechanics through
\begin{equation}\label{statmech_helmholtz}
  F(\beta,N) = -\frac{1}{\beta}\text{ln}\mathcal{Z} = F_{\text{bulk}}(\beta,N) + \mathcal{E}(\beta) = NF_b(\beta) + \mathcal{E}(\beta),
\end{equation}
in which $F_{\text{bulk}}=NF_b$ is the free energy of the bulk of the system proportional to $N$. It is clear that using~\cref{statmech_helmholtz}, we can find the subdivision potential $\mathcal{E}$ by performing a linear regression on $F$. As explained in~\cref{sec:small_thermo} the HMF will be temperature dependent in general. Following the recipe in~\cite{doi:10.1021/acs.jpcb.3c01525} we can write the HMF as a power series in $T$ (or $1/\beta$), and calculate the system's energy in the SC perturbatively.
\begin{align}\label{hmf_perturbation}
  &\hat{H}_S^*(\beta) = \hat{H}_S + \frac{1}{\beta}\hat{H}_S^{(1)}+(\frac{1}{\beta})^2\hat{H}_S^{(2)}+\ldots,\\
  &E_n^* = E_n + \frac{1}{\beta}E_n^{(1)}+ (\frac{1}{\beta})^2E_n^{(2)}+\ldots
\end{align}
In~\cref{hmf_perturbation} the factors related to the dimensionless interaction strength for each term are absorbed into their respective $\hat{H}^{(n)}$ and $E_n^{(n)}$. Let's only take the first order correction of the energy levels. If we can find the partition function expressed using perturbed energy levels, we can find the classical contribution to QFI using the relationship between the partition function and Fisher information~\cref{fisher_free}. Writing the partition function using the modified energy levels and finding the internal energy using~\cref{tdel_internal_energy} we get
\begin{equation}\label{perturbed_energy}
  U' = -\frac{\partial}{\partial \beta}\text{ln}\mathcal{Z}' = \expect{E_n}' - \frac{1}{\beta}\expect{E_n^{(1)}}' = \expect{E_n^{(0)}}',
\end{equation}
in which the prime symbol on $U'$ and $\mathcal{Z}'$ signifies that they are written using the perturbed energy levels up to the first order and $\expect{\bullet}'$ indicates that the averaging is performed using the probabilities given by the eigenenergies $E'_n$ of the perturbed Hamiltonian $\hat{H}'_S$, containing corrections up to first order. Explicitly we have $p'_n= \text{exp}(-\beta (E_n + \frac{1}{\beta}E_n^{(1)}))/\mathcal{Z}'$. From~\cref{perturbed_energy} we can see that, due to cancellation of temperature dependent terms, the internal energy $U'$ is equal to the average of unperturbed eigenenergies and using~\cref{tdel_internal_energy} one can verify that for a general TDEL or including higher order corrections in~\cref{hmf_perturbation} this cancellation would not happen. However, the averaging is done using $p'_n$ which contain unknown energy corrections and therefore we can't directly find $\expect{E_n}'$. However, using the unperturbed energy levels we can write internal energy as
\begin{equation}\label{unperturbed_energy}
 U = -\frac{\partial}{\partial \beta}\text{ln}\mathcal{Z} = \expect{E_n},
\end{equation}
in which Identifying $\expect{\bullet}$ indicates that the averaging is performed using the probabilities given by the unperturbed energy levels of the system Hamiltonian $p_n = \text{exp}(-\beta E_n)/\mathcal{Z}$. Since we know the unperturbed eigenenergies, we can find $\expect{E_n}$. If the probability distributions $p_n$ and $p'_n$ are equal or reasonably close, we can write $\expect{E_n} = \expect{E_n}'$. For this, the Kullback–Leibler divergence (relative entropy) for the two distributions should be small.
\begin{equation}\label{kl_divergence_prob}
  D_{\text{KL}}(p_n||p'_n) = \text{ln}\left(\frac{\mathcal{Z}'}{\mathcal{Z}} \right) + \sum_n p_n E_n^{(1)}.
\end{equation}
For~\cref{kl_divergence_prob} to be small, the energy shifts should be sufficiently smaller than the unperturbed energies. Our calculations show that the results for $\mathcal{E}$ are small enough for this condition to be satisfied. At first sight this might seem contradictory with the fact that for FS systems $E$ and $\mathcal{E}$ are in general comparable, but considering that we chose only a first order correction to the energy levels (the only case where the problem is tractable), subdivision potential being reasonably small compared to the bare energy is expected. Therefore, for our problem it doesn't matter which probabilities we use for taking expected value and we can disregard the prime symbol in $\expect{\bullet}'$ until the end of this section. Now that we can approximate the internal energy $U=\expect{E_n}$ and identifying $\mathcal{E} = (1/\beta)\expect{E_n^{(1)}}$, we can write the energy of the FS system as $E = U + \mathcal{E}$. The effective contribution of the energy shifts due to SC are contained in $\mathcal{E}$. Using this we need to find $\mathcal{Z}'$ but this is not possible because there is only 1 equation but the number of unknowns is equal to the number of eigenenergies. Hence, we need to find an approximation to $\mathcal{Z}'$. If we assume all energy levels are shifted by the same amount, the probability distribution will not change at all. To be able to proceed we assume
\begin{equation}\label{energy_assumption}
  a = \frac{\mathcal{E}}{U} = \frac{\frac{1}{\beta}\braket{E_n^{(1)}}}{\braket{E_n}} = \frac{\frac{1}{\beta}E_n^{(1)}}{E_n}.
\end{equation}
In~\cref{energy_assumption} we assumed that the energy level shifts are not state dependent and that they only depend on the bare energies via a multiplicative factor relating the internal energy $U$ and subdivision potential $\mathcal{E}$. This assumption is not valid in general and one needs to find state dependent shifts to fully capture corrections in the SC regime. However, it can be thought of as a uniform renormalization of system's energy. Using~\cref{energy_assumption} we can approximate the partition function as
\begin{equation}\label{approx_partition}
  \mathcal{Z}'(\beta,h) \approx \sum_n e^{-\beta(E_n +\frac{1}{\beta}E_n^{(1)})} = \sum_n e^{-\beta(1+a)E_n} = \mathcal{Z}(\beta(1+a),h).
\end{equation}
\cref{approx_partition} states that we can approximate $\mathcal{Z}'$ at $(\beta,h)$ by calculating $\mathcal{Z}$ at $(\beta(1+a),h)$. Additionally, using the linearity of derivative and expected value, we can write for the effective Hamiltonian including corrections up to the first order $\hat{H}'_S$
\begin{equation}\label{qfi_edge}
  \braket{\frac{\partial^2(\beta \hat{H}'_S)}{\partial \alpha^2}} = \frac{\partial^2}{\partial \alpha^2}\braket{\beta\hat{H}'_S} = \frac{\partial^2 (\beta\mathcal{E})}{\partial \alpha^2}.
\end{equation}
Finally, substituting~\cref{qfi_edge} and~\cref{approx_partition} into~\cref{fisher_free} we find the QFI as
\begin{equation}\label{qfic_finite_size}
  \mathcal{F}'(\alpha) = \frac{\partial^2\text{ln}\mathcal{Z}'}{\partial \alpha^2} + \frac{\partial^2 (\beta\mathcal{E})}{\partial \alpha^2}
\end{equation}

\section{Derivation of the QFI for the Equilibrium Probe State}
\label{sec:qfi_calculation}
After finding the eigenvalues and eigenvectors of the equilibrium state, we can proceed to calculate the QFI. As stated in the main body, QFI can be decomposed into its classical and quantum contributions. The classical contribution is due to the parameter dependence of the eigenvalues of the probe state and both positive and negative parity sectors contribute to it. A crucial point to emphasize is that, we diagonalize the Hamiltonian via consecutive application of Jordan-Wigner, Fourier and Bogoliubov-Valatin transformations (all of which are unitary for fermionic systems), the total diagonalizing unitary becomes parameter dependent because Bogoliubov-Valatin transform depends on the parameter $\alpha$.
\begin{equation}\label{tot_transform}
  \hat{U}_{\text{tot}}(\alpha) = \hat{U}_{\text{B}}(\alpha)\hat{U}_{\text{F}}\hat{U}_{\text{J}}.
\end{equation}
This unitary transformation preserves the eigenvalues of the original Hamiltonian but transforms eigenstates and also influences parametric derivatives of the observables including the Hamiltonian.
\begin{align}
&\ket{n}_{\eta} = \hat{U}_{\text{tot}}\ket{n},\label{unitary_eigenstate}\\ 
&\frac{\partial}{\partial \alpha}\hat{\mathcal{H}} = \frac{\partial}{\partial \alpha}\left( \hat{U}_{\text{tot}}\hat{H}\hat{U}_{\text{tot}}^{\dag}\right).\label{unitary_derivative}
\end{align}
Looking at the second term at the right hand side of~\cref{fisher_free}, we need the parametric second derivative of the original Hamiltonian. We can write
\begin{equation}\label{d2bh_avg}
  \braket{\frac{\partial^2(\beta \hat{H})}{\partial \alpha^2}} = \text{Tr}\left[\hat{\varrho}_{\text{eq}}\frac{\partial^2(\beta \hat{H})}{\partial \alpha^2}\right] = \text{Tr}\left[\hat{U}_{\text{tot}}^{\dag}\hat{\rho}_{\text{eq}}\hat{U}_{\text{tot}} \frac{\partial^2(\beta \hat{H})}{\partial \alpha^2}\right] = \text{Tr}\left[\hat{\rho}_{\text{eq}}\hat{U}_{\text{tot}}\frac{\partial^2(\beta \hat{H})}{\partial \alpha^2}\hat{U}_{\text{tot}}^{\dag}\right] = \braket{\hat{U}_{\text{tot}}\frac{\partial^2(\beta \hat{H})}{\partial \alpha^2}\hat{U}_{\text{tot}}^{\dag}}_{\eta}.
\end{equation}
Note that in~\cref{d2bh_avg}, for $\hat{H}=\hat{H}_S$, $\hat{\varrho}_{\text{eq}}=\text{exp}(-\beta \hat{H})/\mathcal{Z}$ signifies the equilibrium state in the original qubit basis but for $\hat{H}=\hat{H}^{\flat}_S$ it denotes the equilibrium state in the polaron basis. $\expect{\bullet}_{\eta}$ denotes averaging using the diagonalized equilibrium density matrix $\hat{\rho}_{\text{eq}}$ which is written in the basis of ${\eta}$-fermions. Essentially for a consistent use of~\cref{fisher_free} when a parameter dependent unitary is involved for diagonalization, one needs to take the parametric derivative in the original basis, apply the same operator on the resulting expression which is used for diagonalizing the Hamiltonian and finally average it using the equilibrium state written in the basis of the diagonalized Hamiltonian. Now we can use~\cref{fisher_free} to obtain the classical contribution. 
\begin{equation}\label{qfic}
  \mathcal{F}^c = \frac{\partial^2\psi}{\partial \alpha^2} + \braket{\hat{U}_{\text{tot}}\frac{\partial^2(\beta \hat{H})}{\partial \alpha^2}\hat{U}_{\text{tot}}^{\dag}}_{\eta} = \frac{\partial^2\psi}{\partial \alpha^2} + \tilde{\mathcal{F}}^c,
\end{equation}
in which $\psi = \text{ln}\mathcal{Z}$ is calculated for the total partition function~\cref{partition_total}. We named the second term at the right hand side of~\cref{qfic} as $\tilde{\mathcal{F}}^c$. At WC for~\cref{xy_hamiltonian}, this term identically vanishes if we take the second derivative with respect to both $h$ and $\beta$. At SC for~\cref{pol_hamiltonian}, the term vanishes for parameter $h$, but doesn't for $\beta$. For this case we provided the details of derivation for $\tilde{\mathcal{F}}^c$ in~\cref{sec:subdiv_fisher_extra}.\\

Now let's turn our attention to the quantum contribution to the QFI. This contribution is due to the parameter dependence of the eigenvectors (via the Bogoliubov angle). We can construct the eigenvectors of the probe in the diagonal basis of the Hamiltonian by acting on the $\eta$-vacuum by the Bogoliubov creation operators. For the positive parity subspace the eigenstates can be written as
\begin{equation}\label{eigenstate_eq}
  \ket{n}_{\eta} = \prod_{k\in S_n}\hat{\eta}_k^{\dag}\ket{0^+}_{\eta} = \hat{\eta}_n^{\dag}\ket{0^+}_{\eta}.
\end{equation}
There are $2^{N-1}$ subsets $S_n$ of $\mathcal{P}_{e}(\mathbf{K^+})$ with even number of elements and therefore $2^{N-1}$ eigenstates with positive parity for $\hat{\mathcal{H}}^+$. Following any one to one labeling procedure we can assign a number $n$ from $1$ to $2^{N-1}$ to these eigenstates based on the number $n$ assigned to the set $S_n$. For convenience we show the $n$'th eigenstate of  $\hat{\mathcal{H}}^+$ in positive parity subspace by $\ket{n}_{\eta}$ and also represent the set of applied creation operators to construct this eigenstate by $\hat{\eta}_n^{\dag}$. Since diagonalization is achieved via a parameter dependent unitary, for a consistent calculation of QFI we need to utilize~\cref{unitary_eigenstate} and use the actual eigenstates of the system $\ket{n}$ in our calculations. For convenience let's call the term $\inner{m}{\dot{n}}$ in~\cref{qfi} ``derivative overlap''. Since $\eta$-vacuum, the Bogoliubov operators and $\hat{U}_{\text{tot}}$ are all parameter dependent we can write the derivative overlap term in QFI as
\begin{equation}\label{overlap}
  \inner{m}{\dot{n}} = \bra{m}\partial_{\alpha}(\hat{U}_{\text{tot}}^{\dag}\hat{\eta}_n^{\dag}\ket{0^+}_{\eta})= \braopket{m}{\hat{U}_{\text{tot}}^{\dag}(\partial_{\alpha}\hat{\eta}_n^{\dag})}{0^+}_{\eta} +
  \braopket{m}{(\partial_{\alpha}\hat{U}_{\text{tot}}^{\dag})\hat{\eta}_n^{\dag}}{0^+}_{\eta} +
  \braopket{m}{\hat{U}_{\text{tot}}^{\dag}\hat{\eta}_n^{\dag}\partial_{\alpha}}{0^+}_{\eta}.
\end{equation}
It is straightforward to show that a parameter independent unitary does not change the QFI. Therefore we can simplify the calculations by considering the effect of parameter dependence of $\hat{U}_B(\alpha)$ on QFI and disregarding the parameter independent unitaries $\hat{U}_J$ and $\hat{U}_F$. Hence we can substitute $\hat{U}_{\text{tot}}$ with $\hat{U}_B(\alpha)$ and the eigenstates $\ket{n}$ with $\ket{\tilde{n}}$ in~\cref{overlap} and get exactly the same result as~\cref{overlap}, in which $\ket{n}_c$ are the eigenstates of the Hamiltonian in the basis of $c$-fermions. 
\begin{equation}\label{c_fermion_eig}
  \ket{n}_{\eta} = \hat{U}_B\ket{n}_c = \hat{U}_{\text{tot}}\ket{n}.
\end{equation}
Implementing the mentioned changes in~\cref{overlap} and substituting in~\cref{qfi}, the quantum contribution to the QFI in the positive parity sector becomes
\begin{align}\label{qfiq_expand}
&\mathcal{F}^{q+} = 2\sum_{n,m} \frac{(p_n - p_m)^2}{p_n + p_m} \left[|{}_{\eta}\braopket{m}{\partial_\alpha\hat{\eta}_n^{\dag}}{0^+}_{\eta}|^2 
+ |{}_{c}\braopket{m}{(\partial_\alpha\hat{U}_{B}^{\dag})\hat{\eta}_n^{\dag}}{0^+}_{\eta}|^2 
+ |{}_{\eta}\braopket{m}{\hat{\eta}_n^{\dag}\partial_\alpha}{0^+}_{\eta}|^2 \right. \notag \\  &\quad\left.+2\mathfrak{Re}\Bigl({}_{\eta}\braopket{m}{\partial_\alpha\hat{\eta}_n^{\dag}}{0^+}_{\eta} {}_{c}\braopket{m}{(\partial_\alpha\hat{U}_{B}^{\dag})\hat{\eta}_n^{\dag}}{0^+}_{\eta}\Bigr)
+ 2\mathfrak{Re}\Bigl({}_{c}\braopket{m}{(\partial_\alpha\hat{U}_{\text{tot}}^{\dag})\hat{\eta}_n^{\dag}}{0^+}_{\eta} {}_{\eta}\braopket{m}{\hat{\eta}_n^{\dag}\partial_\alpha}{0^+}_{\eta}\Bigr)\right. \notag \\  
&\quad\left.+ 2\mathfrak{Re}\Bigl({}_{\eta}\braopket{m}{\partial_\alpha\hat{\eta}_n^{\dag}}{0^+}_{\eta} {}_{\eta}\braopket{m}{\hat{\eta}_n^{\dag}\partial_\alpha}{0^+}_{\eta}\Bigr)
\right].
\end{align}
\cref{qfiq_expand} contains 6 terms. Let's call them $\mathcal{F}_{i}^{q+}$, $i=1,...,6$ by their order of appearance in~\cref{qfiq_expand}. For calculating $\mathcal{F}_{1}^{q+}$ (which is due to the parameter dependence of the creation operators), we need to know the effect  of the derivative 
operator on the Bogoliubov operators. Writing $\hat{\eta}_k^{\dag}$ in terms of $\hat{c}_k^{\dag}$ and $\hat{c}_{-k}$ we get
\begin{equation}\label{der_eta_dag}
  \frac{\partial\hat{\eta}_k^{\dag}}{\partial\alpha} = -\frac{1}{2}(\frac{\partial\theta_k}{\partial\alpha})\text{sin}(\frac{\theta_k}{2})\hat{c}_k^{\dagger}  
+ \frac{i}{2}(\frac{\partial\theta_k}{\partial\alpha})\text{sin}(\frac{\theta_k}{2})\hat{c}_k^{\dagger}
\end{equation}
Using the inverse Bogoliubov-Valatin transformation $\hat{c}_k=\text{cos}(\theta_k/2)\hat{\eta}_k + i\text{sin}(\theta_k/2)\hat{\eta}_{-k}^{\dag}$, we can 
re-express the right hand side of~\cref{der_eta_dag} in terms of $\eta$-fermions. Using this back transformation, the parametric derivative of the annihilation operator for the mode 
$k$ can be expressed in terms of $\eta$-fermions.
\begin{align}\label{der_eta_fin}
  &\frac{\partial\hat{\eta}_k^{\dag}}{\partial\alpha} = \frac{i}{2}\frac{\partial\theta_k}{\partial\alpha}\hat{\eta}_{-k},\\
  &\frac{\partial\hat{\eta}_k}{\partial\alpha} = -\frac{i}{2}\frac{\partial\theta_k}{\partial\alpha}\hat{\eta}_{-k}^{\dag}.
\end{align}
Any eigenstate $\ket{n}_{\eta} = \hat{\eta}_n^{\dag}\ket{0^+}_{\eta}$ mentioned in~\cref{eigenstate_eq} can be implemented as $\hat{\eta}_{k_{\mathfrak{m}}}^{\dag}...\hat{\eta}_{k_{\mathfrak{M}}}^{\dag}\ket{0^+}_{\eta}$, in which $\mathfrak{m}$ and $\mathfrak{M}$ indicate the minimum and maximum of indices of the modes the present in $S_n$. Here we have assumed that all $k\in\mathbf{K^+}$ are assigned index numbers from $1$ to $N$ in ascending order from the minimum to maximum value in $\mathbf{K^+}$ and that elements of $S_n$, which are sorted from the smallest to largest $k$ values, use these indices. Based on~\cref{der_eta_fin} and anticommutativity of the fermionic operators it is clear that upon expansion, any derivative of a such product of creation operators $\partial_{\alpha}(\hat{\eta}_{k_{\mathfrak{m}}}^{\dag}...\hat{\eta}_{k_{\mathfrak{M}}}^{\dag})\ket{0^+}_{\eta}$ will vanish unless for the terms in which the mode of the creation operator whose derivative is taken is equal to the negative of one of the modes in the sequence of creation operators constituting the eigenstate. This statement can be casted in a mathematical form. Assume that we are taking the derivative of the creation operator of the $\eta$-fermion corresponding to the mode $k_r$ within the sequence of modes present in $S_n$ corresponsing to the $n$'th eigenstate. We can write
\begin{equation}\label{der_eig_condition}
  \hat{\eta}_{k_{\mathfrak{m}}}^{\dag}...(\partial_{\alpha}\hat{\eta}_{k_r}^{\dag})...\hat{\eta}_{k_{\mathfrak{M}}}^{\dag}\ket{0^+}_{\eta} = \hat{\eta}_{k_{\mathfrak{m}}}^{\dag}...(\frac{i}{2}\frac{\partial\theta_k}{\partial\alpha})\hat{\eta}_{-k_r}^{\dag}...\hat{\eta}_{k_{\mathfrak{M}}}^{\dag}\ket{0^+}_{\eta}
  \neq 0 \iff \exists l\neq r: k_l = -k_r.
\end{equation}
The derivative overlap term in $\mathcal{F}_{1}^{q+}$ can be written as
\begin{equation}\label{qfiq_term1_overlap}
  |{}_{\eta}\braopket{m}{\partial_\alpha(\prod_{i=k_{\mathfrak{m}}}^{k_{\mathfrak{M}}}\hat{\eta}_i^{\dag})}{0^+}_{\eta}|^2 = |{}_{\eta}\braopket{0^+}{\prod_{j=k_{\mathfrak{m'}}}^{k_{\mathfrak{M'}}}\hat{\eta}_j\partial_\alpha(\prod_{i=k_{\mathfrak{m}}}^{k_{\mathfrak{M}}}\hat{\eta}_i^{\dag})}{0^+}_{\eta}|^2
  =\frac{1}{4}|{}_{\eta}\braopket{0^+}{\prod_{j=k_{\mathfrak{m'}}}^{k_{\mathfrak{M'}}}\hat{\eta}_j\sum_{i=k_{\mathfrak{m}}}^{k_{\mathfrak{M}}}\frac{\partial\theta_i}{\partial\alpha}
  \hat{\eta}_{-i}\prod_{\substack{i'=k_{\mathfrak{m}}\\i'\neq i}}^{k_{\mathfrak{M}}}\hat{\eta}_{i'}^{\dag}}{0^+}_{\eta}|^2.
\end{equation}
Notice that we discarded the imaginary unit given in~\cref{der_eta_fin} when writing~\cref{qfiq_term1_overlap} because in the end we take the square of the modulus of the derivative overlap. For the same reason we don't need to take into account the $-1^{\text{idx}_{S_n}(k_r)}$ factor due to anticommutation of the fermionic operators when we bring $\hat{\eta}_{-k_r}$ in front of the sum, as we did in~\cref{qfiq_term1_overlap}. Here, the operator $\text{idx}_{S_n}(k_r)$ gives the index (position) number of the element $k_r$ within the ordered set $S_n$. In short, for the derivative overlap~\cref{qfiq_term1_overlap} not to vanish, given that the calculation is performed in the positive parity sector, the following conditions must hold.
\begin{enumerate}
  \item If we take a derivative of $\hat{\eta}_{k_l}^{\dag}$ with respect to $\alpha$, $\hat{\eta}_{-k_l}^{\dag}$ also needs to be in the product.
  \item Total number of excitations for both $\ket{m}_{\eta}$ and $\ket{n}_{\eta}$ must be even.
  \item $\ket{m}_{\eta}$ must contain all the remaining excitations of $\ket{n}_{\eta}$ except for $k_l$ and $-k_l$.
\end{enumerate}
The conditions stated above imply that for each choice of $\ket{m}_{\eta}$ and $\ket{n}_{\eta}$ which adhere to these conditions, upon expanding $\partial_\alpha(\prod_{i=k_{\mathfrak{m}}}^{k_{\mathfrak{M}}}\hat{\eta}_i^{\dag})\ket{0^+}_{\eta}$ in~\cref{qfiq_term1_overlap} using the derivative of the product rule, only one term will have a non-vanishing contribution. Since $|\mathbf{K^+}|=N$, the total number of eigenstates with non-vanishing contributions to $\mathcal{F}_{Q1}^{q+}$ can be calculated as follows. We choose a pair of modes ($k_l$ and $-k_l$) for each eigenstate $\ket{n}_{\eta}$ and then we choose an even number of modes from $\mathbf{K^+}\setminus \{k_l,-k_l \}$ for the remaining excitations in $\ket{n}_{\eta}$. Since there are a total of $N/2$ possible pairs, we get
\begin{equation}\label{contributing_eigs}
  \frac{N}{2}\binom{N-2}{0}+\binom{N-2}{2}+ ... + \binom{N-2}{N-2} = \frac{N}{2}2^{N-3}
\end{equation}
Because of the conditions explained above, the expression~\cref{qfiq_term1_overlap} reduces to
\begin{equation}\label{qfiq_term1_over_fin}
  \frac{1}{4}|{}_{\eta}\braopket{0^+}{\prod_{j=k_{\mathfrak{m'}}}^{k_{\mathfrak{M'}}}\hat{\eta}_j\sum_{i=k_{\mathfrak{m}}}^{k_{\mathfrak{M}}}\frac{\partial \theta_k}{\partial \alpha}\hat{\eta}_{-k_r}\prod_{\substack{i'=k_{\mathfrak{m}}\\i'\neq i}}^{k_{\mathfrak{M}}}\hat{\eta}_{i'}^{\dag}}{0^+}_{\eta}|^2=\frac{1}{4}(\frac{\partial\theta_k}{\partial\alpha})^2.
\end{equation}
Therefore, we can write $\mathcal{F}_{1}^{q+}$ as
\begin{equation}\label{qfiq_term1}
  \mathcal{F}_{1}^{q+} = \sum_{n,m}\frac{1}{4}\frac{(p_n-p_m)^2}{p_n+p_m}(\frac{\partial\theta_k}{\partial\alpha})^2
\end{equation}
Since $p_n=\text{exp}\left(-\beta(E_0^+ + \epsilon_n)\right)/\mathcal{Z}$ and $p_m=\text{exp}\left(-\beta(E_0^+ + \epsilon_m)\right)/\mathcal{Z}$ with 
$\epsilon_n = \epsilon_m+\epsilon_k+\epsilon_{-k}=\epsilon_m+2\epsilon_k$, we can write
\begin{equation}\label{qfiq_prob_factor}
  \frac{(p_n-p_m)^2}{p_n+p_m} = \frac{e^{-\beta E_0^+}\text{exp}(-\beta\sum_{\vec{m}^+}\epsilon_m)}{\mathcal{Z}}\frac{(e^{-2\beta\epsilon_k}-1)^2}{e^{-2\beta\epsilon_k}+1},
\end{equation}
in which $\vec{m}^+$ is the vector containing all excitations except $k$ and $-k$. Due to the conditions listed above, the summation in $\text{exp}(-\beta\sum_{\vec{m}^+}\epsilon_m)$ 
is performed over even values of $\sum_{j}m_j=m$. Considering that $\vec{n}^+$ and $\vec{m}^+$ only differ by two vector elements ($n_k$ and $n_{-k}$) 
we can write.
\begin{equation}\label{exp_ground}
  e^{-\beta E_0^+} = \prod_{j}e^{-\beta\epsilon_j/2} = e^{-\beta\epsilon_k/2}e^{-\beta\epsilon_{-k}/2}\prod_{j\setminus \{k,-k\}}e^{-\beta\epsilon_j/2}.
\end{equation}
Using~\cref{exp_ground} along with the method used in the previous section to derive the expression for the partition function we can write
\begin{equation}\label{qfiq_term1_prob}
  e^{-\beta E_0^+}\text{exp}(-\beta\sum_{\vec{m}^+}\epsilon_m) =
  \frac{e^{-\beta\epsilon_k}}{2}\left[ \prod_{j\in\mathbf{K^+}\setminus \{k,-k\}}2\text{cosh}(\frac{\beta\epsilon_j}{2}) + \prod_{j\in\mathbf{K^+}\setminus \{k,-k\}}2\text{sinh}(\frac{\beta\epsilon_j}{2}) \right].
\end{equation}
Defining the expression inside the square brackets in~\cref{qfiq_term1_prob} as $\mathds{Z}_k^+$, we can write
\begin{equation}\label{qfiq_term1_fin}
  \mathcal{F}_{1}^{q+}=\frac{1}{4\mathcal{Z}}\sum_{k\in\mathbf{K_+}}\frac{e^{-\beta\epsilon_k}}{2}\mathds{Z}_k^+\frac{(e^{-2\beta\epsilon_k}-1)^2}
  {e^{-2\beta\epsilon_k}+1}(\frac{\partial\theta_k}{\partial\alpha})^2.
\end{equation}
Now we aim to calculate the second term in~\cref{qfiq_expand}, $\mathcal{F}_{2}^{q+}$ which is due to $\alpha$ dependence of $\hat{U}_{B}$. We can write  $\hat{U}_{B}$ as
\begin{equation}\label{bogoliubov_tot}
   \hat{U}_{B} = \text{exp}\left(i\sum_{k}\theta_k \hat{G}_k \right)=\prod_{k}\hat{U}_{B_k},\quad \hat{G}_k=\hat{c}_{k}^{\dag}\hat{c}_{-k}^{\dag}+\hat{c}_{-k}\hat{c}_{k},
\end{equation}
in which $\hat{G}_k$ is the generator for the transformation for each mode pair $(k,-k)$ and the transformation for this pair is $\hat{U}_{B_k}$. This operator establishes a transformation between $c$ and $\eta$-fermions.
\begin{equation}\label{eta_c_rel}
  \hat{U}_{B_k}\hat{c}_k\hat{U}_{B_k}^{\dag} = \hat{\eta}_k.
\end{equation}
Since the only $\alpha$-dependence of $\hat{U}_{B_k}$ is in $\theta_k(\alpha)$ we can write
\begin{equation}\label{ubk_der}
  \frac{\partial}{\partial \alpha}\hat{U}_{B_k}^{\dag} = \frac{\partial \theta_k}{\partial \alpha}\frac{\partial}{\partial \theta_k}e^{-i\theta_k\hat{G}_k} = -i\frac{\partial \theta_k}{\partial \alpha}\hat{G}_k\hat{U}_{B_k}^{\dag}.
\end{equation}
To calculate $\mathcal{F}_{2}^{q+}$ we proceed to calculate for a single mode $k$
\begin{equation}\label{qfiq_term2_k}
  \left(\frac{\partial}{\partial\alpha}\hat{U}_{B_k}^{\dag}\right)\hat{\eta}_k^{\dag} = -i\frac{\partial \theta_k}{\partial \alpha}\hat{G}_k\hat{U}_{B_k}^{\dag}\hat{\eta}_k^{\dag}=-i\frac{\partial \theta_k}{\partial \alpha}\hat{U}_{B_k}^{\dag}\hat{G}_k\hat{\eta}_k^{\dag}.
\end{equation}
In the last equality we have used the fact that $[\hat{G}_k,\hat{U}_{B_k}^{\dag}]=0$. It is straightforward to show that in the basis of $\eta$-fermions we can express $\hat{G}_k$ as
\begin{equation}\label{generator_eta}
  \hat{G}_k = \hat{\eta}_{k}^{\dag}\hat{\eta}_{-k}^{\dag}+\hat{\eta}_{-k}\hat{\eta}_{k}.
\end{equation}
Substituting~\cref{generator_eta} in~\cref{qfiq_term2_k} and simplifying we get
\begin{equation}\label{qfiq_term2_replace}
  \left(\frac{\partial}{\partial\alpha}\hat{U}_{B_k}^{\dag}\right)\hat{\eta}_k^{\dag} = -i\frac{\partial \theta_k}{\partial \alpha}\hat{U}_{B_k}^{\dag}\hat{\eta}_{-k}(1-\hat{n}_k),
\end{equation}
in which $\hat{n}_k=\hat{\eta}_k^{\dag}\hat{\eta}_k$ is the number operator for the $\eta$-fermion at mode $k$. Now that we have this result for a single $k$ mode we can write for the derivative overlap term in $\mathcal{F}_{Q2}^{q+}$ as
\begin{equation}\label{qfiq_term2_overlap}
  |{}_{c}\braopket{m}{(\partial_\alpha\hat{U}_{B}^{\dag})\prod_{i=k_{\mathfrak{m}}}^{k_{\mathfrak{M}}}\hat{\eta}_i^{\dag}}{0^+}_{\eta}|^2 =|{}_{c}\braopket{m}{\sum_{j=k_1}^{k_{N/2}}\frac{\partial\theta_j}{\partial\alpha}\hat{U}_{B_j}^{\dag}\hat{G}_j\prod_{\substack{j'=k_1\\j'\neq j}}^{k_{N/2}}\hat{U}_{B_{j'}}^{\dag}\prod_{i=k_{\mathfrak{m}}}^{k_{\mathfrak{M}}}\hat{\eta}_i^{\dag}}{0^+}_{\eta}|^2 =
  |{}_{\eta}\braopket{m}{\sum_{j=k_1}^{k_{N/2}}\frac{\partial\theta_j}{\partial\alpha}\hat{G}_j\prod_{i=k_{\mathfrak{m}}}^{k_{\mathfrak{M}}}\hat{\eta}_i^{\dag}}{0^+}_{\eta}|^2.
\end{equation}
In the first equality of~\cref{qfiq_term2_overlap} we have disregarded the imaginary unit coming from~\cref{ubk_der} because it will not effect the final result once the modulus square is taken in the final step. Also the summation runs for each $(-k,k)$ pair so the upper limit is $k_{N/2}$. In the second equality, we have used the fact that for $j\neq j'$ $[\hat{G}_j,\hat{U}_{B_j}^{\dag}]=0$ and utilized~\cref{c_fermion_eig} to transform $\bra{\tilde{m}}$ to $\bra{m'}$. Using~\cref{generator_eta} it is easy to prove that the operators $\hat{G}_k$ and $\hat{\eta}_{k'}^{\dag}$ satisfy the following commutation relation.
\begin{equation}\label{gk_etak_commute}
  [\hat{G}_k,\hat{\eta}_{k'}^{\dag}] = (\delta_{k,k'}-\delta_{k,-k'})\hat{\eta}_{-k'}
\end{equation}
Using~\cref{gk_etak_commute} in~\cref{qfiq_term2_overlap} we can show that
\begin{equation}\label{qfiq_term2_simp}
  |{}_{\eta}\braopket{m}{\sum_{j=k_1}^{k_{N/2}}\frac{\partial\theta_j}{\partial\alpha}\hat{G}_j\prod_{i=k_{\mathfrak{m}}}^{k_{\mathfrak{M}}}\hat{\eta}_i^{\dag}}{0^+}|^2
  =|{}_{\eta}\braopket{0^+}{\prod_{j=k_{\mathfrak{m'}}}^{k_{\mathfrak{M'}}}\hat{\eta}_j\sum_{i=k_{\mathfrak{m}}}^{k_{\mathfrak{M}}}\frac{\partial\theta_i}{\partial\alpha}
  \hat{\eta}_{-i}\prod_{\substack{i'=k_{\mathfrak{m}}\\i'\neq i}}^{k_{\mathfrak{M}}}\hat{\eta}_{i'}^{\dag}}{0^+}_{\eta}|^2 = (\frac{\partial\theta_k}{\partial\alpha})^2,
\end{equation}
in which the last equality is established by comparing~\cref{qfiq_term2_simp} and~\cref{qfiq_term1_overlap}. Since the derivative overlap term for $\mathcal{F}_{1}^{q+}$ and $\mathcal{F}_{2}^{q+}$ differs by a factor of 4, we can write the final result for $\mathcal{F}_{2}^{q+}$ as
\begin{equation}\label{qfiq_term2_fin}
    \mathcal{F}_{2}^{q+}=\frac{1}{\mathcal{Z}}\sum_{k\in\mathbf{K_+}}\frac{e^{-\beta\epsilon_k}}{2}\mathds{Z}_k^+\frac{(e^{-2\beta\epsilon_k}-1)^2}
  {e^{-2\beta\epsilon_k}+1}(\frac{\partial\theta_k}{\partial\alpha})^2.
\end{equation}
Using~\cref{qfiq_term1_overlap} and~\cref{qfiq_term2_simp} we can immediately write an expression for $\mathcal{F}_{Q4}^{q+}$.
\begin{equation}\label{qfiq_term2}
  \mathcal{F}_{4}^{q+}=\frac{1}{\mathcal{Z}}\sum_{k\in\mathbf{K_+}}\frac{e^{-\beta\epsilon_k}}{2}\mathds{Z}_k^+\frac{(e^{-2\beta\epsilon_k}-1)^2}
  {e^{-2\beta\epsilon_k}+1}(\frac{\partial\theta_k}{\partial\alpha})^2.
\end{equation}
Now we turn our attention to $\mathcal{F}_{3}^{q+}$. Knowing that the ground state can be written in a product form as indicated in~\cref{bcs_ground}, we can the following identities.
\begin{equation}\label{qfiq_term3_identities}
\begin{aligned}
  &\hat{\eta}_k^{\dag}\ket{0_k^+}_{\eta} = \ket{1_{k}0_{-k}}_c,\quad \hat{\eta}_k\ket{0_k^+}_{\eta} = 0,\quad {}_{\eta}\braopket{0_k^+}{\hat{\eta}_k\partial_\alpha}{0_k^+}_{\eta}=0,\quad {}_{\eta}\braopket{0_k^+}{\hat{\eta}_k^{\dag}}{0_k^+}_{\eta}=0,\quad {}_{\eta}\braopket{0_k^+}{\partial_\alpha}{0_k^+}_{\eta}=0,\\ &{}_{\eta}\braopket{0_k^+}{\hat{\eta}_k^{\dag}\partial_\alpha}{0_k^+}_{\eta}=0,\quad \hat{\eta}_k^{\dag}\partial_\alpha\ket{0_k^+}_{\eta} = -\frac{\partial\theta_k}{\partial\alpha}\text{sin}(\frac{\theta_k}{2})\text{cos}(\frac{\theta_k}{2})\hat{c}_k^{\dag}\ket{0_{k}0_{-k}}_c,
  \quad \hat{\eta}_k\partial_\alpha\ket{0_k^+}_{\eta} = -\frac{i}{2}\frac{\partial\theta_k}{\partial\alpha}\text{cos}^2(\frac{\theta_k}{2})\hat{c}_{-k}^{\dag}\ket{0_{k}0_{-k}}_c.
\end{aligned}
\end{equation}
We proceed to calculate the contribution of the parameter dependence of the ground state $\ket{0^+}_{\eta}$ to the QFI. The derivative overlap term in $\mathcal{F}_{3}^{q+}$ becomes
\begin{equation}\label{qfiq_term3_overlap}
  |{}_{\eta}\braopket{m}{\hat{\eta}_n^{\dag}\partial_\alpha}{0^+}_{\eta}|^2 = |{}_{\eta}\braopket{0^+}{\prod_{j=k_{\mathfrak{m'}}}^{k_{\mathfrak{M'}}}\hat{\eta}_j\prod_{i=k_{\mathfrak{m}}}^{k_{\mathfrak{M}}}\hat{\eta}_i^{\dag}\partial_\alpha}{0^+}_{\eta}|^2
  =|{}_{\eta}\braopket{0^+}{\prod_{j=k_{\mathfrak{m'}}}^{k_{\mathfrak{M'}}}\prod_{i=k_{\mathfrak{m}}}^{k_{\mathfrak{M}}}\delta_{ij}(1-\hat{n}_i)\partial_\alpha}{0^+}_{\eta}|^2
\end{equation}
In~\cref{qfiq_term3_overlap} it is clear that for the expression not to be trivially zero, for each annihilation operator $\hat{\eta}_j$ there should be a corresponding creation operator $\hat{\eta}_i$ for the same mode and the number of creation and annihilation operators should be equal ($\ket{n}_{\eta}$ and $\ket{m}_{\eta}$ should contain the same excitations). In this case, the probabilities will satisfy $p_n=p_m$ and using~\cref{qfiq_expand} we can say that $\mathcal{F}_{3}^{q+}=0$. We can also deduce that $\mathcal{F}_{5}^{q+}$ vanishes without resorting to the factor containing the probabilities and only by looking at the derivative overlap. Simplifying~\cref{qfiq_term3_overlap} based on the arguments provided and using the fact that the ground state can be written as a product we can write
\begin{equation}\label{qfiq_term3_fin}
  |{}_{\eta}\braopket{m}{\hat{\eta}_n^{\dag}\partial_\alpha}{0^+}_{\eta}|^2 = |\prod_{l\in \mathbf{K^+}}{}_{\eta}\bra{0_l^+}\sum_{q\in \mathbf{K^+}}\partial_\alpha\ket{0_q^+}_{\eta}\prod_{k\in \mathbf{K^+}\setminus \{q\}}\ket{0_k^+}_{\eta}|^2 = 0.
\end{equation}
In~\cref{qfiq_term3_fin} we used an identity from~\cref{qfiq_term3_identities}. This immediately suggests that $\mathcal{F}_{5}^{q+}=\mathcal{F}_{6}^{q+}=0$ as well.
\begin{equation}\label{qfiq_terms34}
  \mathcal{F}_{3}^{q+}=\mathcal{F}_{5}^{q+}=\mathcal{F}_{6}^{q+}=0.
\end{equation}
Doing the same calculations for the negative parity sector we find the total quantum contribution to the QFI. All derivation steps in the negative parity sector are similar to the ones that we have shown for the positive parity sector. The only difference is that in the negative parity sector, quantum contribution to the QFI for modes $0$ and $\pi$ vanishes because for these two modes the Hamiltonian in the basis of $c$-fermions is already diagonal and Bogoliubov-Valatin transform is not required (for $k=0$ and $k=\pi$ their corresponding pair $-k$ does not exist). In other words, in~\cref{diag_hamiltonian}, $\hat{c}_0=\hat{\eta}_0$ and $\hat{c}_{\pi}=\hat{\eta}_{\pi}$. Finally, taking both parity sectors into consideration, we get
\begin{equation}\label{qfiq_tot}
  \mathcal{F}^q = \frac{9}{4\mathcal{Z}}\left[\sum_{k\in\mathbf{K^+}}\frac{e^{-\beta\epsilon_k}}{2}\mathds{Z}_k^+\frac{(e^{-2\beta\epsilon_k}-1)^2}{e^{-2\beta\epsilon_k}+1}(\frac{\partial\theta_k}{\partial\alpha})^2
+\sum_{k\in\mathbf{K^-}\setminus\{0,\pi\}}\frac{e^{-\beta\epsilon_k}}{2}\mathds{Z}_k^-\frac{(e^{-2\beta\epsilon_k}-1)^2}{e^{-2\beta\epsilon_k}+1}(\frac{\partial\theta_k}{\partial\alpha})^2\right].
\end{equation}

\section{Calculating $\tilde{\mathcal{F}}^c$ and Microscopic Subdivision Potential}
\label{sec:subdiv_fisher_extra}
In this section, we calculate the second term in~\cref{qfic}, $\tilde{\mathcal{F}}^c$. After that we will calculate the ``microscopic'' subdivision potential defined as 
\begin{equation}\label{micro_subdivision}
  \mathcal{E}^m = -\beta\braket{\frac{\partial\hat{H}_S^{\flat}}{\partial \beta}}_{\eta},
\end{equation}
and compare it with the phenomenological $\mathcal{E}$ extracted using a linear ansatz as given in~\cref{statmech_helmholtz}. Note that in this section, unlike the previous ones, we used the superscript ``$m$'' to differentiate between the phenomenological and microscopic definitions of the subdivision potential.\\

As argued in~\cref{sec:qfi_calculation}, $\tilde{\mathcal{F}}^c$ is only non-zero in SC for $\hat{H}_S^{\flat}$ given in~\cref{pol_hamiltonian} and only if the differentiation is done with respect to $\beta$. Upon differentiating $\beta\hat{H}_S^{\flat}$ twice with respect to $\beta$ we get
\begin{equation}\label{beta_ham_2diff}
  \frac{\partial^2 (\beta\hat{H}_S^{\flat})}{\partial \beta^2} = -\frac{J}{2}\frac{\partial^2 (\beta\expect{\hat{\mathcal{C}}}^2)}{\partial \beta^2} \sum_{n=1}^{N}(1-\gamma)\hat{\sigma}_n^{y}\hat{\sigma}_{n+1}^{y} - h\frac{\partial^2 (\beta\expect{\hat{\mathcal{C}}})}{\partial \beta^2}\hat{\sigma}_n^{z} = -\frac{J_{\beta}^{(2)}}{2}\sum_{n=1}^{N}(1-\gamma)\hat{\sigma}_n^{y}\hat{\sigma}_{n+1}^{y} - h_{\beta}^{(2)}\hat{\sigma}_n^{z},
\end{equation}
in which $J_{\beta}^{(2)}=\partial_{\beta}^2(\beta\expect{\hat{\mathcal{C}}}^2)J$ and $h_{\beta}^{(2)}=\partial_{\beta}^2(\beta\expect{\hat{\mathcal{C}}})h$. Note that both $\hat{H}_S^{\flat}$ and the operator given in~\cref{beta_ham_2diff} preserve parity (they commute with the parity operator $\hat{\Pi}$). Now, we apply the transformation $\hat{U}_{\text{tot}}(\alpha)$~\cref{tot_transform}, containing the Jordan-Wigner, Fourier and Bogoliubov-Valatin transforms on~\cref{beta_ham_2diff}. We get for the following result after the transformation.
\begin{equation}\label{beta_ham_trans}
  \hat{U}_{\text{tot}}\frac{\partial^2 (\beta\hat{H}_S^{\flat})}{\partial \beta^2}\hat{U}_{\text{tot}}^{\dag} = \sum_{k\notin \{0,\pi \}} \mathcal{A}_k (\hat{\eta}_k^{\dag}\hat{\eta}_k + \hat{\eta}_{-k}^{\dag}\hat{\eta}_{-k}) + \mathcal{B}_k(\hat{\eta}_k^{\dag}\hat{\eta}_{-k}^{\dag} + \hat{\eta}_{-k} \hat{\eta}_k) + \mathcal{C}_k + \sum_{k\in \{0,\pi\}}\Delta_k \hat{\eta}_k^{\dag}\hat{\eta}_k - h_{\beta}^{(2)}N.
\end{equation}
The $k$ values in the right hand side of~\cref{beta_ham_trans} depend on the choice of parity sector. Note that the Bogoliubov transformation used in~\cref{beta_ham_trans} is the same transformation that diagonalizes~\cref{pol_hamiltonian} and not a new transformation aimed for diagonalizing~\cref{beta_ham_2diff}. It is important not to neglect the constant term $-h_{\beta}^{(2)}N$ in~\cref{beta_ham_trans} since it depends on the parameter $\beta$ and it will influence the result of our calculations. The transformed operator given in~\cref{beta_ham_trans} commutes with the parity operator because Bogoliubov transform preserves parity. Therefore, we can express it more accurately by writing it in a way that explicitly shows its parity symmetry. Calling the total transformed operator in~\cref{beta_ham_trans} $\hat{\mathds{H}}$, we can write 
\begin{equation}\label{beta_ham_parity}
\begin{aligned}
  &\hat{\mathds{H}}^+ = \sum_{k\in \mathbf{K^+}} \mathcal{A}_k (\hat{\eta}_k^{\dag}\hat{\eta}_k + \hat{\eta}_{-k}^{\dag}\hat{\eta}_{-k}) + \mathcal{B}_k(\hat{\eta}_k^{\dag}\hat{\eta}_{-k}^{\dag} + \hat{\eta}_{-k} \hat{\eta}_k) + \mathcal{C}_k - Nh_{\beta}^{(2)}\hat{\Pi}^+,\\
  &\hat{\mathds{H}}^- = \sum_{k\in \mathbf{K^-}\setminus \{0,\pi \}} \mathcal{A}_k (\hat{\eta}_k^{\dag}\hat{\eta}_k + \hat{\eta}_{-k}^{\dag}\hat{\eta}_{-k}) + \mathcal{B}_k(\hat{\eta}_k^{\dag}\hat{\eta}_{-k}^{\dag} + \hat{\eta}_{-k} \hat{\eta}_k) + \mathcal{C}_k + \sum_{k\in \{0,\pi\}}\Delta_k \hat{\eta}_k^{\dag}\hat{\eta}_k - Nh_{\beta}^{(2)}\hat{\Pi}^-.
\end{aligned}  
\end{equation}
The parity symmetry implies that $\hat{\mathds{H}}$ can be written as a direct sum of its positive and negative parity sector components $\hat{\mathds{H}} = \hat{\mathds{H}}^+ \oplus \hat{\mathds{H}}^-$. The coefficients introduced in~\cref{beta_ham_trans} and~\cref{beta_ham_parity} are as follows.
\begin{equation}\label{beta_ham_coeffs}
\begin{aligned}
  &\Delta_k = 2(h_{\beta}^{(2)} - \zeta_k),\quad \zeta_k = \frac{J_{\beta}^{(2)}}{2}(1-\gamma)\text{cos}(k),\quad \mathcal{A}_k = \Delta_k\text{cos}(\theta_k^{\flat})-\zeta_k\text{sin}(\theta_k^{\flat}),\\
  &\mathcal{B}_k = i\Delta_k\text{sin}(\theta_k^{\flat}) + \zeta_k (i-\text{cos}(\theta_k^{\flat})\text{tan}(k)),\quad \mathcal{C}_k = \Delta_k(1-\text{cos}(\theta_k^{\flat})) + \zeta_k\text{sin}(\theta_k^{\flat})
\end{aligned}
\end{equation}
Expected value of $\hat{\mathds{H}}$ for the equilibrium state~\cref{eq_state} can be calculated as
\begin{equation}\label{beta_expected_thermal}
\begin{aligned}
  \tilde{\mathcal{F}}^c =& \text{Tr}\left[\hat{\mathds{H}}\frac{e^{-\beta\hat{\mathcal{H}}^{\flat}}}{\mathcal{Z}^{\flat}} \right] = \frac{1}{\mathcal{Z}^{\flat}}\text{Tr}\left[\hat{\Pi}^+(\hat{\mathds{H}}^{+}e^{-\beta\hat{\mathcal{H}}^{\flat+}})\hat{\Pi}^+ + \hat{\Pi}^-(\hat{\mathds{H}}^{-}e^{-\beta\hat{\mathcal{H}}^{\flat-}})\hat{\Pi}^- \right] = \frac{1}{\mathcal{Z}^{\flat}}\text{Tr}\left[\hat{\Pi}^+\hat{\mathds{H}}^{+}\prod_{k\in \mathbf{K^+}}e^{-\beta\epsilon_k^{\flat}(\hat{\eta}_k^{\dag}\hat{\eta}_k-\frac{1}{2})}\right]\\
  &+\frac{1}{\mathcal{Z}^{\flat}}\text{Tr}\left[\hat{\Pi}^-\hat{\mathds{H}}^{-}\prod_{k\in \mathbf{K^-}}e^{-\beta\epsilon_k^{\flat}(\hat{\eta}_k^{\dag}\hat{\eta}_k-\frac{1}{2})}\right].
\end{aligned}
\end{equation}
Although the Bogoliubov transformation doesn't preserve the total particle number, it preserves parity. We can express the parity operator introduced in~\cref{parity_operator} as
\begin{equation}\label{parity_jw}
  \hat{\Pi} = (-1)^{\hat{N}},\quad \hat{\Pi}^{\pm} = \frac{1}{2} \pm \frac{1}{2}\text{exp}\left(i\pi\hat{N}\right),
\end{equation}
in which $\hat{N} = \sum_n \hat{c}_n^{\dag}\hat{c}_n$ is the fermionic number operator and $\hat{c}_n$ is the fermionic annihilation operator at site $n$. Writing the number operator for each mode $k$ in the Fourier space in the basis of $c$-fermions we obtain 
\begin{equation}\label{num_op_fourier}
  \hat{N}_k = \hat{c}_{k}^{\dag}\hat{c}_{k} + \hat{c}_{-k}^{\dag}\hat{c}_{-k}.
\end{equation}
Now we write $\hat{N}_k$ in the basis of Bogoliubov quasiparticles. We get
\begin{equation}\label{num_op_bogoliubov}
  \hat{N}_k = \text{cos}(\theta_k^{\flat})(\hat{\eta}_{k}^{\dag}\hat{\eta}_{k} + \hat{\eta}_{-k}^{\dag}\hat{\eta}_{-k}) + 2i\text{sin}(\theta_k^{\flat})(\hat{\eta}_{k}^{\dag}\hat{\eta}_{-k}^{\dag} + \hat{\eta}_{k}\hat{\eta}_{-k}) + 1 - \text{cos}(\theta_k^{\flat}).
\end{equation}
Since the parity operator only depends on the total number of fermions modulo 2, and the terms $\hat{\eta}_{k}^{\dag}\hat{\eta}_{-k}^{\dag}$ and $\hat{\eta}_{k}\hat{\eta}_{-k}$ change the particle number by 2, their presence does not affect the total parity. Therefore, for simplicity we can ignore them along with the constant term. Using this, we can write the parity operator in the basis of $\eta$-fermions as
\begin{equation}\label{parity_bogoliubov}
  \hat{\Pi}^{\pm} = \frac{1}{2} \pm \frac{1}{2}\text{exp}\left(i\pi\sum_{k\in \mathbf{K^{\pm}}}\hat{\eta}_k^{\dag}\hat{\eta}_k \right),\quad \hat{\Pi} = \frac{1}{2}\left[\text{exp}\left(i\pi\sum_{k\in \mathbf{K^+}}\hat{\eta}_k^{\dag}\hat{\eta}_k \right) + \text{exp}\left(i\pi\sum_{k\in \mathbf{K^-}}\hat{\eta}_k^{\dag}\hat{\eta}_k \right) \right].
\end{equation}

\begin{figure}[htbp!]
  \centering
  \includegraphics[width=0.4\linewidth]{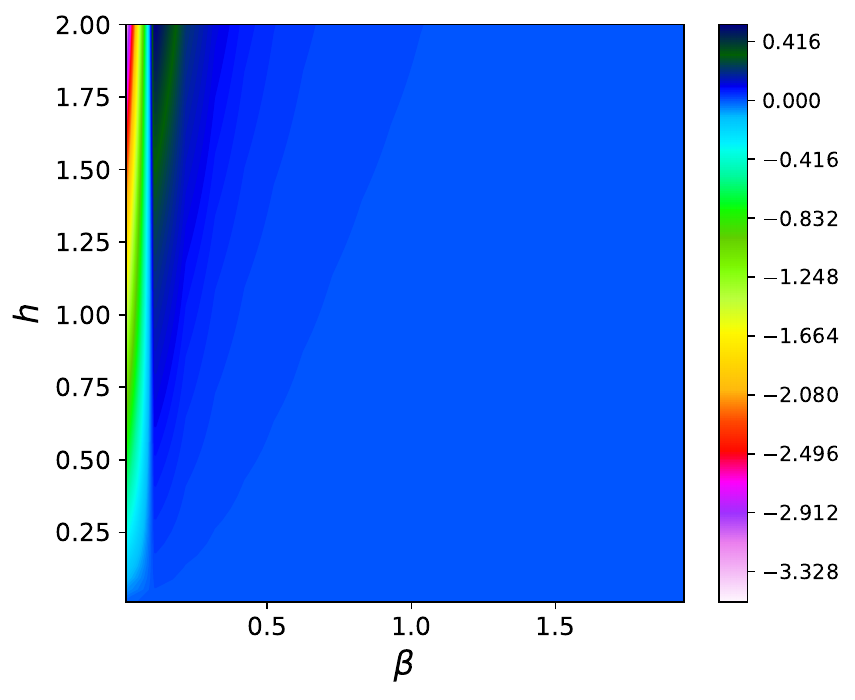}
  \caption{$\tilde{\mathcal{F}}^c(\beta)$ for different values of $\beta$ and $h$. Parameters are $J=1$, $N=8$, $g=0.2$, $\omega_c=50$ and $\gamma = 0.25$}\label{fig:fisher_tilde}
\end{figure}

Using~\cref{parity_bogoliubov} and~\cref{beta_ham_parity} we can calculate the expected value in~\cref{beta_expected_thermal}. We start our calculation in the positive parity sector with the first term of $\hat{\mathds{H}}_{\beta}$.
\begin{equation}\label{term1_beta_expect}
\begin{aligned}
    \frac{1}{\mathcal{Z}^{\flat}}\text{Tr}\left[\hat{\Pi}^+\sum_{j\in \mathbf{K^+}}\mathcal{A}_j\hat{\eta}_j^{\dag}\hat{\eta}_j\prod_{k\in \mathbf{K^+}}e^{-\beta\epsilon_k^{\flat}(\hat{\eta}_k^{\dag}\hat{\eta}_k-\frac{1}{2})}\right] =& \frac{1}{2\mathcal{Z}^{\flat}}\text{Tr}\left[\sum_{j\in \mathbf{K^+}}\mathcal{A}_j\hat{\eta}_j^{\dag}\hat{\eta}_j\prod_{k\in \mathbf{K^+}}e^{-\beta\epsilon_k^{\flat}(\hat{\eta}_k^{\dag}\hat{\eta}_k-\frac{1}{2})}\right]\\
   &+ \frac{1}{2\mathcal{Z}^{\flat}}\text{Tr}\left[e^{i\pi\sum_{k\in \mathbf{K^{\pm}}}\hat{\eta}_k^{\dag}\hat{\eta}_k}\sum_{j\in \mathbf{K^+}}\mathcal{A}_j\hat{\eta}_j^{\dag}\hat{\eta}_j\prod_{k\in \mathbf{K^+}}e^{-\beta\epsilon_k^{\flat}(\hat{\eta}_k^{\dag}\hat{\eta}_k-\frac{1}{2})}\right].
\end{aligned}  
\end{equation}
For the first term in the right hand side of~\cref{term1_beta_expect}, after pulling $\mathcal{A}_j$ out of the trace we can write
\begin{equation}\label{term1p1_beta_expect}
\begin{aligned}
\tilde{\mathcal{F}}^c_1 =& \frac{1}{2\mathcal{Z}^{\flat}}\sum_{j\in \mathbf{K^+}} \mathcal{A}_j \text{Tr}\left[\hat{\eta}_j^{\dag}\hat{\eta}_j \prod_{k\in \mathbf{K^+}}e^{-\beta\epsilon_k^{\flat}(\hat{\eta}_k^{\dag}\hat{\eta}_k-\frac{1}{2})}\right] = \frac{1}{2\mathcal{Z}^{\flat}}\sum_{j\in \mathbf{K^+}} \mathcal{A}_j e^{\frac{-\beta\epsilon_j^{\flat}}{2}} \prod_{k\neq j} 2\text{cosh}\left(\frac{\beta\epsilon_k^{\flat}}{2}\right)\\
&=\frac{1}{2\mathcal{Z}^{\flat}}\left(\sum_{k\in \mathbf{K^+}}\frac{\mathcal{A}_k}{e^{\beta\epsilon_k^{\flat}}+1}\right) \left(\prod_{k\in \mathbf{K^+}}2\text{cosh}\left(\frac{\beta\epsilon_k^{\flat}}{2}\right)\right) = \frac{1}{2\mathcal{Z}^{\flat}}\left(\sum_{k\in \mathbf{K^+}}\frac{\Delta_k\text{cos}(\theta_k^{\flat})}{e^{\beta\epsilon_k^{\flat}}+1}\right) \left(\prod_{k\in \mathbf{K^+}}2\text{cosh}\left(\frac{\beta\epsilon_k^{\flat}}{2}\right)\right).
\end{aligned}  
\end{equation}
For the second term in the right hand side of~\cref{term1_beta_expect}, after pulling $\mathcal{A}_j$ out of the trace and using commutativity of the operators in the arguments of the two exponentials we can write
\begin{equation}\label{term1p2_beta_expect}
\begin{aligned}
\tilde{\mathcal{F}}^c_2 =&\frac{1}{2\mathcal{Z}^{\flat}}\sum_{j\in \mathbf{K^+}} \mathcal{A}_j \text{Tr}\left[\hat{\eta}_j^{\dag}\hat{\eta}_j \prod_{k\in \mathbf{K^+}}e^{(i\pi-\beta\epsilon_k^{\flat})\hat{\eta}_k^{\dag}\hat{\eta}_k-\frac{\beta\epsilon_k^{\flat}}{2}}\right] = \frac{1}{2\mathcal{Z}^{\flat}}\sum_{j\in \mathbf{K^+}} -\mathcal{A}_j e^{\frac{-\beta\epsilon_j^{\flat}}{2}} \prod_{k\neq j} 2\text{sinh}\left(\frac{\beta\epsilon_k^{\flat}}{2}\right)\\
& = \frac{1}{2\mathcal{Z}^{\flat}}\left(\sum_{k\in \mathbf{K^+}}-\frac{\mathcal{A}_k}{e^{\beta\epsilon_k^{\flat}}-1}\right) \left(\prod_{k\in \mathbf{K^+}}2\text{sinh}\left(\frac{\beta\epsilon_k^{\flat}}{2}\right)\right) = \frac{1}{2\mathcal{Z}^{\flat}}\left(\sum_{k\in \mathbf{K^+}}-\frac{\Delta_k\text{cos}(\theta_k^{\flat})}{e^{\beta\epsilon_k^{\flat}}-1}\right) \left(\prod_{k\in \mathbf{K^+}}2\text{sinh}\left(\frac{\beta\epsilon_k^{\flat}}{2}\right)\right).
\end{aligned}  
\end{equation}
It is straightforward to do the same calculations for the negative parity sector. The results are
\begin{equation}\label{qfc_tilde_3_4}
\begin{aligned}
  &\tilde{\mathcal{F}}^c_3 = \frac{1}{2\mathcal{Z}^{\flat}}\left(\sum_{k\in \mathbf{K^-}\setminus \{0,\pi \}}\frac{\Delta_k\text{cos}(\theta_k^{\flat})}{e^{\beta\epsilon_k^{\flat}}+1}+\sum_{k\in \{0,\pi\}}\frac{\Delta_k}{{e^{\beta\epsilon_k^{\flat}}+1}}\right) \left(\prod_{k\in \mathbf{K^-}}2\text{cosh}\left(\frac{\beta\epsilon_k^{\flat}}{2}\right)\right),\\
  &\tilde{\mathcal{F}}^c_4 = \frac{1}{2\mathcal{Z}^{\flat}}\left(\sum_{k\in \mathbf{K^-}\setminus \{0,\pi \}}-\frac{\Delta_k\text{cos}(\theta_k^{\flat})}{e^{\beta\epsilon_k^{\flat}}-1}-\sum_{k\in \{0,\pi\}}\frac{\Delta_k}{{e^{\beta\epsilon_k^{\flat}}-1}}\right) \left(\prod_{k\in \mathbf{K^-}}2\text{sinh}\left(\frac{\beta\epsilon_k^{\flat}}{2}\right)\right).
\end{aligned}
\end{equation}
Using a similar approach, we can show that the contribution of the constant term in~\cref{beta_ham_parity} to~\cref{beta_expected_thermal} for the positive parity sector is 
\begin{equation}\label{const_beta_expect}
\begin{aligned}
  \tilde{\mathcal{F}}^c_5 =& \frac{1}{\mathcal{Z}^{\flat}}\text{Tr}\left[\hat{\Pi}^+\sum_{j\in \mathbf{K^+}}\mathcal{C}_j\prod_{k\in \mathbf{K^+}}e^{-\beta\epsilon_k^{\flat}(\hat{\eta}_k^{\dag}\hat{\eta}_k-\frac{1}{2})}\right] =  \frac{1}{2\mathcal{Z}^{\flat}}\left(\sum_{k\in \mathbf{K^+}}\mathcal{C}_k\right)\left[\left(\prod_{k\in \mathbf{K^+}}2\text{cosh}\left(\frac{\beta\epsilon_k^{\flat}}{2}\right)\right) + \left(\prod_{k\in \mathbf{K^+}}2\text{sinh}\left(\frac{\beta\epsilon_k^{\flat}}{2}\right)\right)\right]\\
  &= \frac{1}{2\mathcal{Z}^{\flat}}\left(\sum_{k\in \mathbf{K^+}}\Delta_k(1-\text{cos}(\theta_k^{\flat}))\right)\left[\left(\prod_{k\in \mathbf{K^+}}2\text{cosh}\left(\frac{\beta\epsilon_k^{\flat}}{2}\right)\right) + \left(\prod_{k\in \mathbf{K^+}}2\text{sinh}\left(\frac{\beta\epsilon_k^{\flat}}{2}\right)\right)\right].
\end{aligned}
\end{equation}
The corresponding contribution for the negative parity sector is 
\begin{equation}\label{wqe}
  \tilde{\mathcal{F}}^c_6 = \frac{1}{2\mathcal{Z}^{\flat}}\left(\sum_{k\in \mathbf{K^-}\setminus \{0,\pi \}} \Delta_k(1-\text{cos}(\theta_k^{\flat}))\right)\left[\left(\prod_{k\in \mathbf{K^-}}2\text{cosh}\left(\frac{\beta\epsilon_k^{\flat}}{2}\right)\right) - \left(\prod_{k\in \mathbf{K^-}}2\text{sinh}\left(\frac{\beta\epsilon_k^{\flat}}{2}\right)\right)\right].
\end{equation}
One can easily show that the contribution to the expected value coming from the $\hat{\eta}_k^{\dag}\hat{\eta}_{-k}^{\dag}$ and $\hat{\eta}_{-k}\hat{\eta}_k$ terms in~\cref{beta_ham_parity} is zero in both positive and negative parity sectors. Finally we can write $\tilde{\mathcal{F}}^c$ as
\begin{equation}\label{qfic_tilde}
  \tilde{\mathcal{F}}^c = \tilde{\mathcal{F}}^c_1 + \tilde{\mathcal{F}}^c_2 + \tilde{\mathcal{F}}^c_3 - \tilde{\mathcal{F}}^c_4 + \tilde{\mathcal{F}}^c_5 + \tilde{\mathcal{F}}^c_6.
\end{equation}
In~\cref{fig:fisher_tilde} we present the result for $\tilde{\mathcal{F}}^c(\beta)$ for a chosen set of parameters. The results clearly show that (compared with $\mathcal{F}^{\flat c}(\beta)$), $\tilde{\mathcal{F}}^c(\beta)$ is only non-negligible for small values of $\beta$, specifically for higher values of $h$.

\begin{figure}[htbp!]
  \centering
  \includegraphics[width=0.8\linewidth]{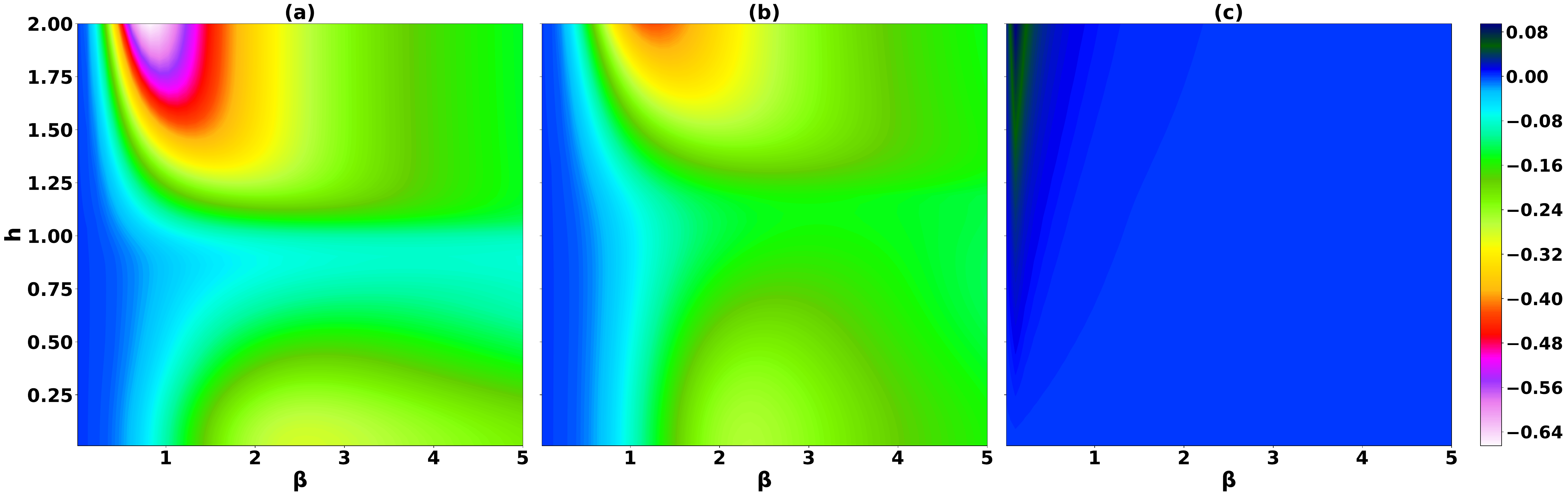}
  \caption{Comparison of subdivision potential calculated for WC and SC using two different methods for different values of $\beta$ and $h$. (a) $\mathcal{E}$, (b) $\mathcal{E}^{\flat}$ and (c)$\mathcal{E}^m$. Parameters are $J=1$, $N=8$ and $\gamma = 0.25$ For calculating $\mathcal{E}^{\flat}$ and $\mathcal{E}^m$ we assumed $g=0.2$ and $\omega_c=50$}\label{fig:subdiv}
\end{figure}
We proceed to calculate the microscopic subdivision potential $\mathcal{E}^m$. As the first step, we calculate the derivative given in~\cref{micro_subdivision}.
\begin{equation}\label{hamiltonian_diff}
  \frac{\partial \hat{H}_S^{\flat}}{\partial \beta} = -\frac{J}{2}\frac{\partial \expect{\hat{\mathcal{C}}}^2}{\partial \beta} \sum_{n=1}^{N}(1-\gamma)\hat{\sigma}_n^{y}\hat{\sigma}_{n+1}^{y} - h\frac{\partial \expect{\hat{\mathcal{C}}}}{\partial \beta}\sigma_n^{z} = -\frac{J_{\beta}^{(1)}}{2}\sum_{n=1}^{N}(1-\gamma)\hat{\sigma}_n^{y}\hat{\sigma}_{n+1}^{y} - h_{\beta}^{(1)}\sigma_n^{z},
\end{equation}
in which $J^{(1)}=\partial_{\beta}\expect{\hat{\mathcal{C}}}^2J$ and $h^{(1)}=\partial_{\beta}\expect{\hat{\mathcal{C}}}h$. The structure of~\cref{hamiltonian_diff} is the same as~\cref{beta_ham_2diff}, the only difference being replacement of $J_{\beta}^{(2)}$ and $h_{\beta}^{(2)}$ with $J^{(1)}$ and $h^{(1)}$ respectively. All calculations from this point onwards is the same as the ones performed starting from~\cref{beta_ham_trans}, but with applying the mentioned parameter replacements. Following this recipe, it is straightforward to carry on the calculations required for obtaining $\mathcal{E}^m$. Below we have provided a comparison between the microscopic subdivision potential $\mathcal{E}^m$ and two other definitions for the subdivision potential. We show them by $\mathcal{E}$ and $\mathcal{E}^{\flat}$ and they are calculated using the linear ansatz given in~\cref{statmech_helmholtz}. For calculating $\mathcal{E}$ and $\mathcal{E}^{\flat}$ we have used $\mathcal{Z}$ and $\mathcal{Z}^{\flat}$ in~\cref{statmech_helmholtz} respectively.

We present our results for the subdivision potential calculated using the mentioned methods in~\cref{fig:subdiv}. The results show that our calculations for phenomenological subdivision potential calculated using the spectrum of the effective Hamiltonian $\mathcal{E}^{\flat}$ and the microscopic subdivision potential $\mathcal{E}^m$ both give larger values across the parameter regime compared with the phenomenological calculation of $\mathcal{E}$ using the spectrum of the bare Hamiltonian. However, comparing~\cref{fig:subdiv}(b) and~\cref{fig:subdiv}(c) we see that even incorporating the temperature-dependent energy levels of the effective Hamiltonian $\hat{H}_S^{\flat}$ in our calculations and utilizing the linear ansatz to calculate $\mathcal{E}^{\flat}$, the microscopic subdivision potential $\mathcal{E}^m$ and $\mathcal{E}^{\flat}$ still yield vastly different results. 
\section{Scaling Ansatz}
\label{sec:ansatz}
In this section, we derive the FS scaling forms for the QFI with respect to the transverse field $h$ (near the quantum critical point) and with respect to the inverse temperature $\beta$ (away from criticality). The derivations are based on the scaling theory of quantum phase transitions and FS scaling~\cite{PhysRev.185.832, PhysRevLett.28.1516,PhysRevB.32.1720,PhysRevB.30.322,PhysRevB.89.094516,ARDOUREL202399,PhysRevB.81.064418, 10.1093/oxfordhb/9780195392043.013.0005, goldenfeld2019lectures}. In general only a term in the classical contribution to the QFI is found using second order parametric derivative of the free entropy~\cref{fisher_free}. However, based on our calculations this term gives the largest contribution to the total QFI and therefore is a suitable starting point for obtaining a scaling ansatz. In the thermodynamic limit, one can write the free energy density (and any other thermodynamic function) as a sum of its regular (analytic) and singular contributions~\cite{goldenfeld2019lectures,10.1093/acprof:oso/9780199577224.001.0001}.
\begin{equation}\label{decomp_n_inf}
 f(T,h) = f_S(T,h) + f_R(T,h).
\end{equation}
For a system in thermodynamic limit and near its critical point, the correlation length diverges according to a power law, which means that the correlation function decays as a power law at the critical point. From $f_S$ one can obtain the critical exponents which govern universal scaling behavior of a macroscopic system near criticality. The anisotropic XY chain under study in this work belongs to the Ising universality class. Therefore, if we were interested in its scaling behavior in the thermodynamic limit, it would suffice to extract the relevant exponents in the Ising universality class. 

However, due to Kadanoff extended singularity theorem~\cite{10.1093/oxfordhb/9780195392043.013.0005,ARDOUREL202399}, $f_S$ cannot be singular unless in the thermodynamic limit where $N\rightarrow\infty$. To put it simply, since the partition function is calculated by summing exponential functions (which are analytic), if the summation is performed only over a finitely many number of exponential functions it is not possible to obtain a singular function as a result. The only way that this can be achieved is in thermodynamic limit where the summation is performed over infinitely many exponentials. Therefore FS systems, even near critical points, do not display singular behavior in their thermodynamic functions. In other words, the FS scaling theory explains how the singularities that would arise in thermodynamic limit, are smoothed out because of the dependence of thermodynamic functions on the ratio of the system size to the intrinsic correlation length. For a spin chain with a fixed lattice spacing, the linear size and number of spins are equivalent $L=N$. We can write the total free energy density as
\begin{equation}\label{decomp_n_fin}
 f(T,h,N) = f_{FS}(T,h,N) + f_R(T,h,N),
\end{equation}
in which $f_{FS}$ is the FS scaling function which in the thermodynamic limit yields $f_S$. According to the FS scaling theory, $f_{FS}$ takes the following form.
\begin{equation}\label{fs_scaling_form}
 f_{FS}(t,\mathfrak{h},N) = b^{-d+z}\Phi(tb^z,\mathfrak{h}b^{1/\nu},ub^{-\omega},N^{-1}b).
\end{equation}
In~\cref{fs_scaling_form} we expressed $f_{FS}$ as a function of reduced temperature and magnetic field $t=(T-T_c)/T_c$ and $\mathfrak{h}=h-h_c$. $\Phi$ is the scaling function near a continuous phase transition which satisfies a homogenous scaling relation under length rescaling by a factor of $b$. $z$ and $\nu$ are the dynamical and correlation length exponents respectively. $u$ is the leading irrelevant coupling, which causes small corrections in the free energy for finite $N$, and $\omega$ is its corresponding exponent (not to be confused with the transition frequency). For the quantum spin chain in our problem the spatial dimension is $d=1$ and the critical exponents are $z=\nu=1$ and $\omega=2$. By choosing $b=N$, the last argument of $\Phi$ becomes constant and after multiplying both sides by $N$ to get the free energy from the free energy density, we can express $F_{FS}$ as 
\begin{equation}\label{fs_scaling_final}
 F_{FS}(t,\mathfrak{h},N) = N^{-1}\Phi(tN,\mathfrak{h}N,uN^{-2}).
\end{equation}
Up to a constant factor, the contribution of $F_{FS}$ to the magnetic susceptibility (equivalently to $\mathcal{F}(h)$ is found as
\begin{equation}\label{fs_scaling_fisher}
 \mathcal{F}_{FS}(h) = -\frac{\partial^2 F_{FS}}{\partial h^2} = N\Phi_{hh}(tN,\mathfrak{h}N,uN^{-2}).
\end{equation}
In~\cref{fs_scaling_fisher} $\Phi_{hh}$ denotes the second partial derivative of $\Phi$ with respect $h$. Note that in the left hand side of~\cref{fs_scaling_fisher}, although $\mathcal{F}_{FS}$ is a function of $h$,$T$ and $N$, we wrote it as $\mathcal{F}_{FS}(h)$ just to show that the Fisher information is calculated with respect to $h$. We will use a similar approach for the argument of QFI for the remainder of this subsection, to be consistent in terms of notation with the rest of the paper. Upon performing Taylor expansion of $\Phi_{hh}$, first with respect to $1/tN$ and then with respect to $uN^{-2}$, for a non-zero $t$ and a large but finite $N$ we get
\begin{equation}\label{fs_taylor_fisher}
 \mathcal{F}_{FS}(h) = \sum_{k=0}^{\infty}\sum_{m=0}^{\infty}\frac{u^m}{T^k}N^{1-k-2m}A_k^{(m)}(\frac{\mathfrak{h}}{T}),
\end{equation}
in which $A_k^{(m)}$ are functions of $\mathfrak{h}/T$. We can see that the largest exponent for $N$ is 1, which occurs for $m=k=0$. The next two largest exponents for $N$ are $0$ and $-1$, respectively. 

The regular part is a non-universal analytic function of $v=1/N$ and therefore can be Taylor expanded as
\begin{equation}\label{regular_taylor}
 f_R(T,h,N) = f_R^{\infty}(T,h) + \sum_{k=1}^{\infty}\frac{1}{k!}\frac{\partial^k f_R}{\partial v^k}\Bigr|_{v=0}v^k.
\end{equation}
Therefore, the contribution to the QFI from this term will be
\begin{equation}\label{regular_taylor_fisher}
 \mathcal{F}_R(T,h,N) = NB_R^{\infty}(T,h) + B_1(h,T)+\frac{B_2(h,T)}{N}+O(N^{-2}).
\end{equation}
Using~\cref{fs_taylor_fisher} and~\cref{regular_taylor_fisher}, we can see that the total scaling function can be written as
\begin{equation}\label{fisher_taylor_beta}
 \mathcal{F}(h) = NC^{\infty}(T,h)+C_1(h,T)+\frac{C_2(h,T)}{N}+O(N^{-2}).
\end{equation}
An important point to consider is that at zero temperature, the free energy reduces to the ground state energy and the expansion~\cref{fs_taylor_fisher} is not valid. In fact, in this case the scaling function will be  from scaling theory, the ground state fidelity susceptibility diverges as
\begin{equation}\label{zero_temp_scaling}
 \mathcal{F}(h)\bigr|_{T=0} \sim N^{\frac{2}{\nu}} = N^2.
\end{equation}
Therefore, for our problem at zero temperature, unlike the case of finite non-zero temperature, it is possible to saturate the Heisenberg limit in magnetometry. Now we consider the scaling function for QFI with respect to the inverse temperature $\beta$ for a system away from the critical point. Away from criticality, in the gapped phase, the free energy is analytic and its FS corrections can be expanded in powers of $1/N$. From the analyticity of the partition function for a finite system, we have
\begin{equation}\label{fs_taylor_nocritical}
 F(T,h,N) = Nf^{\infty}(T,h)+f_1(h,T)+\frac{f_2(h,T)}{N}+O(N^{-2}).
\end{equation}
We find the corresponding contribution of the free energy~\cref{fs_taylor_nocritical} to the QFI as
\begin{equation}\label{fisher_taylor_beta}
 \mathcal{F}(\beta) = ND^{\infty}(T,h)+D_1(h,T)+\frac{D_2(h,T)}{N}+O(N^{-2}).
\end{equation}

\section{Additional Results}
\label{sec:add_res}
\begin{figure}[htbp!]
  \centering
  \includegraphics[width=0.8\linewidth]{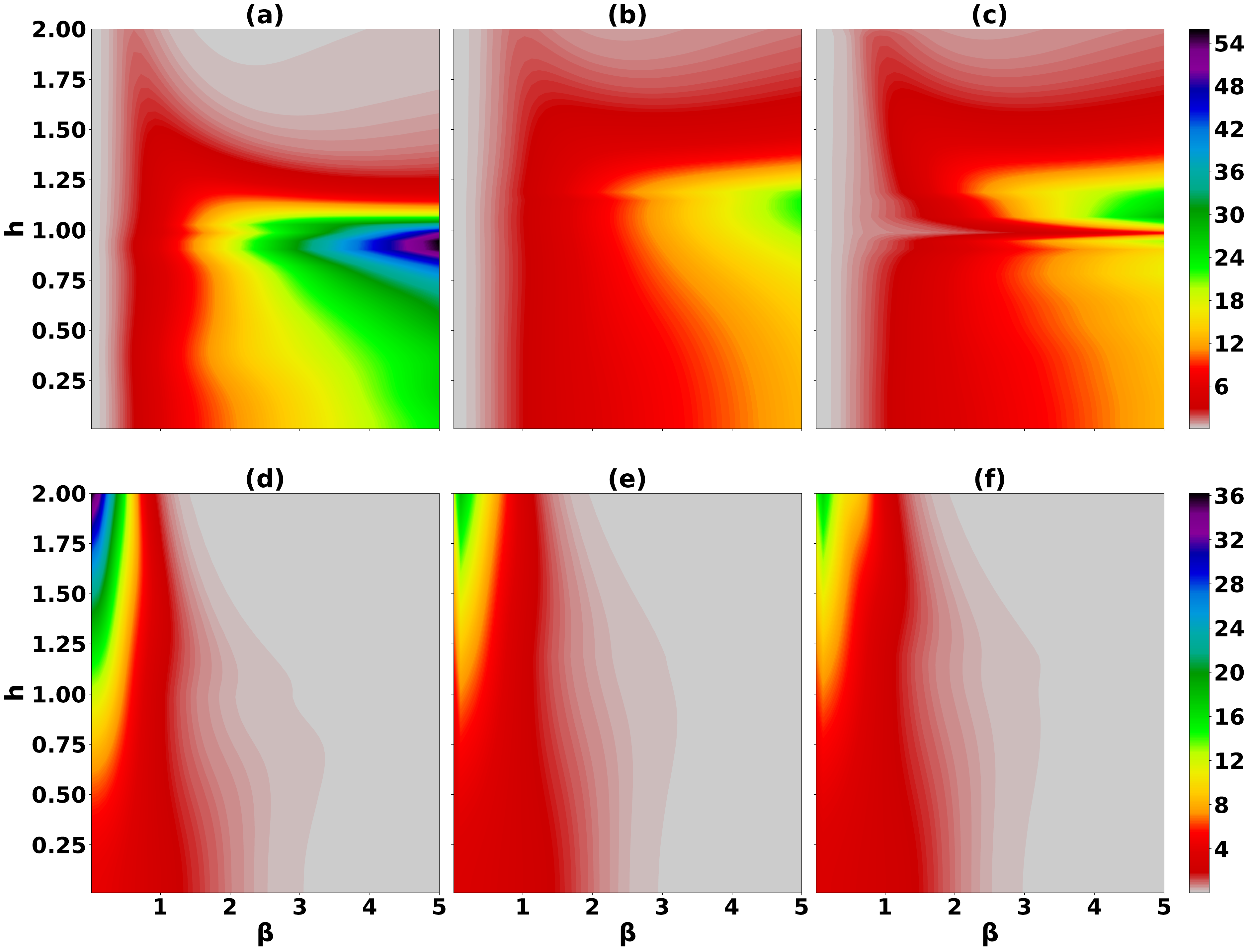}
  \caption{Comparison between QFI calculated for 3 cases. Top panel: QFI calculated for $h$ (a) at WC $\mathcal{F}(h)$, (b) at SC $\mathcal{F}^{\flat}(h)$, (c) using the phenomenological approach $\mathcal{F}^{\flat'}(h)$ incorporating temperature-dependent eigenvalues of the effective Hamiltonian. Bottom panel: QFI calculated for $\beta$ (d) at WC $\mathcal{F}(\beta)$, (e) at SC $\mathcal{F}^{\flat}(\beta)$, (f) using the phenomenological approach $\mathcal{F}^{\flat'}(\beta)$ incorporating temperature-dependent eigenvalues of the effective Hamiltonian. The parameters are $N=8$, $J=1$, $\gamma=0.25$. For calculating $\mathcal{F}^{\flat}(h)$ and $\mathcal{F}^{\flat}(\beta)$ we set $g=0.2$.}\label{fig:full_qfi_sm}
\end{figure}

In this section we provide additional results supporting a number of our assertions in the main body of the manuscript. In~\cref{fig:full_qfi_sm} we present the QFI in WC and SC calculated using microscopically derived effective Hamiltonian along with the phenomenologically calculated QFI. However, as opposed to~\cref{fig:full_qfi}, here we utilize the spectrum of the temperature-dependent effective Hamiltonian given in~\cref{pol_hamiltonian} to calculate the partition function $\mathcal{E}^{\flat}$ in order to obtain the subdivision potential using the linear ansatz~\cref{statmech_helmholtz} via $F^{\flat} = NF_b^{\flat}+\mathcal{E}^{\flat}$. Moreover, the effective partition function introduced in~\cref{approx_partition} is calculated using the temperature-dependent spectrum as $\mathcal{Z}^{\flat'}=\mathcal{Z}^{\flat}(\beta(1+a),h)$. All parameters used in~\cref{fig:full_qfi_sm} is identical to the ones used in~\cref{fig:full_qfi}.
Comparing~\cref{fig:full_qfi_sm}(b) with~\cref{fig:full_qfi_sm}(c) and~\cref{fig:full_qfi_sm}(e) with~\cref{fig:full_qfi_sm}(f) we see that (unlike what we observe in~\cref{fig:full_qfi}) incorporating the spectrum of microscopically derived effective Hamiltonian into our phenomenological calculations forces the outcomes of QFI for both magnetometry and thermometry problems to largely agree at SC. However, it is clear that the results for $\mathcal{F}^{\flat}(h)$ and $\mathcal{F}^{\flat'}(h)$ differ around the phase transition line. Additionally $\mathcal{F}^{\flat}(\beta)$ and $\mathcal{F}^{\flat'}(\beta)$ yield different values in the intermediate values of $h$ and low to intermediate range of $\beta$. In general, the advantage of phenomenological methods is that even without requiring microscopic derivation from first principles, they can effectively describe the behavior of the system. In our problem, as shown in~\cref{fig:full_qfi}), not incorporating the microscopic details of the system at SC into our phenomenological model, results in a considerable disagreement between the accurate results obtained for $\mathcal{F}^{\flat}(\alpha)$ and the phenomenological ones $\mathcal{F}'(\alpha)$. However, if we the HMF or the temperature-dependent effective Hamiltonian is known, as shown in~\cref{fig:full_qfi_sm} one can directly use the microscopic approach to obtain more accurate results.
\begin{figure}[htbp!]
  \centering
  \includegraphics[width=0.5\linewidth]{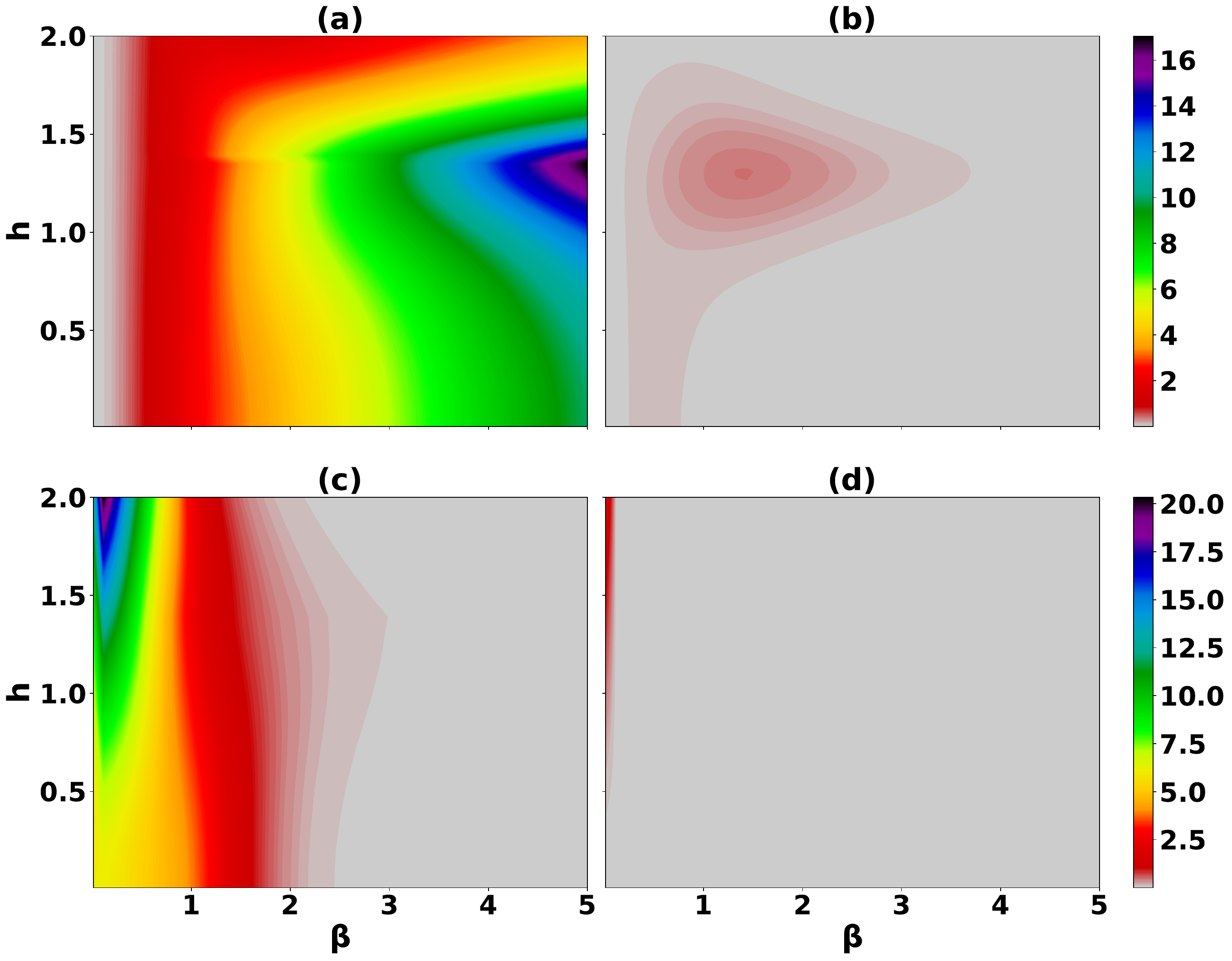}
  \caption{Classical and quantum contributions to the QFI at SC for $\gamma=0.75$, $N=8$, $J=1$. For calculating $\mathcal{F}^{\flat}(h)$ and $\mathcal{F}^{\flat}(\beta)$ we set $g=0.2$. In (a) and (b) $\mathcal{F}^{\flat c}(h)$ and $\mathcal{F}^{\flat q}(h)$ are shown respectively. In (c) and (d) $\mathcal{F}^{\flat c}(\beta)$ and $\mathcal{F}^{\flat q}(\beta)$ are shown respectively.}\label{fig:qfi_sc_sm}
\end{figure}
In~\cref{fig:qfi_sc_sm} we show the results for classical and quantum contributions to the QFI at SC, for a larger value of the anisotropy parameter $\gamma$ compared with the case $\gamma=0.25$ that we assumed in~\cref{fig:qfi_cq1} (all other parameters are identical). Comparing~\cref{fig:qfi_sc_sm}(a) with~\cref{fig:qfi_cq1}(a) and~\cref{fig:qfi_sc_sm}(b) with~\cref{fig:qfi_cq1}(b) we see that the maximum value for both $\mathcal{F}^{\flat c}(h)$ and $\mathcal{F}^{\flat q}(h)$ within the parameter regime of investigation has decreased by increasing $\gamma$, with a shift in the maximum value attained towards a higher value of $h$. This can be understood from the fact that for a fixed value of $\beta$, the expression for $J^{\flat}$ given in~\cref{new_params} increases for larger values of $\gamma$, which pushes the phase transition point $h^{\flat}=J^{\flat}$ towards a larger values of $h$. More specifically, for~\cref{fig:qfi_sc_sm}(a), we see that increasing $\gamma$ results in a reduction/increase in $\mathcal{F}^{\flat c}(h)$ for points in the $(\beta,h)$ plane below/above the phase transition line. Inspecting~\cref{fig:qfi_sc_sm}(c) and~\cref{fig:qfi_sc_sm}(d) and comparing them with their counterparts in~\cref{fig:qfi_cq1} we see that the classical contribution $\mathcal{F}^{\flat c}(\beta)$ increases/decreases for lower/higher ranges of $\beta$. The change in the quantum contribution $\mathcal{F}^{\flat q}(\beta)$ is not significant.  

\begin{figure}[htbp!]
  \centering
  \includegraphics[width=0.5\linewidth]{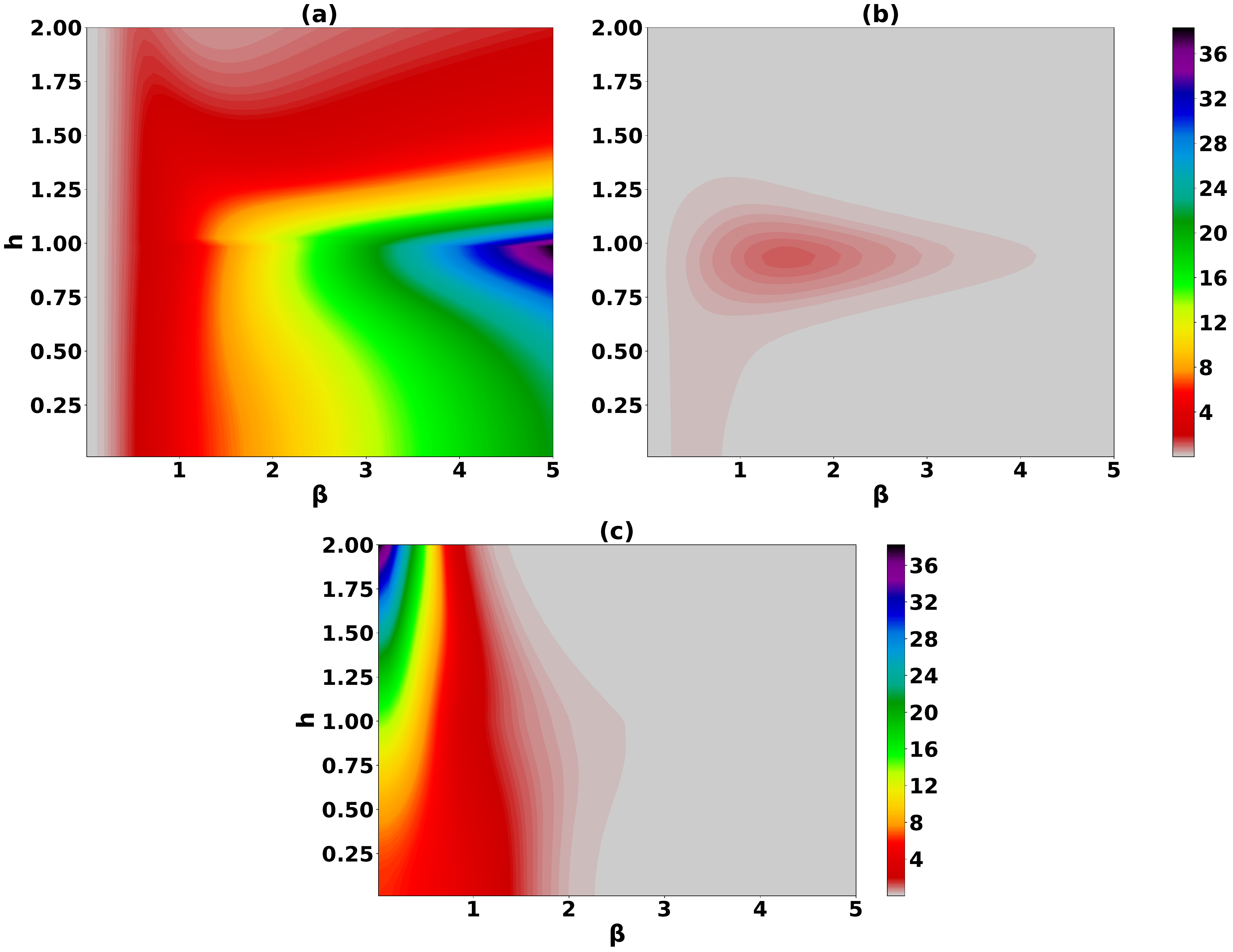}
  \caption{Classical and quantum contributions to the QFI at WC for $\gamma=0.75$, $N=8$, $J=1$. In (a) and (b) $\mathcal{F}^{c}(h)$ and $\mathcal{F}^{q}(h)$ are shown respectively. In (c) $\mathcal{F}^{c}(\beta)$ is shown.}\label{fig:qfi_wc_sm}
\end{figure}

\begin{figure}[htbp!]
  \centering
  \includegraphics[width=\linewidth]{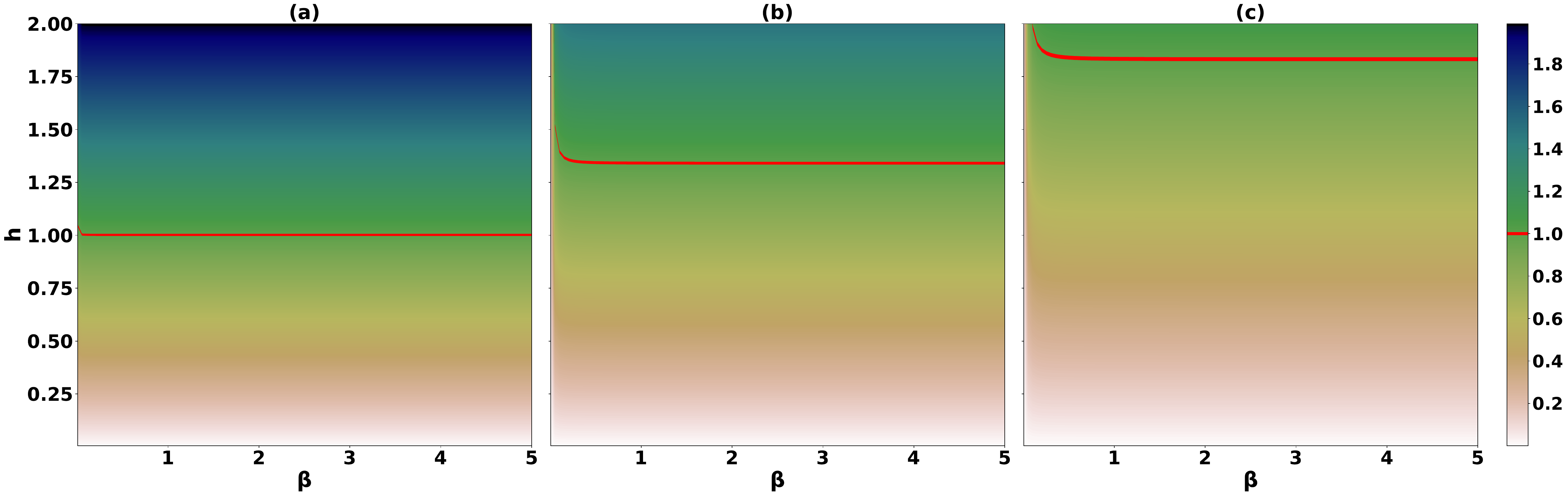}
  \caption{Dependence of the second-order phase transition point on temperature and system-bath coupling strength. The points on the red line correspond to the phase transition points for the given set of parameters. The coupling strength values are (a) $g=0.01$, (b) $g=0.3$, (c) $g=0.5$. The remaining parameters are $\gamma=0.25$, $N=8$, $J=1$, $\omega_c = 50$.}\label{fig:pt_loc}
\end{figure}

For completeness, in~\cref{fig:qfi_wc_sm} results for the classical and quantum contributions to the QFI in the WC limit are shown. Since for the case of WC, the Bogoliubov angles are independent of $\beta$, we have $\mathcal{F}^q(\beta)=0$. Comparing~\cref{fig:qfi_sc_sm}(a) and~\cref{fig:qfi_wc_sm}(a) we see that the peak value of both classical and quantum contributions to the QFI calculated for $h$ are considerably larger in WC limit than the case of SC. Additionally, in general for a vast range $h$ and $\beta$ the values obtained for QFI in both thermometry and magnetometry problems yield a higher value in WC limit than the SC regime. However, we can also see that for small temperature magnetometry, SC with bath can be advantageous. Similarly, comparing~\cref{fig:qfi_sc_sm}(c) and~\cref{fig:qfi_wc_sm}(c) we see that the peak value for QFI calculate for $\beta$ is larger in WC. However, in SC this peak shifts to the larger $\beta$ values, providing advantage in small-temperature thermometry. Comparing~\cref{fig:qfi_wc_sm} and~\cref{fig:qfi_cq2} we see a similar pattern between the classical and quantum contributions of the QFI that we explained for~\cref{fig:qfi_sc_sm}.\\
Finally, we analyze the phase transition of the spin chain at SC with the bath. SC with the local bath causes the phase transition point to change as 
\begin{equation}\label{phase_trans}
  \frac{h}{J} \rightarrow \frac{h^{\flat}}{J^{\flat}} = \frac{\expect{\hat{\mathcal{C}}}h}{\frac{J}{2}\left((1+\gamma)+\expect{\hat{\mathcal{C}}}^2(1-\gamma) \right)},
\end{equation}
in which $\expect{\hat{\mathcal{C}}}$ is defined in~\cref{decay_mag}. The results for three different values of $g$ are given in~\cref{fig:pt_loc}.
The points for which $h^{\flat}/J^{\flat}=1$ are shown in red in~\cref{fig:pt_loc}. The results clearly show that for larger values of $g$, for a fixed value of $\beta$ considered in the $\beta$-$h$ plane, the phase transition point shifts to a higher value of $h$. Therefore, for larger values of system-bath coupling, phase transition to the paramagnetic order occurs for higher magnetic field and lower temperatures. 

\end{document}